\documentclass[edeposit,tocnosub,noragright,centerchapter,12pt]{uiucecethesis09}

\makeatletter
\usepackage{svg}
\usepackage[textsize=scriptsize, colorinlistoftodos,prependcaption]{todonotes}
\usepackage{xargs}  
\usepackage{diagbox}
\usepackage{subcaption}

\usepackage[utf8]{inputenc}
\usepackage{xcolor}
\usepackage{setspace}
\usepackage{epsfig}  %
\usepackage{amsmath}  %
\usepackage{lscape}  %
\usepackage{caption}	%

\usepackage[shortlabels]{enumitem}
\usepackage{bbding}
\usepackage{comment}
\usepackage{booktabs}
\usepackage{multirow}
\usepackage{soul}
\usepackage[linesnumbered,ruled,vlined,resetcount,algochapter]{algorithm2e}
\usepackage{listings}

\makeatletter
\let\old@lstKV@SwitchCases\lstKV@SwitchCases
\def\lstKV@SwitchCases#1#2#3{}
\makeatother
\usepackage{lstlinebgrd}
\makeatletter
\let\lstKV@SwitchCases\old@lstKV@SwitchCases

\lst@Key{numbers}{none}{%
    \def\lst@PlaceNumber{\lst@linebgrd}%
    \lstKV@SwitchCases{#1}%
    {none:\\%
     left:\def\lst@PlaceNumber{\llap{\normalfont
                \lst@numberstyle{\thelstnumber}\kern\lst@numbersep}\lst@linebgrd}\\%
     right:\def\lst@PlaceNumber{\rlap{\normalfont
                \kern\linewidth \kern\lst@numbersep
                \lst@numberstyle{\thelstnumber}}\lst@linebgrd}%
    }{\PackageError{Listings}{Numbers #1 unknown}\@ehc}}
\makeatother

\newcommand\hide[1]{}
\SetCommentSty{mycommfont}
\usepackage{cite}
\usepackage[hyphens]{url}
\usepackage[]{hyperref}         %

\definecolor{ballblue}{rgb}{0.13, 0.67, 0.8}
\definecolor{bleudefrance}{rgb}{0.19, 0.55, 0.91}
\definecolor{altred}{rgb}{0.98,0.14,0.56}
\definecolor{deepred}{rgb}{0.7,0,0}
\definecolor{deepgreen}{rgb}{0,0.5,0}
\definecolor{lightgreen}{rgb}{0,0.6,0}

\makeatletter
\let\orig@lstnumber=\thelstnumber

\newcommand\lstsetnumber[1]{\gdef\thelstnumber{#1}}
\newcommand\lstresetnumber{\global\let\thelstnumber=\orig@lstnumber}
\makeatother

\usepackage[capitalize]{cleveref}
\usepackage{rotating}

\newif\ifsubmit
\submittrue

\ifsubmit
\newcommand{\ssl}[1]{{}}
\newcommand{\kwc}[1]{{#1}}
\newcommand{\kwa}[1]{{}}
\else
\newcommand{\ssl}[1]{{\color{blue}SSL: #1}}
\newcommand{\kwc}[1]{{\color{blue}#1}}
\newcommand{\kwa}[1]{{\color{blue}#1}}
\fi

\usepackage[htt]{hyphenat}

\usepackage[justification=raggedright]{caption}	%

\usepackage{colortbl}
\phdthesis

\title{Code Generation and Runtime Techniques for Enabling Data-Efficient Deep Learning Training on GPUs}
\author{Kun Wu}
\department{Electrical and Computer Engineering}
\degreeyear{2024}

\advisor{Wen-mei Hwu}

\committee{Professor Wen-mei Hwu, Chair\\
        Professor Deming Chen\\
        Associate Professor Steven S. Lumetta\\
        Professor Sanjay Patel}%

\begin{document}

\maketitle

\parindent 1em%

\frontmatter

\setcounter{page}{2}

\begin{abstract}

As deep learning models scale, their training cost has surged significantly. Due to both hardware advancements and limitations in current software stacks, the need for data efficiency has risen. Data efficiency refers to the effective hiding of data access latency and the avoidance of unnecessary data movements. Significant challenges arise from the growing disparity between GPU memory bandwidth and computational throughput, imminent GPU memory capacity limitations, and inefficiencies in the PyTorch software stack, including a lack of device-specific PCIe transfer optimizations and high-level domain-specific abstractions.

To effectively mitigate these data inefficiencies for deep learning training, this dissertation analyzes data inefficiency in representative deep training tasks, specifically in graph neural networks (GNNs) and large language models (LLMs). It then proposes novel runtime and code generation techniques to mitigate these challenges and implements these optimizations seamlessly within the PyTorch stack while maintaining strong programmability and interoperability.

First, Hector intermediate representation (IR) and its code generator are \kwc{devised} to introduce domain-specific high-level abstraction and systematically address memory-intensive performance challenges for relational graph neural networks (RGNNs). The performance challenges stem from RGNN's inherent memory intensiveness, the gap between the programming interface and the kernel APIs, and the high kernel optimization cost due to kernel coupling with layout and heterogeneity. Using a general matrix multiply~(GEMM) template and a traversal template, Hector achieves up to a 43.7$\times$ speed-up in training and inference compared to the state-of-the-art systems. Linear operator reordering and compact tensor materialization further achieve up to 3.8$\times$ speed-up compared to the Hector unoptimized code.

Second, PyTorch-Direct is introduced to incorporate the GPU-centric PCIe data transfer paradigm in PyTorch for GNN training. PyTorch-Direct significantly reduces CPU utilization, resulting in higher end-to-end training performance. For the input datasets and GNN architectures evaluated, PyTorch-Direct decreases the overall training time by up to 38.2\%.

Finally, in LLM training, \kwc{the throughput has been} increasingly constrained by GPU memory \kwc{capacity}. \kwc{To mitigate this}, the SSDTrain \kwc{offloading} framework is \kwc{designed and implemented}. Since activations take most of the GPU memory, SSDTrain offloads activations to \kwc{Non-Volatile Memory Express (NVMe)} SSDs with a direct GPU–SSD data path and good interoperability.  The evaluation shows that SSDTrain reduces activations peak memory use by up to 47\% with negligible overhead. \kwc{We further analyze how the reduced activation memory use may be leveraged to increase throughput by increasing micro-batch size and reducing pipeline parallelism bubbles.}

\kwc{Together, these contributions demonstrate that c}ode generation and runtime techniques can systematically mitigate the data \kwc{management} bottlenecks in deep learning training, which stem from the data-intensive nature of workloads and the oversimplification inherent in the deep learning training software stack.

\end{abstract}

\begin{dedication}
To my parents, for their unconditional love and support.
\end{dedication}

\begin{acknowledgments}

First and foremost, I want to express my heartfelt gratitude to my advisor, Prof.\ Wen-mei Hwu. I was incredibly fortunate to receive an offer from him despite my initial challenges with English and inexperience. From the very beginning, he kept his door open. He incisively urged us to conduct profound research by identifying fundamental and scientific problems in real-world systems. He is a role model, demonstrating how to work as a scholar in all aspects. He showcases the invaluableness of commitment and perseverance. His unwavering support, wisdom, benevolence, and compassion have been constant sources of inspiration. Wen-mei generously provided research assistantship throughout our Ph.D.\ programs. He expanded our networks for collaboration and future careers. I am especially grateful that Wen-mei enabled us to explore potential problems freely and that he is patient with me even though I did not pursue many of the potential projects we found.

I am grateful to my final exam and prelim exam committees for their thoughtful feedback, which strengthened the dissertation.
I thank Prof.~Vikram Adve, Prof.\ Deming Chen, Prof.\ Steve Lumetta, and Prof.\ Sanjay Patel for serving on these committees.

Next, I would like to thank Dr.\ Dejan Milojicic, my internship manager and our collaborator. Like Prof.\ Izzat El Hajj, my internship at HP Labs was a turning point in my Ph.D.\ study. With Dejan's unyielding support and guidance, I practiced driving the research agenda, maintaining focus on key topics, connecting the dots, delivering convincing presentations, coordinating with other teams, etc. All these skills have been instrumental in conducting impactful research and pursuing a Ph.D.\ degree. In addition to his great empowerment, Dejan's insightful vision, whether shared in meetings or public presentations, has been a tremendous source of learning and inspiration.

Special thanks to Prof.\ Sitao Huang, Dr.\ Xiang Song, Dr.\ Xiaofan Zhang, Prof.\ Steve Lumetta, and Dr.\ Seung Won Min for their tremendous help during my search for dissertation topics.
In particular, Sitao led me into compiler research in the Pylog project. Through the PyTorch-Direct project led by Seung Won, I began optimizing the PyTorch stack for graph neural networks. This has been the starting point of our collaboration with the Amazon DGL team, which finally yielded the Hector work.
Without Sitao's and Seung Won's support, this dissertation would not have been possible. 
Xiang provided his utmost support during the ideation and execution of the Hector project, as did Xiaofan and Steve during the SSDTrain project. Their help was critical in laying the foundation for this dissertation.
Moreover, Steve, Xiaofan, and Sitao regularly provided constructive feedback on my dissertation writing, significantly elevating the quality of the final work. Additionally, I am grateful for help from Dr.\ Mert Hidayeto\u{g}lu, Dr.\ Zaid Qureshi, and Dr.\ Vikram Sharma Mailthody during my dissertation research.

Among the many invaluable opportunities that Wen-mei opens up and grants us, one is the chance to collaborate with many highly energetic and extraordinary scholars, including members of the Illinois Microarchitecture Project using Algorithms and Compiler Technology~(IMPACT) Group, professors at the University of Illinois, and industrial scientists.
I enjoyed and learned much from my close collaborators, e.g., Dr.\ Jeongmin Park, Dr.\ Da Zheng, Dr.\ Sai Rahul Chalamalasetti, and Dr.\ Israt Nisa.

I thank my software engineer intern hosts: Dr.\ Aart Bik, Dr.\ Penporn Koanantakool, Jerry Zheng, Dr.\ Howard Chen, James Player, Qingwei Lin, and Bo Qiao. The experience informed me of the focus of companies and, accordingly, how academic research could help or differentiate. Especially my time at Google helped me get started in large language models.

Before embarking on my Ph.D.\ journey, I was lucky to have been guided and mentored by many extraordinary scholars at Tsinghua and the University of California, Santa Barbara. Their insights and support equipped me to face the diverse challenges that arose during my dissertation research. I owe many thanks to Prof.\ Guohao Dai, Prof.\ Xing Hu, Dr.\ Shuangchen Li, Dr.\ Xinfeng Xie, Dr.\ Jie Fu, Prof.\ Yu Wang, Prof.\ Yuan Xie, and Prof.\ Guoqi Li.

Finally, I thank my girlfriend Peizhen Wu, who is also on her way to earning her Ph.D. Her enthusiastic support, belief in me, and knack for pacifying me when I am in trouble have been invaluable. More importantly, she suggests and exemplifies how to navigate the interpersonal aspects of research. 

\end{acknowledgments}

\tableofcontents

\chapter{LIST OF ABBREVIATIONS}

\begin{symbollist*}
\item[AI] Artificial Intelligence
\item[ALU] Arithmetic Logic Unit
\item[API] Application Programming Interface
\item[AT(en)] The ``A TENsor'' Library
\item[B] Byte
\item[BERT] Bidirectional Encoder Representations From Transformers
\item[BMM] Batched Matrix Multiplication
\item[BLAS] Basic Linear Algebra Subroutines
\item[CNN] Convolutional Neural Network
\item[COO] Coordinate Format
\item[CPU] Central Processing Unit
\item[CSR] Compressed Sparse Row
\item[CUDA] Compute Unified Device Architecture
\item[DGL] Deep Graph Library
\item[DMA] Direct Memory Access
\item[DLRM] Deep Learning Recommendation Model
\item[DRAM] Dynamic Random-Access Memory
\item[DSL] Domain-Specific Language
\item[DWPD] Disk Writes Per Day
\item[ETL] Extract, Transform, and Load
\item[FIFO] First In, First Out
\item[FLOP] Floating-Point Operations
\item[FP] Floating Point
\item[G] Billion
\item[GAT] Graph Attention Network
\item[GDS] GPUDirect Storage
\item[GELU] Gaussian Error Linear Unit
\item[GEMM] General Matrix-Matrix Multiplication
\item[GEMV] General Matrix-Vector Multiplication
\item[GIL] Global Interpreter Lock
\item[GNN] Graph Neural Network
\item[GPT] Generative Pre-Trained Transformer
\item[GPU] Graphics Processing Unit
\item[GraphSAGE] The ``Graph SAmple and AggreGatE'' Algorithm
\item[g-SpMM] Generalized SpMM
\item[g-SDDMM] Generalized SDDMM
\item[HBM] High Bandwidth Memory
\item[HGT] Heterogeneous Graph Transformer
\item[HPC] High-Performance Computing
\item[I/O] Input/Output
\item[IPC] Instructions Per Cycle
\item[IR] Intermediate Representation
\item[ISA] Instruction Set Architecture
\item[JEDEC] Joint Electron Device Engineering Council
\item[JESD] JEDEC Standard
\item[JIT] Just-In-Time
\item[K] Thousand
\item[KV] Key-Value
\item[L1/TEX] The Level-One / Texture Cache
\item[Layer Norm] Layer Normalization
\item[LLAMA] Large Language Model Meta AI
\item[LLM] Large Language Model
\item[LSU] Load-Store Unit
\item[M] Million
\item[MLIR] Multi-Level Intermediate Representation
\item[MLC] Multi-Level Cell
\item[MLP] Multilayer Perceptron
\item[MM] Matrix Multiplication
\item[ms] Milliseconds
\item[MoE] Mixture-of-Experts
\item[MSE] Mean Squared Error
\item[NAND] Not-And
\item[NVMe] Non-Volatile Memory Express
\item[NVMe-oF] NVMe over Fabrics
\item[OOM] Out-Of-Memory
\item[P] Quadrillion
\item[PBW] Petabytes Writes
\item[PCIe] Peripheral Component Interconnect Express
\item[P/E] Program/Erase
\item[PEP] Python Enhancement Proposal
\item[PTX] Parallel Thread Execution
\item[PyG] PyTorch Geometric
\item[RAID] Redundant Array of Independent Disks
\item[RAPIDS] Rapid Analytics on Platforms In Data Science
\item[RELU] Rectified Linear Unit
\item[RGAT] Relational Graph Attention Network 
\item[RGCN] Relational Graph Convolutional Network
\item[RGNN] Relational Graph Neural Network
\item[RNG] Random Number Generator
\item[ROK] Recompute-Offload-Keep
\item[s] Seconds
\item[SASS] Streaming ASSembler
\item[SDDMM] Sampled Dense-Dense Matrix Multiplication
\item[SIMT] Single Instruction, Multiple Threads
\item[SLC] Single-Level Cell
\item[SM] Streaming Multiprocessor
\item[SPEC] Standard Performance Evaluation Corporation
\item[SPEC HPG] SPEC High-Performance Group
\item[SPEC OSG] SPEC Open System Group
\item[SPMD] Single Program, Multiple Data
\item[SpMM] Sparse Matrix Dense Matrix Multiplication
\item[SQL] Structured Query Language
\item[SSD] Solid-State Drive
\item[SXM] Server PCIe Module
\item[T] Trillion
\item[T5] Text-To-Text Transfer Transformer
\item[TACO] The Tensor Algebra Compiler
\item[Tanh] Hyperbolic Tangent Function
\item[TC] Tensor Core
\item[TCO] Total Cost of Ownership
\item[THP] TorcH Python
\item[TLC] Triple-Level Cell
\item[TPU] Tensor Processing Unit
\item[TVM] Tensor Virtual Machine
\item[us] Microseconds
\item[UVM] Unified Virtual Memory
\item[WAF] Write Amplification Factor
\item[XLA] Accelerated Linear Algebra
\item[XU] Transcendental and Data Type Conversion Unit
\item[ZeRO] Zero Redundancy Optimizer
\item[ZNS] Zoned Namespaces
\end{symbollist*}

\mainmatter

\chapter{Introduction}

In recent years, deep learning models have demonstrated remarkable capabilities in learning from vast amounts of data, leading to broad adoption and superior performance in various real-world applications, e.g., recommender systems~\cite{Pinterest,naumovDeepLearningRecommendation2019}, content creation~\cite{openaiChatGPT2022,midjourneyMidjourney2022}, etc. As these models continue to achieve transformative results, their scale has proliferated to further enhance their competence. However, this increase in model complexity has caused a significant rise in training cost: The training cost of frontier models has grown at 2.4$\times$ annually in the last eight years. For example, training GPT-4, one of the largest models to date, incurred approximately US\$100 million. Notably, the growth in training \kwc{compute} cost has outpaced that of inference, \kwc{with the former being nearly double the latter}~\cite{villalobosTradingComputeTraining2023}.

As a result, deep learning training has increasingly posed pressing data efficiency challenges to the software computing stack. Data efficiency means effectively hiding data access latency and avoiding unnecessary data accesses. Several factors contribute to the growing importance of data efficiency. 

\begin{figure}[]
    \centering
    \includegraphics[width=0.8\linewidth]{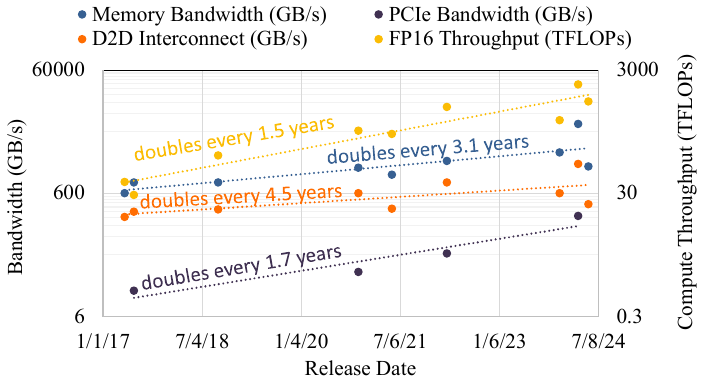}
    \caption{Trend of recent GPUs for deep learning. We collect the inter-device~(D2D) bandwidth, PCIe bandwidth, memory bandwidth, and floating-point throughput of Nvidia 100-level GPUs since Kepler and Google TPUs~\cite{epochParameterComputeData,techpowerupGPUSpecsDatabase2024,jouppiTPUV4,timothyprickettmorganLotsQuestionsGoogle2024,smithNVIDIABlackwellArchitecture2024,TensorProcessingUnit2017,jouppiInDatacenterPerformanceAnalysis2017}. }
    \label{fig:gpu_trend_various}
\end{figure}

\begin{figure}[]
    \centering
    \includegraphics[width=0.8\linewidth]{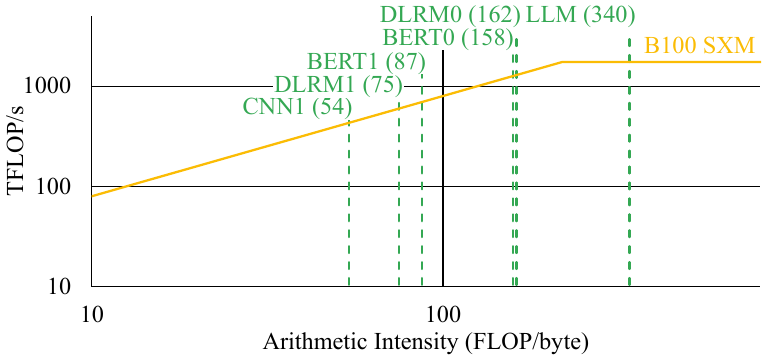}
    \caption{Comparison of the memory bandwidth and FP16 throughput of Nvidia B100 SXM~\cite{smithNVIDIABlackwellArchitecture2024} with the arithmetic intensity of Google internal production workloads~\cite{jouppiTPUV4}. }
    \label{fig:roofline}
\end{figure}

First, the rapid growth of large-scale models has driven an increase in GPU computational power that far outpaces improvements in data transfer bandwidth. As shown in Figure~\ref{fig:gpu_trend_various}, for recent GPUs for deep learning, FP16 throughput~(yellow dotted line, right vertical axis) has increased not only faster than memory bandwidth~(blue dotted line, left vertical axis), but also faster than the device's PCIe bandwidth~(purple dotted line, left vertical axis) and device-to-device interconnect bandwidth~(orange dotted line, left vertical axis). Moreover, hardware-accelerated lower-precision computation has enlarged the gap between the computational throughput and memory bandwidth: FP16 multiply costs 70\% less energy compared with FP32 multiply~\cite{billdallyDeepLearningHardware2022}. At the same time, FP16 data transfers only reduce energy usage by 50\% due to their proportionality to transfer size. Consequently, the training processes of most deep learning models are memory-bound. This is illustrated in Figure~\ref{fig:roofline}, where the characteristics of Nvidia B100 are compared with the deep learning training production workloads reported by Google TPU architects~\cite{jouppiTPUV4}. All models except the reported LLM workload are bound by memory. Even in LLM training, memory-intensive operations account for a significant portion of the overall execution time~\cite{liLLMAnalysisLatencyMemory2023,yuanLLMInferenceUnveiled2024}.

Second, GPU memory capacity alone cannot sustain the growth in computational throughput, necessitating additional buffer and domain-specific PCIe transfer optimizations. Due to the limited capacity of GPU memory, deep learning training predominantly relies on mini-batches, where the entire training dataset is stored outside the GPU, and only a small subset is transferred during each step. For graph neural networks~(GNNs), the generic mini-batch transfer scheme---where the CPU prepares the mini-batch input and initiates direct memory access (DMA) PCIe transfer---introduces significant performance overhead due to fine-grained, gather-style random accesses. This can even lead to the loss of scalability~\cite{minLargeGraphConvolutional2021}. As EMOGI demonstrates~\cite{minEMOGIEfficientMemoryaccess2020,minLargeGraphConvolutional2021,minFinegrainedMemoryAccess2022}, an optimized GPU-centric transfer scheme avoids these issues by programming the GPU to use zero-copy techniques, allowing it to gather features and perform PCIe transfers simultaneously.

\begin{figure}[]
    \centering
    \includegraphics[width=0.8\linewidth]{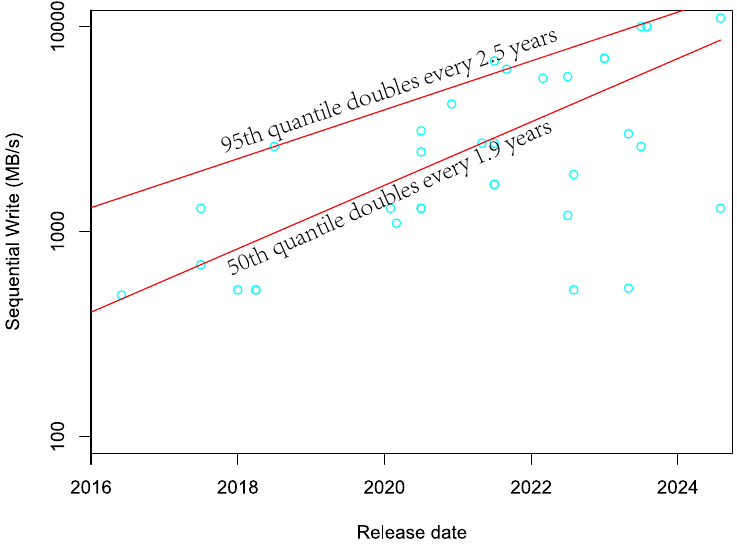}
    \caption{The trend of enterprise SSD sequential write bandwidth~\cite{techpowerupEnterpriseSSDDatabase2024}. For each SSD model, only the data of the variant with maximal capacity is collected. Red lines show the growth rates predicted by quantile regression. The visualization code is adapted from Derek Jones's work~\cite{derekjonesShapeCodeMemory2020}. }
    \label{fig:ssd_trend_bandwidth}
\end{figure}

For large language models~(LLMs), the growth in GPU memory capacity and main memory capacity has struggled to keep up with the increasing demands driven by GPU throughput. In contrast, SSDs offer large storage capacity, and their growth has kept up with these demands. Section~\ref{sec:llm_scaling} details the reasoning. Given this, we choose to offload tensors to SSDs for LLM training to overcome GPU memory limitations. Nevertheless, SSD bandwidth is limited, and the gap between SSD bandwidth growth and GPU computational throughput growth continues to widen, as shown in Figures~\ref{fig:gpu_trend_various} and~\ref{fig:ssd_trend_bandwidth}. It is essential to carefully manage data transfers to prevent training throughput from being constrained by SSD bandwidth. We propose techniques for selecting which tensors to offload and hiding transfer latency. To evaluate the trade-off between performance and memory savings, we compare different strategies, i.e., tensor recomputation, offloading, and keeping tensors in GPU memory. These are detailed in Chapter~\ref{ch:ssdtrain}.

\kwc{Despite the importance of data efficiency, several obstacles exist to address it within the current PyTorch-based deep learning training software stack.}
As one of the most popular deep learning frameworks, PyTorch offers an intuitive interface through the dynamic Python language. It abstracts away the complexity of CUDA-accelerated systems, making them user-friendly and fostering a robust ecosystem in the deep learning community. However, PyTorch is by no means a ``silver bullet''~\cite{brooksNoSilverBullet1987}. Instead, the PyTorch stack design is largely compute-oriented. This focus creates significant challenges when attempting to tackle data efficiency, a new paradigm requiring optimizing data access alongside computation.

\kwc{One significant challenge posed by PyTorch's compute-centric design, which we encountered while integrating the EMOGI PCIe transfer scheme}~\cite{minEMOGIEfficientMemoryaccess2020,minLargeGraphConvolutional2021,minFinegrainedMemoryAccess2022} \kwc{for GNNs, is its assumption that both the input and output of each operator reside on the same device to which the operator is dispatched.} However, the optimized transfer scheme uses the GPU to gather node features and perform PCIe transfer simultaneously as an integral process, demanding the input to be in the host memory.
To adopt such a data-efficient scheme and retain the PyTorch programming interface, the PyTorch runtime code has to be recompiled with the addition of a full-fledged new tensor type, its special host memory allocator, and its set of new dispatch rules. 
Chapter~\ref{ch:pytorch_direct} details our solution for incorporating EMOGI PCIe transfer into PyTorch.

Second, although PyTorch incorporates high-performance math libraries and uses them for corresponding operators~\cite{CuBLAS,CUTLASS2022}, it lacks high-level abstraction to capture domain-specific semantics. This limitation makes it difficult to safely optimize the code to eliminate redundant data movement and achieve more efficient execution schedules. For example, in relational GNNs~(RGNNs), a common operation is producing a per-edge vector, edge message, by multiplying the source node features with a weight matrix specific to the edge type. Since edges with the same edge type and source node will get the same edge message as the result, repetitive computation and output footprint can be eliminated. However, existing frameworks with generic GNN abstraction cannot leverage these optimization opportunities because they lack the necessary abstraction to capture and track edge-type-specific information. Chapter~\ref{ch:hector} details how a code generator with domain-specific intermediate representation~(IR) enables the optimization discussed above, compact materialization.

\kwc{In this thesis, we show that c}ode generation and runtime techniques can systematically mitigate the data \kwc{management} bottlenecks in deep learning training, which stem from the data-intensive nature of workloads and the oversimplification inherent in the deep learning training software stack.

To prove the dissertation statement, the dissertation examines the data inefficiency in representative scenarios in GNNs and LLMs, proposes runtime and code generation techniques to mitigate such inefficiency, and implements transparent incorporation into the PyTorch stack with good programmability and interoperability. 
The contributions of this dissertation are as follows:

\begin{itemize}
\item Hector IR and code generator for end-to-end RGNN training and inference~\cite{wuHectorEfficientProgramming2024}. RGNN execution faces significant performance challenges due to inherent memory intensiveness, the gap between the programming interface and the kernel APIs, and the high kernel optimization cost due to kernel coupling with layout and heterogeneity. To systematically address these issues, we present Hector. \kwc{Hector generates optimized CUDA kernels to eliminate redundant data movement within GPU and reduces GPU memory footprint.} The IR design decouples the model semantics, data layout, and operator-specific schedule and expresses these opportunities to integrate them into the design space. Based on a general matrix multiply~(GEMM) template and a traversal template, Hector already achieves up to 43.7$\times$ speed-up in training and inference compared to state-of-the-art systems. Linear operator reordering and compact tensor materialization obtain up to 3.8$\times$ speed-up compared to the Hector unoptimized code. Chapter~\ref{ch:hector} details Hector.
\item PyTorch-Direct, a GPU-centric data access paradigm for GNN training~\cite{min2021pytorchdirect,minLargeGraphConvolutional2021,minGraphNeuralNetwork2022}. Training GNNs on large graphs that do not fit in GPU memory suffers from significant throughput and CPU utilization overhead. By enabling GPUs to efficiently access complicated data structures in host memory directly without CPU intervention, PyTorch-Direct significantly reduces CPU utilization in GNN training, resulting in higher end-to-end training performance. For the input datasets and GNN architectures evaluated, PyTorch-Direct decreases the overall training time by up to 38.2\%. One of its key advantages is the minimal required programmer effort: Users can take full advantage of the benefits that PyTorch-Direct provides by modifying at most two lines of their original code. Chapter~\ref{ch:pytorch_direct} details PyTorch-Direct.
\item SSDTrain activations\footnote{In deep learning, activations are the tensors produced in forward propagation to be used for gradient computation in the backward propagation.} offloading framework for LLM training~\cite{wuTBAFasterLarge2024}. \kwc{After mitigating the data inefficiency in CUDA kernels and PCIe transfers, we take the next step to address higher-level data inefficiency, particularly challenges in overlapping kernels and transfers. Notice that} LLM training systems are increasingly constrained by GPU memory, with activations being one of the primary culprits. We propose SSDTrain to address this by offloading activations to Non-Volatile Memory Express (NVMe) SSDs. We demonstrate its viability in large-scale systems by modeling. We incorporate into SSDTrain a direct GPU–SSD data path and good interoperability. To fully overlap computation with data transfer, SSDTrain features asynchronous data transfer, tensor deduplication, forwarding, and adaptive offloading. The evaluation shows SSDTrain reduces the activations peak memory use by up to 47\% with negligible overhead. We introduce the recompute-offload-keep (ROK) curve to show runs with SSDTrain's offloading are on the efficient frontier in the design space. \kwc{We further analyze how the reduced activation memory use may lead to increased throughput by increasing micro-batch size and reducing pipeline parallelism bubbles.} Chapter~\ref{ch:ssdtrain} details SSDTrain.
\end{itemize}

The remaining chapters serve the following purposes:

\begin{itemize}
    \item Chapter~\ref{ch:background} introduces the background of this dissertation, involving GNNs, LLMs, the Nvidia GPU architecture and programming model, and the PyTorch computing stack.
    \item Chapter~\ref{ch:future_work} \kwc{presents a final discussion on PyTorch-Direct, Hector, and SSDTrain. Then, it explains the future work on top of this dissertation.}
    \item Chapter~\ref{ch:conclusion} concludes this dissertation.
\end{itemize}

\chapter{Background}
\label{ch:background}

\kwc{This chapter provides the background knowledge necessary for understanding the subsequent chapters. 
Readers may choose one or more sections to read for a particular chapter or skip the sections they are familiar with.
Section~\ref{sec:bg_gnn} introduces GNNs, which are the focus of \cref{ch:hector,ch:pytorch_direct}.
Section~\ref{sec:bg_transformers} delves into transformers, offering context for Chapter~\ref{ch:ssdtrain}.
Section~\ref{sec:bg_cuda} covers the architecture, programming model, programming interface and compilation flow for Nvidia GPUs.
Section~\ref{sec:bg_python} overviews the Python language.
Section~\ref{sec:bg_pytorch} introduces the PyTorch computing stack.

}

\section{Graph Neural Networks}
\label{sec:bg_gnn}
Inspired by the success of convolutional neural networks~(CNNs)~\cite{CNN0}, people devised GNNs as a new type of neural network that applies similar filters to graphs~\cite{lecunGCN, hamilton2017inductive, Pinterest, GCNPierre, kipf2016semi, kipf2016variational, pmlr-v48-niepert16}.
While CNNs excel at extracting features from grid-like data such as images, GNNs are designed to propagate and transform features according to the structure of graphs, allowing them to retain relational information between entities represented by nodes and edges.
GNNs are increasingly applied in diverse domains, including social network analysis and recommender systems~\cite{hamilton2017inductive,Pinterest,kipf2016semi}, etc.

GNNs have shown significant advantages in graph representation learning~\cite{HamiltonYL17, hamilton2017inductive, Pinterest}, where the goal is to embed graph-structured information into low-dimensional dense vectors.
The trained model can produce vectors for specified nodes or edges. Then, tasks\kwc{, e.g., node classification, link prediction, etc.,} can be performed by only relying on them rather than all the raw data in the graph, e.g., the adjacency list, node features, etc. 
As Hamilton et al.\ \cite{HamiltonYL17} noted, traditional algorithms, e.g., DeepWalk~\cite{DeepWalk} and node2vec~\cite{node2vec}, cannot generalize to perform inference on unseen nodes or edges during training, and their representation power is limited.
In comparison, GNNs offer a more powerful and flexible approach, capable of addressing these limitations and enabling inductive learning for new graph data.
\kwa{(Paragraph moved.)}

A widely-used GNN model is graph convolutional network~(GCN)~\cite{kipf2016semi}.
Formally, a GCN layer is defined as 
$    \overrightarrow{{h}^{(l+1)}} = \sigma\left(A^{*}\overrightarrow{{h}^{(l)}}W^{(l)}\right)$
, where $W^{(l)}$ denotes the trainable weight matrix of the $l$-th layer, $\sigma$ is a non-linear activation function and $\overrightarrow{{h}^{(l)}}$ is the node representation at layer $l$. In particular, the node input features are denoted as $\overrightarrow{h^{(0)}}$. $A^{*}$ is the adjacency matrix normalized by node degrees:

\begin{equation*}
A^{*}_{i,j}=
\begin{cases}
\frac{1}{\sqrt{d_{out,j}}\cdot\sqrt{d_{in,i}}}, &\text{if there is an edge from } j \text{ to } i\\
0, &\text{otherwise}
\end{cases}
\end{equation*}

where $d_{out,j}$ is node $j$'s out degree and $d_{in,i}$ is the in degree of node $i$.

\begin{figure}[!tb]
\centering
\includegraphics[width=0.8\linewidth]{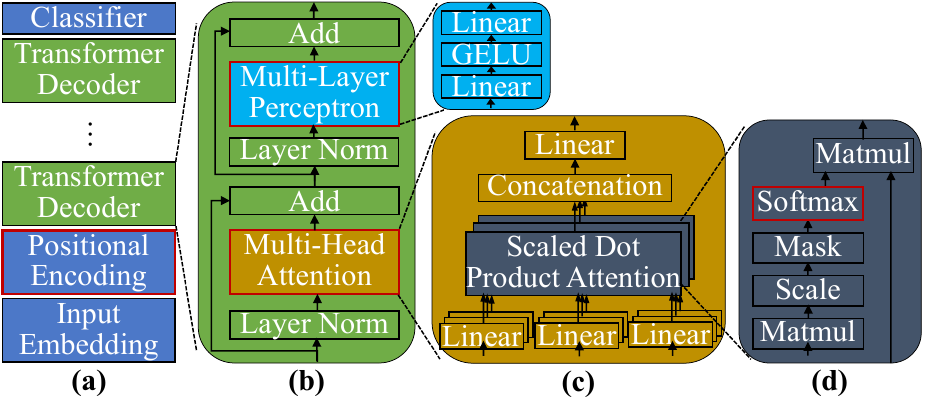}
\caption{\label{fig:transformer} Hierarchical breakdown of the GPT model. In training, dropout is applied to the output of each layer with red borders.}
\end{figure}

\section{Transformer-Based Large Language Models}
\label{sec:bg_transformers}

LLMs now drive a wide range of applications, including chatbots~\cite{openaiChatGPT2022}, search~\cite{BingChatMicrosoft2023}, content generation~\cite{midjourneyMidjourney2022}, reasoning~\cite{langchainLangChain2022}, etc. These models, when sufficiently large in size, demonstrate emergent abilities~\cite{weiEmergentAbilitiesLarge2022} and thus the ability to handle complicated tasks. \kwc{Consequently, LLMs today can be as large as containing hundreds of billions of parameters. Furthermore,} model designers \kwc{are driven to} continue to scale up the size of LLMs, carrying more parameters. %

Most LLM architectures, including GPT~\cite{radfordLanguageModelsAre2019},  are transformer-based~\cite{vaswaniAttentionAllYou2017}. As Figure~\ref{fig:transformer}(a) shows, the GPT model consists mainly of multiple transformer layers. Before transformer layers, GPT takes in the tokenized text and maps the tokens into dense vectors with positional information. The task determines the last part of the model architecture.
For instance, a classifier could be added for text classification tasks.
Figure~\ref{fig:transformer}(b) shows that each transformer layer is primarily made up of an attention block and a multi-layer perception~(MLP) block. Attention blocks~(Figure~\ref{fig:transformer}(c)) compute a weight, called attention, for each token pair, and produce dense vectors for each token via weighted summation. The MLP blocks transform the vector of each token into a new vector.

GPT is a decoder-only model because it only involves transformer decoder layers. A transformer encoder layer has the same structure as the transformer decoder layer except that the latter imposes causality on the attention mask in Figure~\ref{fig:transformer}(d): the causal mask ensures that the new vectors produced by the attention block for each token depend only on vectors of tokens, not after this token. By this categorization, transformer models are classified as (1)~encoder-only, e.g., BERT~\cite{devlinBERTPretrainingDeep2019}, (2)~decoder-only, e.g., GPT, Llama~\cite{touvronLlamaOpenFoundation2023}, and (3)~encoder-decoder, e.g., T5~\cite{raffelExploringLimitsTransfer2023}. In encoder-decoder models, the transformer decoder layers take in both outputs from the encoders and another text and apply two attention blocks---the self-attention block is applied to the new text, and the cross-attention block is applied among the tokens in the sequence from the encoder and tokens in the new text.

Parallelizing LLM training involves partitioning and/or replicating the model and the data into different GPUs~\cite{xuGSPMDGeneralScalable2021}. Pipeline parallelism, data parallelism, and model parallelism are the three levels of parallelism available to all LLM models and widely adopted in frameworks, e.g., Megatron, DeepSpeed, and PyTorch 2.0~\cite{shoeybiMegatronLMTrainingMultiBillion2020a,rasleyDeepSpeedSystemOptimizations2020,anselPyTorchFasterMachine2024}.
Pipeline parallelism partitions the model into several chunks of layers and places them on different GPUs. In a step, when the GPUs finish their layers, the output is passed to the GPUs owning the next layers.
Data parallelism replicates the models in different groups of GPUs and assigns separate micro-batches to each group.
At the end of a step, the gradients in each group are aggregated to update all the model replicas.
Model parallelism shards a weight tensor and puts shards onto different GPUs. Each GPU performs a part of the computation using its shard for the corresponding operator.
Given the system scale and interconnect, all or a few among the three levels may be used.
Zero Redundancy Optimizer~(ZeRO)~\cite{rajbhandariZeROMemoryOptimizations2020a} further reduces memory use with data parallelism by sharding the optimizer states, and/or optionally the gradients and parameters and stores the shards across these GPUs.

\section{Nvidia GPU Architectures and Programs}
\label{sec:bg_cuda}

\begin{figure}[]
    \centering
    \includegraphics[width=0.45\linewidth]{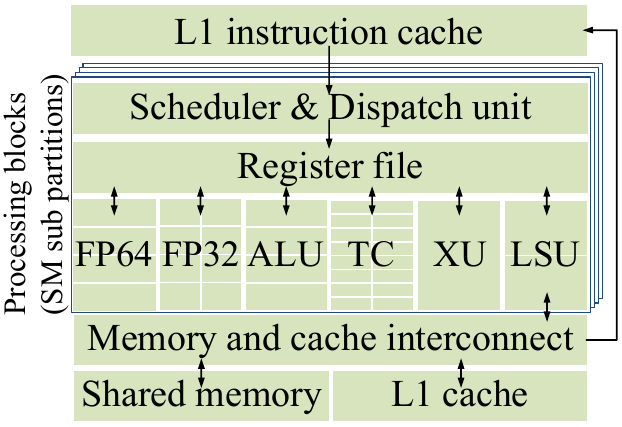}
    \caption{Structure of a streaming multiprocessor~(SM) in an Nvidia Volta V100 GPU~\cite{V100Whitepaper,jiaDissectingNVIDIAVolta2018,nickollsInstructionsManagingParallel2019,davidm.koppelmanEE7722GPU2023,nvidiaKernelProfilingGuide2024}. The execution units include FP64 units, FP32 units, arithmetic logic units~(ALUs), tensor cores~(TCs), transcendental and data type conversion units~(XUs), and load-store units~(LSUs). }
    \label{fig:sm_arch}
\end{figure}

While the CPU is designed to minimize the latency of each operation, the GPU is a massively parallel processor optimized to maximize throughput~\cite{PMPP4}. To support this computational parallelism, each GPU device is equipped with memory that has very high bandwidth, reaching an order of magnitude of TB/s after high-bandwidth memory~(HBM) is adopted. Just as each CPU chip contains multiple cores, each Nvidia GPU contains hundreds of cores, called streaming multiprocessors~(SMs). The structure of an SM is illustrated in Figure~\ref{fig:sm_arch}. In an SM, the scheduler selects instructions ready for execution, which are then dispatched by the dispatcher unit to various function units. The function units include floating-point units, arithmetic logic units~(ALUs), tensor cores, transcendental and data type conversion units~(XUs), and load-store units~(LSUs). LSUs are responsible for transferring data between the register file and memory, while the other function units operate on values stored in registers. For fast on-chip \kwc{memory}, each SM also contains its own L1 cache and a scratchpad, called shared memory.

A common way to program Nvidia GPUs with a parallel computing workload is to create a CUDA C++ program. Functions executed on Nvidia GPUs are called kernels. During the execution of a kernel, a massive number of threads execute the same logic specified in the kernel's CUDA C++ function definition. CUDA C++ well matches Nvidia GPUs' single instruction, multiple threads~(SIMT) execution model: At each time during execution, all threads that are being executed in an SM execute the same instruction. Threads within a CUDA kernel are organized into blocks, and each block is scheduled onto one SM, where it remains until all threads within the block finish executing. To hide latency, programmers typically aim for high occupancy, i.e., ensuring that each SM is assigned a large number of threads.

The Nvidia compiler, nvcc, compiles CUDA kernels into the PTX~(Parallel Thread Execution) intermediate language. The machine code executed by the GPU is in a proprietary instruction set architecture~(ISA), called SASS~(Streaming ASSembler). Translation from PTX to SASS can occur at compile time or runtime via the GPU driver.

\section{The Python Language}
\label{sec:bg_python}

People in the deep learning community use Python extensively. Python is easy to use due to its simplicity, expressiveness, and powerful features. One of the key features of Python is its interpreter, so users do not need to compile their code before execution. Additionally, Python's dynamic typing, known as duck typing, frees programmers from declaring the type of each variable and allows variables to change types during execution, simplifying development. Python also features rich ecosystems with widely-used package managers, e.g., Python's built-in Pip, Anaconda, etc.

To illustrate how friendly Python is, we write a Python program to sort tuples in Listing~\ref{lst:sort_second_python}. In contrast, Listing~\ref{lst:sort_second_cpp} shows the C++ program doing the same job. Both programs sort the \texttt{records} variable according to each tuple's second element. The \texttt{records} variable stores the name and address of each person as a two-string tuple. The tuples will be sorted according to the second string, i.e., the addresses. Both programs execute three steps: 1)~defining the \texttt{records} variable, 2)~sorting, and 3)~printing the sorted records.

As shown, the Python code is more expressive and quicker to develop due to several factors. First, it does not need compilation and an entry point \texttt{main()} function. Second, Python does not require the type declaration of each variable. Third, Python has a simpler lambda function syntax and supports containers in the built-in \texttt{print()} function.

\begin{lstlisting}[float,caption={Python code to sort tuples according to their second elements.},label={lst:sort_second_python}]
# Step (1) Define the records variable
records = [("Alice", "227 CSL"), 
     ("Bob", "1210 Siebel Center"),
     ("Charlie", "2120 ECE Building")]
# Step (2) Sort the records
records.sort(key=lambda x: x[1])
# Step (3) Print results
print(records)
\end{lstlisting}

\begin{lstlisting}[float,language=C++,caption={C++ code to sort tuples by their second elements~\cite{geeksforgeeksSortingVectorTuple2020}.},label={lst:sort_second_cpp}]
#include <bits/stdc++.h>
using namespace std;
int main() {
   // Step (1) Define the records variable
   vector<tuple<string, string> > records{ 
      {"Alice", "227 CSL"}, 
      {"Bob", "1210 Siebel Center"},
      {"Charlie", "2120 ECE Building"}};
   // Step (2) Sort the records
   sort(records.begin(), records.end(), 
      [](auto a, auto b) {return get<1>(a) < get<1>(b);});
   // Step (3) Print results
   for (auto r: records) {
      cout << get<0>(r) << " " << get<1>(r) << "\n";}}
\end{lstlisting}

One of the biggest concerns of Python is the significant serialization penalty caused by the global interpreter lock~(GIL) in multithreading programs. As the most widely-used Python implementation, CPython~\cite{wikipediaCPython2008} uses GIL to ensure thread safety. To mitigate this, frameworks work around GIL. One method is to put performance-critical logic in the C++ framework libraries and release the GIL once the control flow goes outside the Python code~\cite{CanPytorchBypass2019}. Another direction is to remove the GIL from the Python implementation. Although there are some alternate GIL-free Python implementations~\cite{wikipediaIronPython2006} to CPython, many frameworks rely on CPython-specific implementation details, making it challenging to migrate these frameworks to such alternatives. These limitations led to the \kwc{Python Enhancement Proposal (PEP) 703} to make GIL optional~\cite{samgrossPEP703Making2023} in CPython, which has been accepted recently.

\begin{lstlisting}[float,caption={PyTorch code to define model in Figure~\ref{fig:pytorch_graph} and perform a training step. We denote the input hidden dimension as \texttt{IN\_DIM} and the output hidden dimension as \texttt{OUT\_DIM}.},label={lst:pytorch_nested_module}]
import torch
from torch.nn.parameter import Parameter
# Step (1) Define the nested module
class MyLinear(torch.nn.Module):
    def __init__(self):
        super(MyLinear, self).__init__()
        self.x = Parameter(torch.randn(IN_DIM, OUT_DIM))
        self.b = Parameter(torch.randn(OUT_DIM))
    def forward(self, inp):
        return torch.matmul(inp, self.x) + self.b
class MyLinearWithActivation(torch.nn.Module):
    def __init__(self):
        super(MyLinearWithActivation, self).__init__()
        self.linear = MyLinear()
        self.activation = torch.nn.Tanh()
    def forward(self, inp):
        return self.activation(self.linear(inp))
# Step (2) Define the deep learning model
linear_with_activation = MyLinearWithActivation()
loss_fn = torch.nn.MSELoss()
# Step (3) Execute a training step
y = linear_with_activation(x)
loss = loss_fn(y, y_expected)
loss.backward()
\end{lstlisting}

\section{The PyTorch Computing Stack}
\label{sec:bg_pytorch}

\begin{figure}[]
    \centering
    \includegraphics[width=0.65\linewidth]{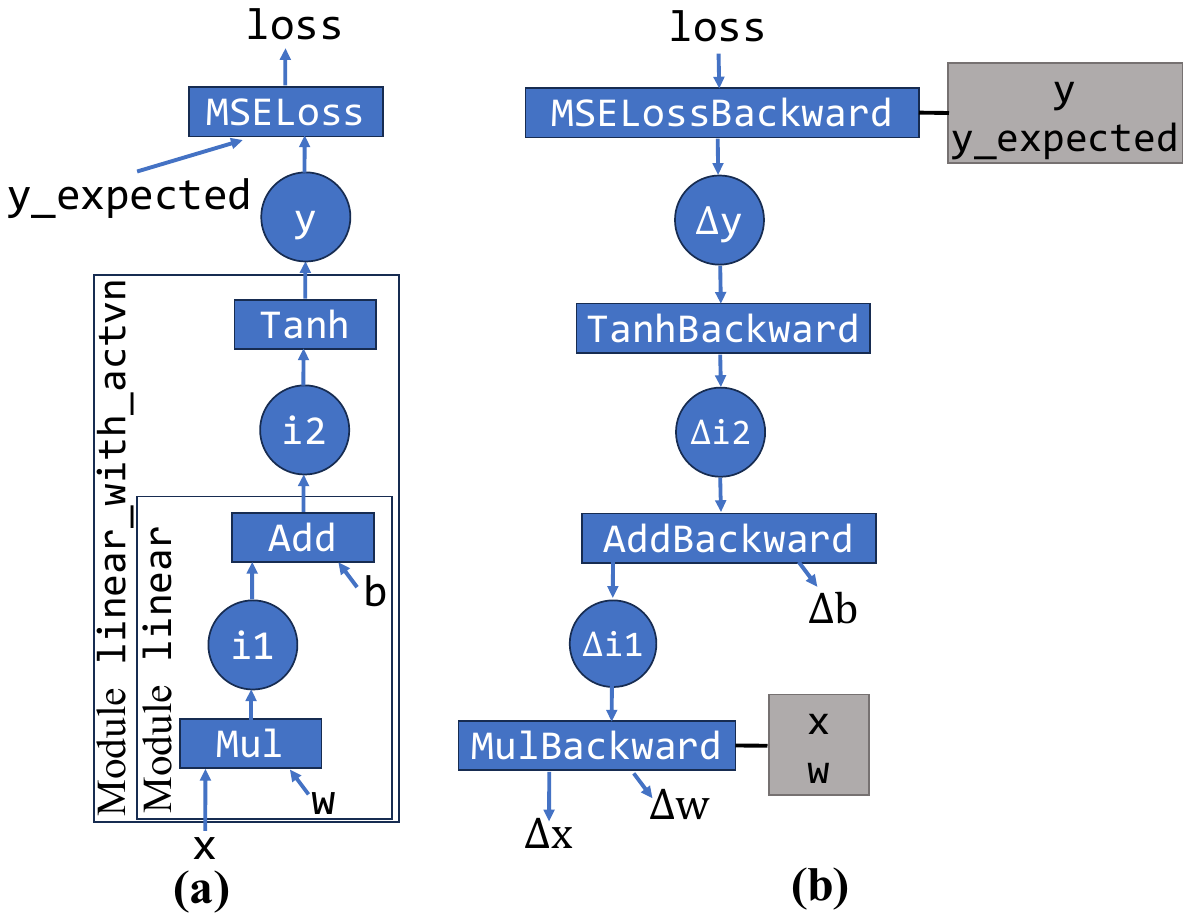}
    \caption{\kwc{Computational graphs of} (a)~forward propagation and (b)~backward propagation  for code in Listing~\ref{lst:pytorch_nested_module}. To compute gradients in backward propagation, dependent tensors are stored in the graph~(gray blocks with black text) during forward propagation. By default, PyTorch operates in eager mode and only constructs and stores the \kwc{graph of} backward propagation in memory during runtime. }
    \label{fig:pytorch_graph}
\end{figure}
Thanks to its intuitive, dynamic, and flexible programming interface, PyTorch is friendly to both users and developers who build packages on top of PyTorch. As a result, PyTorch has gained much popularity since its inception. 

By default, PyTorch executes code eagerly, meaning operations are computed immediately as they are called. For example, Listing~\ref{lst:pytorch_nested_module} defines a simple \kwc{classifier} model with mean squared error~(MSE) as the loss function and executes a training step.  The model contains a nested module, \texttt{linear\_with\_activation}, and a loss function \texttt{loss\_fn()}. \texttt{linear\_with\_activation} is made up of a linear layer and a hyperbolic tangent activation function. Figure~\ref{fig:pytorch_graph} illustrates both the forward-propagation and backward-propagation computational graphs for this example. As shown in step~(1) of Listing~\ref{lst:pytorch_nested_module}, the modules are defined as subclasses of \texttt{torch.nn.Module}. In the class definition, the initialization \kwc{method} \texttt{\_\_init\_\_()} initializes the parameters of layers and submodules. The \kwc{method} \texttt{forward()} defines the forward propagation logic of this module. Step~(2) constructs the model, consisting of the nested module \texttt{linear\_with\_activation} and the MSE loss function \texttt{loss\_fn}. Step~(3) executes a training step, i.e., one forward propagation and one backward propagation pass. \kwc{\texttt{x} is the input data, \texttt{y} is the predicted labels, and \texttt{y\_expected} is the ground-truth labels.}

\begin{figure}[!t]
    \centering
    \includegraphics[width=0.85\linewidth]{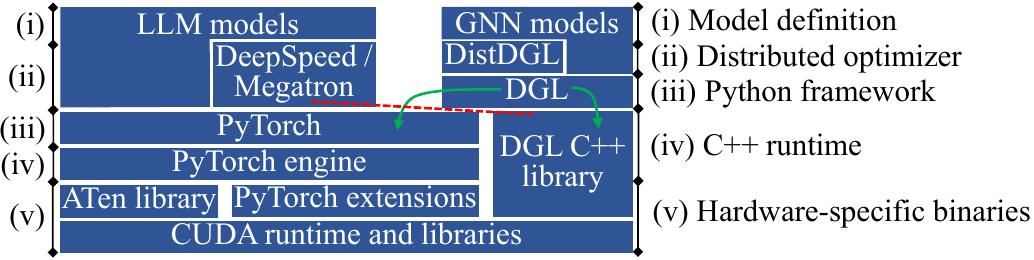}
    \caption{The PyTorch computing stack for GNN workloads and LLM workloads. As the GNN framework, Deep Graph Library~(DGL) calls PyTorch \kwc{functions} and functions provided by DGL C++ libraries~(green arrow). The LLM distributed optimizers, DeepSpeed and Megatron, do not rely on the DGL C++ library~(red dashed line). }
    \label{fig:pytorch_stack}
\end{figure}

Notice that users only need to define the forward propagation logic, as shown in Figure~\ref{lst:pytorch_nested_module}. PyTorch's auto-differentiation mechanism handles the computation of gradients without requiring users to manually specify the backward propagation logic. During forward propagation, PyTorch records \kwc{activations, i.e.,} the intermediate tensors, \kwc{and weights} required for backward propagation and constructs the corresponding computational graph. \kwc{In Figure~\ref{fig:pytorch_graph}, for example, \texttt{y} and \texttt{y\_expected} are the activations stored for  the backward propagation process of \texttt{MSELoss}, \texttt{MSELossBackward}}. During backward propagation, PyTorch executes the backward-propagation computational graph. PyTorch calls precompiled kernels to execute operators in forward propagation and backward propagation.

Another way to provide auto differentiation uses just-in-time~(JIT) compilation. For example, to perform auto differentiation, JAX 1)~captures the forward propagation functions' IR through trace-based JIT, 2)~generates the gradient functions' IR via transformation, and 3)~compiles CUDA binaries using XLA~\cite{joergAutomatedGPUKernel2019}. Similar\kwc{ly,} JIT-based approaches are adopted by PyTorch JIT~\cite{pengwuWorkshopsASPLOS_2024README}, Mathematica~\cite{dakkakCompilingHighlevelScripting2020}, Zygote~\cite{zygoteFluxMLZygoteJl2019}, CLAD~\cite{ioanaifrimGPUAccelerationAutomatic2021}, and Enzyme~\cite{mosesReversemodeAutomaticDifferentiation2021}.

Figure~\ref{fig:pytorch_stack} shows the PyTorch computing stack for GNNs and LLMs, the primary workloads addressed in this dissertation. At the top of the stack are the GNN models and LLM models. Users can use distributed optimizers, e.g., DeepSpeed for LLMs and DGL for GNNs. Both single-GPU execution and distributed optimizers are built on top of PyTorch, although DGL also relies on its library for graph-related operations. At the bottom of the stack are the CUDA runtime and math libraries. The stack \kwc{comprises} five layers from the top to the bottom: (i)~Model definition specifies the model architecture and pre-trained parameters. (ii)~Distributed optimizers provide mechanisms for device-level parallelism and communication. (iii)~Python frameworks offer layer definition, dataloading, and profiling utilities. (iv)~The C++ runtime provides a GIL-free context, auto differentiation mechanism, and functionality of operator dispatching. (v)~Hardware-specific binaries \kwc{provide the binaries to execute operators on devices. These binaries may} leverage vendor-optimized libraries and provide support for new hardware, e.g., tensor cores on Nvidia GPUs. \kwc{At this level, support for new operators can be added to the stack by creating new PyTorch extensions and registering them during runtime. PyTorch extensions use \texttt{pybind11}}~\cite{wenzeljakobPybindPybind112016} \kwc{to allow the new C++ code to interact with the Python runtime.}

\chapter{Hector: An Efficient GPU Programming and Compilation Framework for Relational Graph Neural Networks}
\label{ch:hector}
\section{Introduction\kwa{(Chapter 3 and Chapter 4 are swapped.)}}

\kwa{(Two sentences are removed.)}

GNN-specific machine learning frameworks, e.g., DGL~\cite{wang2019deep} and PyTorch Geometric~(PyG)~\cite{fey2019fast}, \kwc{are optimized specifically for homogeneous graphs.}
\kwc{For example, they} implement several highly-optimized operations, e.g., sparse-dense matrix multiply~(SpMM) and sampled dense-dense matrix multiply~(SDDMM), to speed up the execution~\cite{huFeatGraphFlexibleEfficient2020a}.
Most of these operators and optimizations are for homogeneous graphs~\cite{huangEfficientSparseMatrix2021,yeSparseTIRComposableAbstractions2022,huFeatGraphFlexibleEfficient2020a}.
However, real-world graphs are typically heterogeneous by nature and contain multiple types of nodes and edges.
For example, a citation graph may represent entities involving authors, articles, etc., as nodes of different types;
the edges may model various types of relations, e.g., an article citing the others.
Recently, to incorporate the information provided by such heterogeneity,
RGNNs~\cite{rgcn,hgt} are proposed to define dedicated parameters and data paths for each type.

RGNN poses {three major} challenges to the existing GPU computation stack due to its inherent computation patterns, the gap between the programming interface and the kernel APIs, and the high cost of kernel code optimizations due to its coupling with data layout and heterogeneity.

The first challenge with GNN implementations on GPUs stems from their need to traverse graphs and {scatter/gather tensor data} in order to use high-performance GEMM kernels to implement message passing.
In RGNN, message passing is the procedure in each layer where an edgewise operation is followed by a nodewise aggregation operation. In other words, messages are passed through edges to the destination nodes. We show how message passing works in models in Section~\ref{sec:rgnn_formulation}.
During message passing, the graph structure and data layout significantly impact the memory access patterns and execution throughput~\cite{wangEmpiricalAnalysisPerformance2021, zhengNatureGraphNeural2021}.~(Examples and details are in Section~\ref{sec:design}).
Furthermore, as the connectivity of the input graph is involved in the {gather} computation, the computation patterns of GNNs are affected not only by the model definition but also by the graph. Such data-dependent behavior precludes any one-size-fits-all optimization strategy when executing GNNs. Additionally, RGNN introduces new complications into the design space due to {the need for the operations to account for heterogeneity. We detail this in Section~\ref{sec:background}.}

The second challenge in RGNN implementation stems from the lack of an abstraction layer between the programming interface and kernel APIs, resulting in extra data movement.  A typical example is an edgewise typed linear layer.
We detail the context and cause of the extra data movement in the edgewise typed linear layer in Section~\ref{sec:segmentmm}. 
But essentially, an edgewise typed linear layer multiplies one of the vectors on each edge with the layer weight dedicated to the edge type.
To achieve this, many existing PyTorch-based systems materialize a temporary three-dimensional edgewise weight tensor, where the slice corresponding to each edge is the weight matrix of its edge type.
This temporary weight tensor is huge, causing redundant data access and memory footprint.
Hector avoids such unnecessary copying activities by having typed linear transformations operate on views of tensors, a feature that PyTorch lacks, and decouples the materialization of its operands 
from the source-level expression~(Section \ref{sec:materialization}).

Third, code generation is necessary. 
High-performance neural network implementations have historically been based on pre-built libraries, e.g.,  cuBLAS~\cite{nvidiaCublasgemmBatchedCuBLASDocuemntation}. 
GNNs make this less practical because the number of kernels to optimize is multiplied by the number of adjacency-matrix storage format choices such as Blocked-Ellpack~\cite{PMPP4}.
For instance, cuSPARSE only supports the latter in a few APIs~\cite{AcceleratingMatrixMultiplication}.
The typed edges and nodes of RGNN further exacerbate the problem, which makes the traditional pre-built libraries even less adequate and compels framework developers to either painstakingly develop optimized layers from scratch or settle for slow implementation.
For example, it took more than a month for a full-time engineer to implement and deploy the typed linear layer of RGNN in DGL~\cite{nisaFeatureGatherMm}.
Another consequence is the performance degradation caused by limited kernels due to high implementation costs. For example, the  DGL  \texttt{HeteroConv} operator uses a Python native loop to separately launch kernels for each of the relation types in a heterogeneous graph, leading to serial execution of small GPU kernels that underutilize GPU resources on small graphs.

To systematically address these challenges, we propose Hector, a two-level IR and an associated code generator framework. 
The higher-level IR, called inter-operator level IR, defines the model semantics as sets of operators and expresses layout choices in a decoupled manner. At the lower level, the intra-operator level IR provides the facility to express template specialization and lower them to CUDA kernels. 
We further propose two optimizations, i.e., compact materialization~(Section~\ref{sec:materialization}) and linear operator reordering~(Section~\ref{sec:inter_op_opt}).
We show in the corresponding Sections how these two optimizations are conveniently enabled by the two-level IR design.
\cref{sec:inter_op_ir,sec:intra-op-ir,sec:ir_design} further the design and rationale of the two-level IR.

In short, Hector 1)~represents the key properties of RGNN models to capture opportunities to reduce memory accesses in inter-operator scheduling and materialization, 2)~generates code flexibly with proper data access schemes to eliminate redundant data movement, and 3)~expresses model semantics, data layout, and operator-specific optimization in a decoupled manner to reduce programming effort. To the best of our knowledge, Hector is the first to propose a multi-level IR to capture RGNN-specific opportunities involving cross-relation inter-operator optimizations and tensor data layout with consideration of the type dimension added by RGNNs.
The contribution of Hector is as follows:
\begin{enumerate}
\item We propose the Hector two-level IR and code generation framework to systematically optimize and generate GPU kernels for RGNN training and inference. It bridges the gap between the programming interface and the kernel generation process, decouples models, data layout, and operator-specific schedule from each other, and leverages optimization opportunities from the three aspects.
\item \label{contrb:eval} We devised the Hector code generator based on two generalized CUDA templates, {i.e., a} GEMM template and a node and/or edge traversal template. The generated code {achieves} up to 9.9$\times$ speed-up in inference and up to 43.7$\times$ speed-up in training compared to the {best among the} state-of-the-art systems~\cite{wuSeastarVertexcentricProgramming2021, xieGraphilerCompilerGraph, guiHGLAcceleratingHeterogeneous} when running RGCN, RGAT, and HGT~\cite{rgcn, busbridge2019relational, hgt} on heterogeneous datasets provided by DGL and OGB packages~\cite{huOpenGraphBenchmark2021, aifb, mutag, bgs, am, toutanovaObservedLatentFeatures2015}. Hector also encountered fewer out-of-memory~(OOM) errors, which is significantly alleviated by the optimization mentioned in Contribution~\ref{contrib:dse}. 
{In fact, with compaction enabled, Hector incurs no OOM error for all the datasets tested.}
\item \label{contrib:dse} We devised two optimizations: compact tensor materialization and linear operator reordering.
The best combination of options varies across models and datasets and further obtains up to 3.8$\times$ speed-up in inference and 2.7$\times$ speed-up in training compared to our basic generated code mentioned in Contribution~\ref{contrb:eval}. 
Through profiling, we found that the improved memory efficiency allows Hector to accommodate larger computations and improve GPU hardware utilization for forward propagation. In contrast, backward propagation does not benefit from larger input due to its latency-bound nature caused by atomic updates and outer products. 
\end{enumerate}

\section{Background and Motivation}\label{sec:background}

\subsection{RGNN Formulation and Operators}
\label{sec:rgnn_formulation}

\begin{figure}[!b]
\centering
\includegraphics[width=\linewidth]{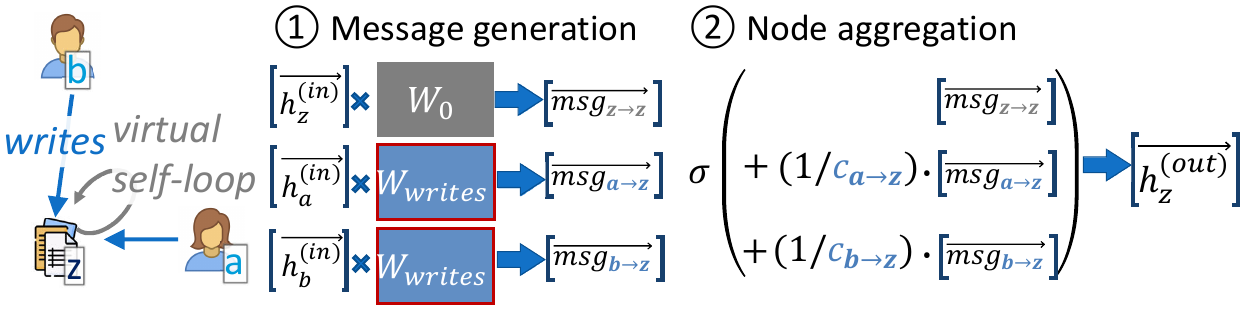}
\caption{\label{fig:rgnn_layer} The forward propagation of an RGCN layer could be divided into \textcircled{1} message generation on edges and \textcircled{2} node aggregation. We focus on paper node $z$ in a large citation graph as an example. $z$ only has two incoming edges, from $a$ and $b$, respectively. $\overrightarrow{{h}^{(in)}}$ and $\overrightarrow{{h}^{(out)}}$ are node features. $W_{writes}$ is the weight for the type ``writes''. $W_{0}$ is the weight for virtual self-loops. $\sigma$ is the activation function. Notably, some runtime implementations may replicate data, e.g., $W_{writes}$. }
\end{figure}

RGNNs extend GNNs to model different node and edge types for relational graph data.
For example, extended from GCN, a relational graph convolutional network~(RGCN) layer is defined as 
\begin{equation} \label{eq:rgcn}
     \overrightarrow{{h}_v^{(l+1)}} = \sigma\left(\overrightarrow{{h}_v^{(l)}}W_0^{(l)}+\sum_{r \in R}\sum_{u \in \mathcal{N}_v^r}\frac{1}{c_{v,r}}\overrightarrow{{h}_u^{(l)}}W_r^{(l)}\right)
\end{equation}
, where $\mathcal{N}_v^r$ denotes neighbors of node $v$ in relation $r \in R$, 
${h}_n^{(l)}$ is the $l$-th layer node representation of $n$. 
$W_r^{(l)}$ is the weight for relation $r$. $c_{v,r}$ is a problem-specific normalization factor.
Figure~\ref{fig:rgnn_layer} shows an example of how output features are produced in the message passing formulation equivalent to Formula~\ref{eq:rgcn}: 
The forward propagation of an RGNN layer could be divided into \textcircled{1} the edge message generation stage and \textcircled{2} the node aggregation stage.
For simplicity, we focus on the output feature $\overrightarrow{h_z^{(out)}}$ of node $z$: To obtain $\overrightarrow{h_z^{(out)}}$, \textcircled{1} a message $\overrightarrow{msg}$ is generated for each incoming edge, and \textcircled{2} the edge messages go through weighted aggregation and an activation function $\sigma$ to produce $\overrightarrow{h_z^{(out)}}$. 
Notably, to obtain the output feature of node $v$, the input feature of $v$ itself is applied to the $W_0^{(l)}$ and added to the transformed neighbor features. We call this a virtual self-loop because it could be seen as if each node now has a new edge to itself.

\begin{figure}[!bt]
\centering
\includegraphics[width=0.8\linewidth]{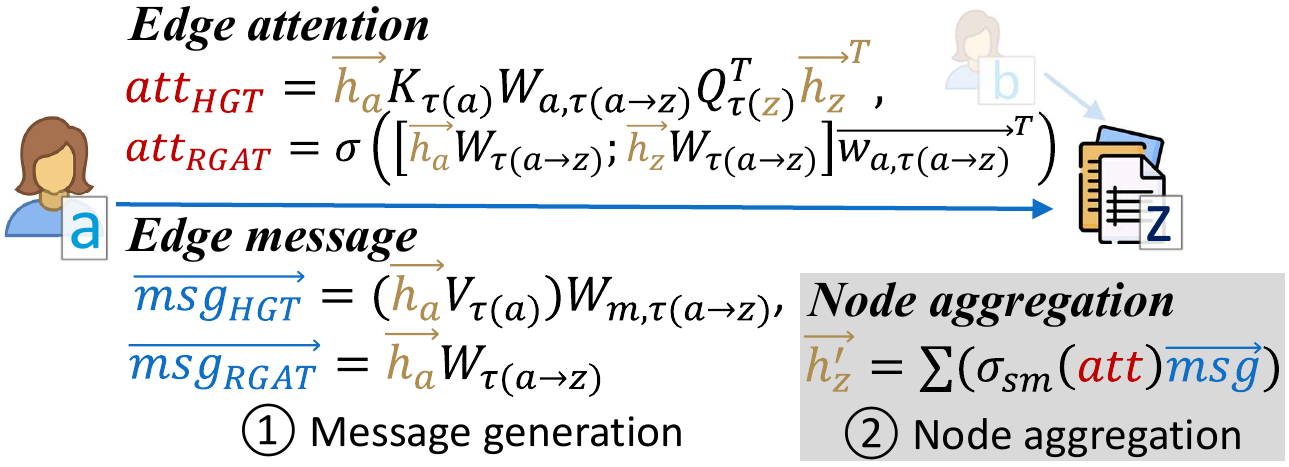}
\caption{\label{fig:rgat_layer} HGT and RGAT layer. $\overrightarrow{{h}_n}$ and $\overrightarrow{{h}_n^\prime}$ are node $n$'s features. Denote the type of edge from $a$ to $z$ as $\tau(a\rightarrow z)$. Weights $W_{a,\tau(a\rightarrow z)}$ differ by edge type $\tau(a\rightarrow z)$: For example, assuming there are two edge types, ``writes'' and ``cites'', $W_{a,\text{``writes''}}$ is a different weight from $W_{a,\text{``cites''}}$. They are defined and learned according to the edge type. $W_{m,\tau(a\rightarrow z)}$ and $\overrightarrow{{w}_{a,\tau(a\rightarrow z)}}$ are in similar situations. Weights $W_{\tau(n)}$ differ by the node type $\tau(n)$ of $n$. $\sigma$ is a leaky rectified linear unit~(ReLU) in the case of RGAT. $\sigma_{sm}$ stands for edge softmax. $[\vec{s};\vec{t}]$ concatenates $\vec{s}, \vec{t}$.}
\end{figure}

Relational graph attention network~(RGAT)~\cite{busbridge2019relational} and heterogeneous graph transformer~(HGT)~\cite{hgt} are shown in Figure~\ref{fig:rgat_layer}. Attention is introduced in these more complex models:
Attention is produced in the message generation stage together with edge messages.
Similar to the normalization factor, it is a scalar that emphasizes the message associated with the same edge during the subsequent node aggregation stage. However, attention is learned, as it is produced by operations among weights and features.

In addition to describing GNNs in two stages---message generation and node aggregation---a popular formulation uses the SpMM and SDDMM pair. The DGL~\cite{wang2019deep} paper has proven that GNN message passing can be expressed as generalized SpMM~(g-SpMM) and generalized SDDMM~(g-SDDMM) operations, with their backward propagation also following the same structure. SpMM computes the product of two matrices, $C=A\times B$, where the left matrix $A$ is sparse and in a sparse matrix format. The right matrix $B$ is dense. Notice that each row in $C$ is a weighted aggregation of specific rows in $B$ according to $A$:

\begin{equation*}
\overrightarrow{c_{i,\cdot}}=\sum_{j\in\{j\mid A_{i,j}\neq 0\}}A_{i,j}\cdot\overrightarrow{b_{j,\cdot}}
\end{equation*}

where $\overrightarrow{c_{i,\cdot}}$ is the vector of $C$'s $i$-th row, and  $\overrightarrow{b_{j,\cdot}}$ is the vector of $B$'s $j$-th row. g-SpMM generalizes SpMM in three ways: (1)~the scalar $A_{i,j}$ is generalized to data corresponding to the edge $j\rightarrow i$, (2)~the product operator $\cdot$ is generalized to a message function that produces a vector after taking as input the data of the edge $j\rightarrow i$ and the $\overrightarrow{b_{j,\cdot}}$ vector of node $j$, and (3)~the summation operator $\sum$ is generalized to a custom aggregation function.

SDDMM selectively computes the product of two dense matrices based on a sparse matrix:

\begin{equation*}
C_{i,j}=A\times B \odot S =
\begin{cases}
\overrightarrow{a_{i,\cdot}}\cdot \overrightarrow{b_{\cdot,j}} \cdot S_{i,j}, &\text{if } S_{i,j}\neq0\\
0, &\text{otherwise}
\end{cases}
\end{equation*}

where $\overrightarrow{a_{i,\cdot}}$ is the vector of $A$'s $i$-th row, $\overrightarrow{b_{\cdot,j}}$ is the vector of $B$'s $j$-th column, and $S$ is the sparse matrix. g-SDDMM generalizes SDDMM in two ways: (1)~the scalar $S_{i,j}$ is generalized to data corresponding to the edge $j\rightarrow i$ and (2)~the two product operators $\cdot$ are generalized to one message function that produces a vector after taking as input the data of the edge $j\rightarrow i$,  the $\overrightarrow{a_{i,\cdot}}$ vector of node $i$, and the $\overrightarrow{b_{\cdot,j}}$ vector of node $j$.

\subsection{RGNN Performance Characteristics}

In {non-graph neural networks,} 
most linear operators, e.g., convolution, can be efficiently implemented with GEMM kernels. 
GEMM takes up most of the execution time due to its cubic complexity.
While some operators can be optimized by transformations, e.g., Winograd for convolution layers~\cite{lavinFastAlgorithmsConvolutional2015}, the operators are still computation-intensive after such computation reduction.
GPUs are excellent at GEMM because the latter's high computation complexity allows leveraging the massive parallel compute units on GPUs\kwc{. At the same time,} the input data could be sufficiently reused to allow the memory bandwidth to keep up with the computation throughput.

In contrast, GNNs spend a much larger portion of their execution time on memory-intensive, non-GEMM operations ~\cite{wangEmpiricalAnalysisPerformance2021, zhengNatureGraphNeural2021}. One major source of memory-intensiveness is the sparsity of graphs: to be not bound by the memory {bandwidth}, Nvidia H100 GPU requires the data reuse of single-precision float to be {at least} 16 times. However, the average degree of a graph often falls below this threshold, e.g., the graph datasets in Table~\ref{tab:datasets}. The heterogeneity of RGNNs further exacerbates the issue due to lowered data reuse by the introduction of dedicated weights to different edge types and node types, as shown in Figure~\ref{fig:rgat_layer}.

\begin{figure}[!t]
\centering
\includegraphics[width=\linewidth]{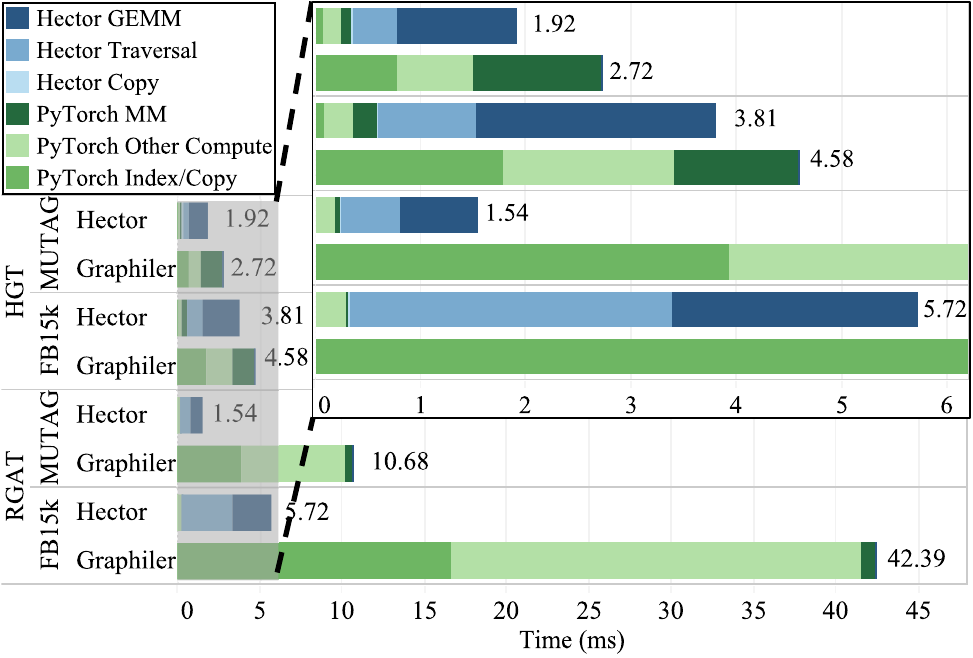}
\caption{\label{fig:bg_breakdown} Breakdown of inference time by Graphiler and Hector. Matrix multiply~(MM) includes SpMM. We categorize PyTorch time not accounted for by kernels as ``PyTorch Other Compute''.} 
\end{figure}

\subsection{Inefficiency in Existing Computation Stack: A Case Study on Edgewise Typed Linear Layers}
\label{sec:segmentmm}

\begin{figure}[!t]
\centering
\includegraphics[width=0.8\linewidth]{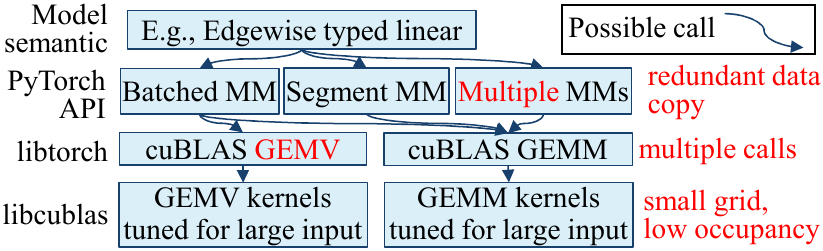}
\caption{\label{fig:inefficient_stack} Inefficiency~(in red) exists in all layers of existing systems.}
\end{figure}

We use an edgewise typed linear layer as an example to walk through the various performance overheads in the existing computation stack, as summarized in Figure~\ref{fig:inefficient_stack}. 
Edgewise typed linear layer applies a typed linear operator on each edge to one of its vectors. The weight of the linear operator used in the computation depends on each edge's type. For example, the edge message in an RGCN layer~(Figure~\ref{fig:rgnn_layer}) or an RGAT layer~(Figure~\ref{fig:rgat_layer}), is produced by a typed linear layer.

A typed linear layer is typically implemented using batched matrix multiply~(BMM) or segment matrix multiply~(segment MM)~\cite{isratsegmm}. 
For example, PyG \texttt{FastRGCNConv} implemented typed linear layers using BMM to unleash parallelism. However, a temporary tensor must be created from the weight tensor due to the lack of support for indirect addressing by PyTorch tensor APIs: the typed linear layer could be denoted as  $Y[i,0,j]:=\sum_k(X[i,0,k]\times W[T[i],k,j])$ where $X[i,0,\cdot]$, $Y[i,0,\cdot]$ and $W[T[i],\cdot,\cdot]$ are input feature, output feature of node $i$ and the weight of node $i$'s type. The middle dimension of $X$ and $Y$ are needed to make the operation a matrix multiply. However, there is currently no support for specifying $T[i]$ as one of the arguments to an operator in PyTorch;  
one must create $W^\prime[i,k,j]:=W[T[i],k,j]$ before the typed linear layer.

Segment MM requires presorting features by types. Then,  the node/edge feature tensor is in the form of segments of features of the same type: the segment MM kernel then applies the corresponding weight tensor of the type to each segment. 
If neither BMM nor segment MM can be employed, one may fall back to multiple matrix multiplies, leading to higher device API overhead and GPU under-utilization.

Another type of inefficiency is suboptimal math library calls. PyTorch has routines to handle various scenarios, e.g., a tensor is strided in memory layout or is \texttt{NestedTensor}, a pack of tensors. Consequently, PyTorch sometimes performs BMM by launching multiple general matrix-vector multiplies~(GEMVs) kernels, which also leads to API overhead and GPU under-utilization.

Lastly,  CUDA math libraries were initially developed for large inputs and may not be {efficient} for small inputs~\cite{nvidiaCublasgemmBatchedCuBLASDocuemntation}.

To better illustrate the points, Figure~\ref{fig:bg_breakdown} breaks down HGT and RGAT inference time on FB15k and MUTAG.
Section~\ref{sec:eval_methodology} details the system configurations and datasets.
This experiment measured Graphiler~\cite{xieGraphilerCompilerGraph}, which executed compiled TorchScript code and delivered the best inference performance among the existing systems tested in Hector.
Figure~\ref{fig:bg_breakdown} shows that indexing and copying take up a significant portion, and the portion of GEMM operations, i.e., MM vs. Other compute,  varied with graphs.
By profiling, we found that the CUDA API overhead is 22\% of the time of the critical path, which is the sum of the API overhead and kernel duration. This is partly due to a huge number of kernel launches caused by 1)~libraries calling a series of kernels to fulfill an API invocation and 2)~some operators calling separate sets of kernels for each types in the graph.

In contrast, Hector 1)~\textbf{lowers more of the logic to GEMM}, 
and 2)~assembles kernels with flexible access schemes to \textbf{gather and scatter data on the fly} to eliminate redundant data movement. Consequently, Hector does not replicate weights during computation. As shown, \textbf{this strategy achieves better performance than using hand-optimized kernels with dedicated functions to data movement, e.g., in Graphiler}.

To address the performance challenges in RGNN systems due to both RGNN's inherent computation pattern and the system design, we propose the Hector IR and code generation framework. By the IR design that \textit{decouples} and \textit{expresses} the model semantics, data layout, and operator-specific schedules, Hector opens up these opportunities and the integration of all three aspects into the design space.
Table~\ref{tab:salesman_matrix} shows the feature comparison of Hector with existing systems. 

\begin{table}[!htbp]
\centering
\begin{tabular}{@{}llllll@{}}
\toprule
&                                & \rotatebox{75}{Graphiler} & \rotatebox{75}{Seastar}   & \rotatebox{75}{HGL}       & \rotatebox{75}{\textbf{Hector}}   \\ \midrule
\multirow{2}{*}{\textbf{Target}}                                                 & \textbf{Inference}             & \Checkmark & \Checkmark &           & \Checkmark       \\
& \textbf{Training}              &           & \Checkmark & \Checkmark & \Checkmark       \\\cline{1-2}
\multicolumn{2}{@{}l}{\textbf{Memory efficiency}}                          & \Checkmark &           & \Checkmark & \textbf{better} \\\cline{1-2}
\multirow{3}{*}{\textbf{\begin{tabular}[c]{@{}l@{}}Design\\ space\end{tabular}}} & \textbf{Data layout}           &           &           &           & \Checkmark       \\
& \textbf{Intra-operator schedule}     &           &           &           & \Checkmark       \\
& \textbf{Inter-operator optimization} & \Checkmark & \Checkmark & \Checkmark & \Checkmark       \\ \bottomrule
\end{tabular}
\caption{Features of Hector and prior~\cite{xieGraphilerCompilerGraph, wuSeastarVertexcentricProgramming2021, guiHGLAcceleratingHeterogeneous} GNN compilers.\label{tab:salesman_matrix}
}
\end{table}

\section{Design and Implementation}\label{sec:design}

\subsection{Overview of Workflow and System Components}\label{sec:hector_overview}
Hector consists of a programming interface, a code generator, and Python modules.
The code generator takes in the model definition and generates both CUDA kernels and host functions that configure and invoke the CUDA kernels.

\begin{figure}[!t]
\centering
\includegraphics[width=\linewidth]{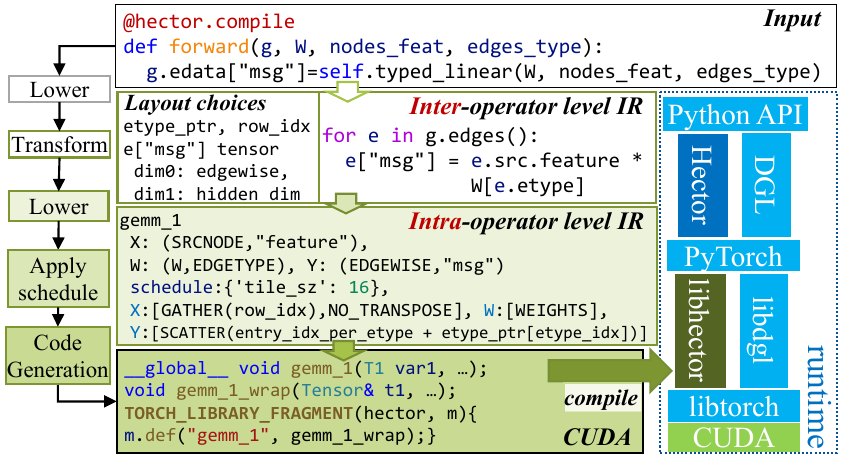}
\caption{\label{fig:runtime_arch} Hector workflow and software architecture. }
\end{figure}

Figure~\ref{fig:runtime_arch} uses an example to illustrate the workflow. The input is an excerpt of DGL code invoking a typed linear layer on the input node features. Applying the \texttt{@hector.compile} decorator triggers a transpiling pass to lower the code into Hector inter-operator level IR. {In this example, the typed linear transformation \texttt{typed\_linear}} can be efficiently implemented as GEMM kernels. {To this end,} Hector lowers the transform to an operator instance derived from the GEMM template at the inter-operator level. {After the analysis and optimizations at the inter-operator level, Hector further lowers the code to a detailed GEMM specification at the intra-operator level.} The GEMM output $A$ collects edge data generated from the node data. The first input $B$ is the weight matrix $W$, and the second input $C$ is the collection of features of all the source nodes of the edges involved. The intra-operator level IR indicates that the GEMM operation should use the default tile width of 16 and be carried out without scatter, gather, or transpose applied to input or output matrices. Eventually, Hector generates a segment MM~(Section~\ref{sec:segmentmm}) kernel, \texttt{gemm\_1}. 
The Layout Choices section of Figure~\ref{fig:runtime_arch} shows the default layout choice. \texttt{etype\_ptr} specifies the offsets of each segment of different type. \texttt{row\_idx} is the source node index array in the COO format. The result tensor \texttt{e["msg"]} has the number of edges as the number of rows, and the number of the columns is the input dimension of the hidden layer. We detail in Section~\ref{sec:materialization} an optimization technique, compact materialization, that is opened up by the decoupled layout choices from the inter-operator level IR.

The generated code is compiled into a shared library where host functions are exported through the \texttt{pybind11} utilities.
Hector falls back to existing routines in PyTorch when certain operators are not yet supported.
During runtime, the precompiled functions are loaded and registered as subclasses of PyTorch \texttt{autograd.Function}.

\subsection{Inter-Operator Level IR}
\label{sec:inter_op_ir}

The inter-operator level IR follows the Python grammar but involves some new constructs, as listed in Table~\ref{tab:ir_constructs}. Listing~\ref{lst:ir_example} illustrates how the attention calculation in a single-headed RGAT layer could be expressed using the inter-operator level IR.
Lines 10-16 shows a code segment that generates attention values for all edges of graph \texttt{g} and then invoke the \texttt{edge\_softmax(g)} function that spans lines 1 through 9. As shown in Listing~\ref{lst:ir_example}, the message generation and aggregation stages are expressed as for-each edge loops starting from line~2, line~8, and line~10, and for-each node loop starting from line~4. To accumulate data from the incoming edges of each node n, the \texttt{n.incoming\_edges()} iterator is used. Notably, the data layout that specifies how to access the input and output data per edge or node as well as the incoming edges associated with each node, is abstracted away in Listing~\ref{lst:ir_example}.

\subsubsection{Programming Interface}
Hector provides a decorator, \texttt{@hector.compile}, to take the existing PyG or DGL forward propagation logic and generate code for it, as exemplified by the input in Figure~\ref{fig:runtime_arch}. The decorator, when applied to a method, invokes a simple transpiling pass that replaces the PyG and DGL method calls, e.g., SpMM/SDDMM,  with an implementation in the inter-operator level IR, and replaces supported constructs from PyG and DGL with expressions in Hector IR.
Similarly to statically-typed compilers in other Python packages~\cite{pytorchTorchScriptPyTorchDocumentation, lamNumbaLLVMbasedPython2015}, the function to compile can use most of the Python features except dynamic ones, e.g., assigning objects of different types to the same variable. We support a few types as the function arguments for heterogeneous graphs, involving \texttt{Tensor} and \texttt{dict[str, Tensor]} objects, i.e., \texttt{dict} objects where the keys are \texttt{str} objects and the values are \texttt{Tensor} objects.

Besides, one can use the Hector inter-operator level IR itself to express the model, as exemplified by Listing~\ref{lst:ir_example}. %

\begin{lstlisting}[caption={Expressing the attention calculation in a single-headed RGAT model using Hector inter-operator level IR.},label={lst:ir_example},language=Python]
def edge_softmax(g):
    for e in g.edges():
        e["att"] = exp(e["att"])
    for n in g.dst_nodes():
        n["att_sum"] = 0.0
        for e in n.incoming_edges():
            n["att_sum"] += e["att"]
    for e in g.edges():
        e["att"] /= e.dst["att_sum"]
for e in g.edges():
    hs = e.src.feature * W[e.etype]
    atts = dot_prd(hs, w_s[e.etype])
    ht = e.dst.feature * W[e.etype]
    attt = dot_prd(ht, w_t[e.etype])  
    e["att"] = leakyrelu(atts + attt)
edge_softmax(g)
\end{lstlisting}

\begin{figure}[!htbp]
\centering
\includegraphics[width=0.4\linewidth]{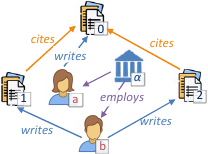}
\caption{\label{fig:opt_example} The citation graph used as the example in Figures~\ref{fig:compact_opt_opt}~and~\ref{fig:linear_opt}. }
\end{figure}

\begin{figure}[!htbp]\captionsetup[subfigure]{font=small}
\centering
\subcaptionbox{GEMM kernel and IRs of RGAT edge message computation with vanilla materialization. The two red squares mark identical terms because \texttt{msg} depends only on source node and edge type. Both schemes in (a) and (b) \textcircled{1} gather the source node’s features into a matrix, \textcircled{2} perform the GEMM computation, and \textcircled{3} scatter the output features to rows in the output tensor. Each dotted square mark a block in \textcircled{2} the GEMM kernel. \texttt{row\_idx} specifies the source node index of each edge, and is used in step \textcircled{1}. \texttt{etype\_ptr} specifies the offsets of edge of each type and is used in step \textcircled{3}.}
[\linewidth]{\includegraphics[scale=1.47]{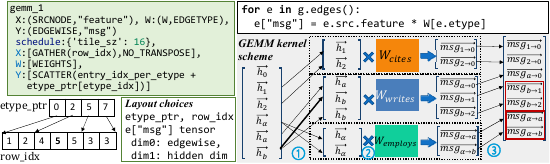}}
\subcaptionbox{GEMM kernel and IRs of RGAT edge message computation with compact materialization. Differences in IRs are marked in orange.}
[\linewidth]{\includegraphics[scale=1.45]{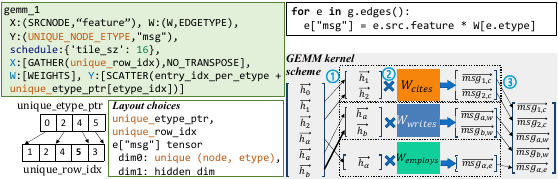}}
\caption{\label{fig:compact_opt_opt} When computing RGAT edge messages, compact materialization could be applied. This figure uses the graph in Figure~\ref{fig:opt_example}. Compared with (a)~vanilla materialization, (b)~compact materialization saves both the memory footprint and the computation. In (a), the leading dimension of the output message tensor accommodates different edges. In (b), it accommodates unique (source node, edge type) pairs. \texttt{unique\_row\_idx}, and \texttt{unique\_etype\_ptr} describes the mapping from (source node index, edge type index) to the unique index.}
\end{figure}

\subsubsection{Compact Tensor Materialization and Data Layout}
\label{sec:materialization}

{The} Hector inter-operator level IR deliberately abstracts away the data layout from the model semantics. As exemplified by Listing~\ref{lst:ir_example}, the IR only expresses the association of variables with nodes or edges, e.g., \texttt{e["att"]} and \texttt{n["att\_sum"]}, without dictating the mapping of elements in the conceptual variable to the memory space. 

In Hector, we devised compact materialization, which is a technique enabled by the decoupling between model semantics and data layout. 
Note that certain edge data are determined by sparse combinations of source node features and edge types, e.g.  $\overrightarrow{{msg}_{HGT}}$ in Figure~\ref{fig:rgat_layer}. Rather than computing and storing such data for each edge, we instead compute and store the data once for each $\left(\text{edge type}, \text{unique node index}\right)$ pair  that actually exists, reducing the resources spent on computing and storing common subexpressions.
As exemplified in Figure~\ref{fig:compact_opt_opt}, the materialized tensor involves seven rows when each row vector corresponds to a \texttt{msg} of an edge.
Alternatively, the system can materialize the tensor with only five rows, where each row vector corresponds to a \texttt{msg} of an $\left(\text{edge type}, \text{unique node index}\right)$ pair.
We call the former vanilla materialization and the latter compact materialization.
For the vanilla scheme, the row number is the edge index specified by the sparse adjacency. For the compact scheme, it is a unique non-negative integer assigned to each $(\text{source node}, \text{edge type})$. We precompute this mapping and store it in a CSR-like format. Hector does not create the temporary weight tensor, as explained in Section~\ref{sec:segmentmm}.
In summary, compact materialization is a technique to eliminate repetitive identical computations and results in edgewise operators. It is applicable when an edgewise operator depends only on the source node data and edge type, and its output has the shape of $(\text{number of edges}, \text{hidden dimension size})$. After this optimization, the output shape is reduced to $(\text{number of unique} \allowbreak (source\ \allowbreak node, edge\ type)\text{ pairs}, \text{hidden dimension size})$, and repetitive computation is eliminated.
Section~\ref{sec:dse_eval} provided further analysis of the effects of compact materialization on memory footprint reduction.

\begin{table}[!htbp]
\centering
\begin{tabular}{@{}l@{}l@{}l@{}l@{}}
\multicolumn{4}{l}{\textbf{Methods of graph variables}}\\\hline
\multicolumn{1}{|l}{node iterator}        & \multicolumn{3}{l|}{\texttt{g.dst\_nodes()}, \texttt{g.src\_nodes()}}                                                      \\\hline
\multicolumn{4}{|l|}{
\noindent{
\begin{tabular}[t]{@{}ll|ll@{}}
\hspace{-2.25pt}edge iterator        & \texttt{g.edges()}  & weight slicing, e.g.,    &\texttt{W[e.etype]} \\
\end{tabular}}}     \\\hline
\multicolumn{1}{|l}{neighbor iterator}    & \multicolumn{3}{l|}{\texttt{n.incoming\_edges()}, \texttt{n.outgoing\_edges()}}\\\hline 
\multicolumn{4}{l}{\rule{0pt}{10pt}\textbf{Attributes}}\\\hline
\multicolumn{4}{|l|}{
\noindent{
\begin{tabular}[t]{@{}ll|ll@{}}
\hspace{-2.25pt}nodes                & \texttt{e.src}, \texttt{e.dst}                       & types                       & \texttt{e.etype}, \texttt{n.ntype}  
\end{tabular}}}     \\\hline
\multicolumn{4}{|l|}{
\noindent{
\begin{tabular}[t]{@{}ll|ll@{}}
\hspace{-2.25pt}input data, e.g.,    & \texttt{n.feature}      &produced data, e.g.,& \texttt{e["att"]}
\end{tabular}}}     \\\hline
\multicolumn{4}{l}{\rule{0pt}{10pt}\textbf{Operators}}\\\hline
\multicolumn{2}{|l}{GEMM-eligible computation, e.g.,}        & \multicolumn{2}{l|}{\texttt{linear()}, \texttt{outer\_prod()}}                                                      \\\hline
\multicolumn{2}{|l}{GEMM-ineligible computation, e.g.,}    & \multicolumn{2}{l|}{\texttt{dot\_prod()}} \\\hline
\multicolumn{2}{|l}{manipulation, e.g.,}    & \multicolumn{2}{l|}{\texttt{reshape()}, \texttt{concat()}} \\\hline
\end{tabular}
\caption{Hector inter-operator level IR constructs. The graph's variable is named as \texttt{g}, node's as \texttt{n}, and edge's as \texttt{e}. }\label{tab:ir_constructs}
\end{table}

Besides tensor materialization, the multi-level IR design also allows data layout optimizations involving 1)~architecture-specific optimizations, e.g., padding, and 2)~various sparse adjacency encoding.
At the inter-operator level, data layout specifications are decoupled from the model semantics and do not influence the transform passes at this level. 
However, they determine the data access scheme and make a difference when generating CUDA code at the intra-operator level.
Hector inter-operator level IR bookkeeps the specifications, which are passed to the intra-operator level during lowering. 
The intra-operator level operator instances choose the data access scheme corresponding to the data layout specifications while assembling the kernel code.
We leave the exploration of data layout optimizations to future work and detail our plan in Section~\ref{sec:future_work}.

\subsubsection{Linear Operator Reordering}
\label{sec:inter_op_opt}
\begin{figure}[!htbp]\captionsetup[subfigure]{font=small}
\centering
\subcaptionbox{The original inter-operator level IR.}
[\linewidth]{\includegraphics[scale=1.8]{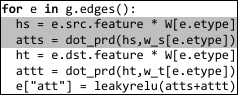}}
\subcaptionbox{After the linear operator reorder, the gray region in (a) is rewritten.}
[\linewidth]{\includegraphics[scale=1.8]{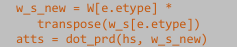}}
\subcaptionbox{Visualization of computing \texttt{atts} in (a). Orange parentheses mark the computation order change after linear operator reorder.}
[\linewidth]{\includegraphics[scale=1.8]{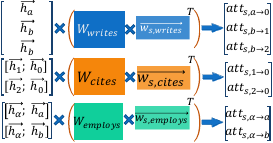}}
\caption{\label{fig:linear_opt} In the example graph in Figure~\ref{fig:opt_example}, when computing edge attention of RGAT, linear operator reordering could be applied. (a) shows the original inter-operator level IR to compute RGAT edge attention. (c) visualizes the computation of the first term, \texttt{atts}, and uses the orange parentheses to mark how the linear operator reordering changes the order of the computation. (b) The transformation rewrites the code.}
\end{figure}

Linear operator reordering is an inter-operator level optimization. When a linear operator, e.g., linear layer and dot product, is followed by another linear operator, their order may be switched. 
For example, for \texttt{atts} as shown in Figure~\ref{fig:linear_opt}(c), we may calculate $W_r\vec{w}_{r}^T$ first instead. Its profitability can be determined by counting the number of multiplication and addition involved in the two GEMMs before and after the order is changed. For now, we implement the pass to switch the orders of two linear operators whenever this produces an operator between weights, because it reduces the complexity by reducing one of its factors, the number of nodes/edges, to the size of hidden dimension. For simplicity, rewritten operator instances use PyTorch BMM to compute the product of weights and apply PyTorch slicing when necessary.

\subsubsection{Graph-Semantic-Aware Loop Transformation}\label{sec:graph_aware_loop}
Loop transformation at this level is augmented with the graph-semantic-specific equivalence rule: a for-each loop over the edges is equivalent to a for-each loop nest iterating over all the incoming/outgoing edges of all destination or source node. Loop transformation is applied during the lowering pass to canonicalize and fuse loops in order to more thoroughly identify kernel fusion opportunities.

\subsubsection{Lowering Inter-Operator Level IR}

To lower the IR to the intra-operator level, Hector greedily lowers every eligible operator to instances derived from GEMM templates~(Section~\ref{sec:two_templates}). Then, it fuses each remaining region and lower them to as few traversal instances~(Section~\ref{sec:two_templates}) as possible.
To achieve this, Hector scans the code three times. Each time, it attempts to lower operators to instances of a specific preference level. During the first pass, it attempts to lower operators to GEMM-template-derived instances. In the next pass, it attempts the traversal-template-derived instances. The third pass will lower all the remaining operators to PyTorch function calls.
During each pass, whenever an operator can be lowered, Hector marks the operator itself, together with all subsequent operators that can be fused into it, with the lowering decision. 
After all the operators have been examined in a pass, the marked operators are lowered and fused. Before the second pass, it canonicalizes the for loops and fuses loop nests whenever possible to discover kernel fusion opportunities.

\subsection{Intra-Operator Level IR}\label{sec:intra-op-ir}

{The} intra-operator level IR serves between the inter-operator level IR and the generated CUDA code. At this level, the IR should encode specifications to emit CUDA code and provide sufficient information specific to each operator invocation to the transform and lowering passes at the inter-operator level. 
The code transformation components at this level provide the methods to generate specialized CUDA code for the operators, to apply operator-specific schedules, and to return necessary information on operator selection and kernel fusion feasibility to the passes at the inter-operator level.

Hector's code generator ultimately lowers the IR to two basic constructs, the GEMM template and the traversal template. 
Algorithms~\ref{algo:gemm_template} and~\ref{algo:traversal_template} illustrate the edge traversal template and the GEMM template. 
The node traversal template is similar to Algorithm~\ref{algo:traversal_template}, and we will revisit it in Section~\ref{sec:op_schedule}.
For simplicity, function template specialization refers to routines specialized for the specific instances derived from the two templates and involve 1)~function arguments, e.g., number of rows, etc., 2)~special registers, e.g., \texttt{threadIdx}, and 3)~loop variables.

\subsubsection{The GEMM Template and the Traversal Template}\label{sec:two_templates}
We base the code generation on GEMM and traversal templates because RGNNs involve not only sparse operations but also multiple dense operations to project vectors across different semantic spaces.
The GEMM template serves edgewise and nodewise linear transformations, as exemplified by the computation of RGAT edge messages in Figure~\ref{fig:compact_opt_opt}. The GEMM template is defined as a matrix multiply augmented with custom gather and scatter schemes. It is formulated as $Y[S] = X[G] \times W[T]$ where $Y$, $X$, $W$ are output, input, and weights, respectively; $S$, $G$, and $T$ are scatter list, gather list, and the type of the nodes or edges, respectively. 
The traversal template performs generic nodewise or edgewise operations. It serves operators that cannot be lowered to GEMM templates, e.g., edgewise dot products.

As shown in Algorithm~\ref{algo:gemm_template}, the GEMM template is based on tiled matrix multiplication. The GEMM template starts with %
the work assignment per block during the \texttt{GetRange<kid>} subroutine~(line 1). 
The \texttt{idxTileRow} and \texttt{idxTileCol} whose range is determined by \texttt{GetRange<kid>} is used to position the workload.
Typically, it is the coordinate of the tile of the output matrix.
Factors that affect $X$'s loading scheme, \texttt{LoadXToShmemIfInRange<kid>}, and $W$'s, \texttt{LoadWToShmemOrRegistersIfInRange<kid>}, involve whether gather lists or transpose needs to be applied on the fly~(lines 4-5).
Gather list $G$ in the Input section is sometimes needed to locate the rows in the source matrix $X$: For example, in Figure~\ref{fig:compact_opt_opt}(a), \texttt{row\_idx} is needed in step \textcircled{1}.
The required information will be passed during the lowering.
The operator instance then accordingly chooses the data access scheme code piece for kernel code generation.
The storing scheme \texttt{StoreCIfInRange<kid>} depends similarly on whether a scatter list will be applied. 
Atomic intrinsics are used in the case of multiple simultaneous updaters.

In the traversal template, as shown in
Algorithm~\ref{algo:traversal_template}, the edge type, node indices retrieval scheme in lines~5-7 depend on the sparse adjacency encoding.
Similarly to the GEMM template, when a row vector needs to be loaded or stored, the tensor materialization scheme determines how the row is located in the materialized tensor.
All statements are initially inserted into the innermost loop. 
After Hector finishes the loop transformations, it then defines work assignment on line~1 in Algorithm~\ref{algo:traversal_template} for the operator instance derived from the traversal template using a simple scheme. For example, if the loop nest is three levels, as exemplified by Algorithm~\ref{algo:traversal_template}, we assign the outermost loop, i.e., \texttt{idxEdge} or \texttt{idxNode} loop, to each thread block and the two inner loops to the multi-dimensional threads in each block.

\begin{algorithm}[!htbp]
{
\KwIn{References of Tensor $Y, X, W$, gather list $G$, etc.}
\texttt{tileRowRange}, \texttt{tileColRange} $\gets$ \textbf{\texttt{GetRange<kid>}}()\;
\ForEach{\texttt{idxTileRow} $\in$ \texttt{tileRowRange}}{
\ForEach{\texttt{idxTileCol} $\in$ \texttt{tileColRange}}{
\textbf{\texttt{LoadXToShmemIfInRange<kid>}}()\;
\textbf{\texttt{LoadWToShmemOrRegistersIfInRange<kid>}}()\;
\texttt{\_\_syncthreads()\;}
\texttt{Y\_reg} $\gets$ \texttt{X\_shmem} $\times$ \texttt{W\_shmem\_or\_reg}\;
\texttt{\_\_syncthreads();}
}
\textbf{\texttt{StoreYIfInRange<kid>}}()\;
}
}
\caption{Hector's GEMM template in pseudo-code. Each instance is assigned a unique identifier \texttt{kid} and gets function template specialization \textbf{\texttt{FuncName<kid>}}.}
\label{algo:gemm_template}
\end{algorithm}

\begin{algorithm}[!htbp]
{
\KwIn{References of input and output tensors. Other necessary data, e.g., adjacency.}
\texttt{eRange}, \texttt{hRange}, \texttt{fRange} $\gets$ \textbf{\texttt{GetRange<kid>}}()\;
\ForEach{\texttt{idxEdge} $\in$ \texttt{eRange}} {
    \ForEach{\texttt{idxHead} $\in$ \texttt{hRange}}{
        \ForEach{\texttt{idxFeat} $\in$ \texttt{fRange}}{
        \texttt{eType} $\gets$ \textbf{\texttt{GetEType<kid>}}()\;
        \texttt{srcIdx} $\gets$ \textbf{\texttt{GetSrcId<kid>}}()\;
        \texttt{dstIdx} $\gets$ \textbf{\texttt{GetDstId<kid>}}()\;
        \tcp{initial insertion point}
        }
    }
}
}
\caption{Hector's edge traversal template in pseudo-code. Similarly to Algorithm~\ref{algo:gemm_template}, each instance gets specialized \textbf{\texttt{FuncName<kid>}}.}
\label{algo:traversal_template}
\end{algorithm}

\subsubsection{Adapting to Different Sparse Adjacency Encoding}
At the intra-operator level, the templates work for any sparse adjacency encoding as long as specific interfaces are implemented. For example, the edge traversal shown in Algorithm~\ref{algo:traversal_template} works as long as the function template specialization \texttt{GetEType<kid>}, \texttt{GetSrcId<kid>}, and \texttt{GetDstId<kid>} are implemented: If the sparse adjacency is COO, \texttt{GetSrcId<kid>} is a subscript operator applied to the row indices array. If it is CSR, then \texttt{GetSrcId<kid>} is a binary search in the row pointer array.

\subsection{Rationale of the Hector Two-Level IR}
\label{sec:ir_design}

Central to the code generator is the two-level IR.
Inter-operator level IR optimizations address the opportunities brought in by heterogeneous relation types. These optimizations manipulate operators and their connections. A high-level IR %
abstracts away the low-level details that can complicate or even hinder the transformations. 
Intra-operator level IR optimizations reduce the data movement by generating access schemes in kernels rather than using specialized kernels and dedicated indexing/copying kernels. These optimizations manipulate low-level data access and schedule details, and thus are better supported by a low-level IR.

The two-level IR enables concerted but decoupled choices of intermediate data layout and compute schedules.
For example, in Figure~\ref{fig:runtime_arch}, the semantics of the model are decoupled from the layout choices.
Hector implements the model semantics and layout choices in intra-operator level IR with specific access schemes.
The next few paragraphs explain how the two-level IR design facilitates operator-specific optimizations, operator selection, and kernel fusion.

\subsubsection{Operator-Specific Schedule}
\label{sec:op_schedule}
Each instance derived from the GEMM template provides the option to apply a coarsening factor in $\{2,4\}$, to choose the tile size, and to apply \texttt{\_\_launch\_bounds\_\_} that limits the number of registers in exchange for more active warps. 
The coarsening factor is the number of elements each thread deals with in the loading, computing, and storing stages. When applied, each block still works on the same assignment, but its number of threads shrinks by the factor~\cite{PMPP4}.
We also allow a per-row scalar to be applied to the tiles of matrix $A$.
This eliminates the extra memory-intensive traversal to perform weighted vector summation by attention or norm.

As for the traversal template, similarly to the discussion in Section~\ref{sec:graph_aware_loop}, we incorporate graph-semantic-aware loop transformation rules that allow Hector to leverage graph semantics to open up the trade-off between more data reuse opportunities and greater parallelism. As mentioned in Section~\ref{sec:two_templates}, initially, all statements are in the innermost loop in each instance derived from the traversal template. 
Loop hoisting is performed to enhance data reuse: The template features insertion points before and after the end of each loop level. For each statement, Hector finds the outermost level where it can be placed before applying the template. 
In addition, the template also provides a partial result aggregation method, which is applied during lowering by default, to reduce global memory traffic by accumulating results within a thread and within a warp before atomically adding them to the data in global memory.

\subsubsection{Operator Selection and Kernel Fusion}
Transformation and lowering passes at the inter-operator level need information about operator instances, specifically operator preference %
and the feasibility of kernel fusion.
Preference level is the mechanism Hector uses to select the operator instance when there are multiple candidates. For example, an operator instance derived from the GEMM template may %
 have an alternative derived from the traversal template but the alternative would lead to lower performance due to much lower data reuse. 
For good performance, operator instances derived from the GEMM template are assigned a higher preference level than those derived from the traversal template unless otherwise specified. Instances that fall back to PyTorch have the lowest preference level.

Operator instances also provide methods to determine the feasible operators to be fused within the IR. %
Operator instances derived from the GEMM template can be fused with the consumer if 1)~the latter multiplies the row vectors in the GEMM output with scalars and 2)~the two operators are in the same loop~(nest). Operator instances derived from the traversal template can be fused with each other as long as they are in the same loop~(nest).
If the inter-operator level pass finds that some temporary variables are created and merely used inside the fused operator, it passes that knowledge to the method so that the variable no longer needs to be created in the global memory.

\subsection{Backward Propagation}
\label{sec:bck_prop}

Similarly to PyTorch, Hector supports auto-differentiation by %
maintaining the backward propagation counterparts of the operators.
Hector first emits the backward propagation via inter-operator level IR and removes unused gradients and their computation.
The lowering and code generation schemes are similar to those in forward propagation.
However, additional processing is needed because the PyTorch auto-differentiation requires the backward propagation methods to be paired with the forward propagation methods in the \texttt{autograd.Function} definitions. To achieve this, Hector bookkeeps the kernel calls in each forward propagation method. For each forward propagation method, Hector puts all the corresponding backward propagation kernel calls in the body of the backward propagation method.

\subsection{Code Generation}
\label{sec:code_gen}
The code generation procedure emits code based on the CUDA kernel specifications detailed in the form of intra-operator IR. Kernel code generation is fairly straightforward and is implemented using a template-based approach. 
Hector then emits the host functions that configure grids and blocks, gets raw pointers from the \texttt{libtorch} \texttt{at::Tensor} references, and launches the corresponding kernel. The host functions are exported via \texttt{pybind11} utilities. 

The Hector performs a pass that scans all the functions generated to collect a list of preprocessing required for the input dataset, involving transposition, converting COO to CSR, etc. The code generator then emits the preprocessing code.

\subsection{Applicability of the Optimizations to GNNs.}\label{sec:gnn_applicability}
Linear operator reordering and compact materialization are specific to RGNNs. Linear operator reordering is specific to RGNNs because RGNNs typically require linear projections from different semantic spaces, introduced by the heterogeneity of node types and edge types, to a common space before further operations. 
Compact materialization is specific to RGNNs because of the additional tensor dimension brought in by different node types and edge types. 

Some of the intra-operator IR optimizations could benefit ordinary GNNs, which can be treated as a special case of RGNNs whose relation type number is one. Intra-operator level IR allows specification of both data access schemes and schedules, thus allowing flexible code generation to accommodate different dense or sparse tensor layouts, a need that often arises from compact materialization. However, the ability to generate code for different data access schemes and schedules can be beneficial when compiling ordinary GNNs.

\section{Evaluation}
\label{sec:eval}

We evaluate Hector with the following questions to answer.

\begin{enumerate}[Q1.]
\item How does the performance of Hector compare with state-of-the-art systems? How does Hector achieve it?
\item How much improvement do the two optimizations detailed in \cref{sec:materialization,sec:inter_op_opt}, compaction materialization and linear operator reordering, make? 
\item Any architectural insights for GPU for RGNNs?
\end{enumerate}

\cref{sec:baseline_eval} answers Q1. \cref{sec:dse_eval} answers Q2 and further analyzes the performance implications of the two optimizations through a case study. \cref{sec:arch_analysis} addresses Q3.

\begin{table}[!htbp]
\centering
\begin{tabular}{lll|lll}
\toprule
\textbf{Name} & \multicolumn{1}{l}{\textbf{\begin{tabular}[c]{@{}l@{}}\#\ nodes\\(\#\ types)\end{tabular}}} & \multicolumn{1}{l|}{\textbf{\begin{tabular}[c]{@{}l@{}}\#\ edges\\(\#\ types)\end{tabular}}}  & \textbf{Name} & \multicolumn{1}{l}{\textbf{\begin{tabular}[c]{@{}l@{}}\#\ nodes\\(\#\ types)\end{tabular}}} & \multicolumn{1}{l}{\textbf{\begin{tabular}[c]{@{}l@{}}\#\ edges\\(\#\ types)\end{tabular}}}    \\ \midrule
aifb                           & 7.3K~(7)    & 49K~(104) & fb15k                           & 15K~(1)  & 620K~(474) \\ 
am                           & 1.9M~(7) & 5.7M~(108) & mag                           & 1.9M~(4) & 21M~(4) \\ 
bgs                              & 95K~(27)   & 673K~(122)&  mutag                          & 27K~(5)   & 148K~(50) \\     
biokg                           & 94K~(5)   & 4.8M~(51)&   wikikg2                     & 2.5M~(1) & 16M~(535) \\ \bottomrule 
\end{tabular}
\caption{Heterogeneous graph datasets~\cite{aifb, mutag, bgs, am, huOpenGraphBenchmark2021, toutanovaObservedLatentFeatures2015} used in our evaluation. The numbers reflect the default preprocessing by the OGB and DGL packages, e.g., adding inverse edges.}\label{tab:datasets}
\end{table}

\subsection{Experimental Setup}
\label{sec:eval_methodology}

To assess performance, we measure the inference and training time of Hector and other systems on a single-GPU computer. Its hardware components include one Intel Core i9-9900K CPU, 128 GB dual-channel memory, and one Nvidia RTX 3090 GPU with 24 GB memory. The operating system is Ubuntu 18.04.5, with kernel version 5.4.0-135. The CUDA and driver versions are 12.1 and 530.30.02, respectively. PyTorch and DGL versions are 2.0.1 and 1.1.1, respectively.

As shown in Table~\ref{tab:datasets}, we use public datasets from DGL~\cite{wang2019deep} and OGB~\cite{huOpenGraphBenchmark2021}.
We measure (1)~inference and (2)~training time on three RGNN models,  RGCN~\cite{rgcn}, RGAT~\cite{busbridge2019relational}, and HGT~\cite{hgt}, comparing with previous systems, involving DGL~\cite{wang2019deep}, PyG~\cite{fey2019fast}, Seastar~\cite{wuSeastarVertexcentricProgramming2021}, Graphiler~\cite{xieGraphilerCompilerGraph}, and HGL~\cite{guiHGLAcceleratingHeterogeneous}.
We ported Seastar artifacts to the same version of CUDA and Python packages as Hector depends on because one of Seastar's dependencies, dgl 0.4, used an API deprecated since CUDA~11.

For RGCN, RGAT, and HGT, excluding comments, Hector took in 51 lines in total and produced more than 3K lines of CUDA kernel code, 5K lines of other C++ code to define host functions, and 2K lines of Python code to define subclasses of PyTorch \texttt{autograd.Function}. The implementation also involves 2K lines of Python code providing common utilities.

To best align with the hyper-parameters prior work used in its evaluation, we set the input and output feature dimensions as 64 and the number of heads as 1. 
We measure the inference and training time of the single layer used. 
In training, to obtain a loss, we compute the negative log-likelihood loss by comparing the output with a precomputed random label tensor. 
For each case, we run the full graph inference and training for at least 10 epochs and average the elapsed time.
To align with the existing system, nodes are presorted to enable segment MM for typed linear layers.

\subsection{Comparison with Prior Work}
\label{sec:baseline_eval}

For the performance of DGL and PyG, we measure all public implementations of these models from DGL, PyG, and Graphiler artifacts.
PyG provides two RGCN convolution layers: \texttt{RGCNConv} places nodes in segments of the same type but launches separate kernels for each of the node types, leading to device underutilization.
\texttt{FastRGCNConv} replicates weights and uses \texttt{bmm()}. It is consistently faster than the \texttt{RGCNConv} implementation.
Similarly, DGL's built-in segmentMM-based RGCN layer is faster than other DGL implementations.
For HGT, the DGL segmentMM-based \texttt{HGTConv} primitive generally has the best performance.
In the cases where some variants encounter OOM errors, we choose the best among those that run without issues.
Some cases are missing due to insufficient operator support, such as HGL on HGT and Graphiler on training. We do not measure HGL in inference because it is designed to optimize training.

\begin{figure}[!htbp]\captionsetup[subfigure]{font=small}
\centering
\subcaptionbox{Training time}
[\linewidth]{\includegraphics[scale=0.8]{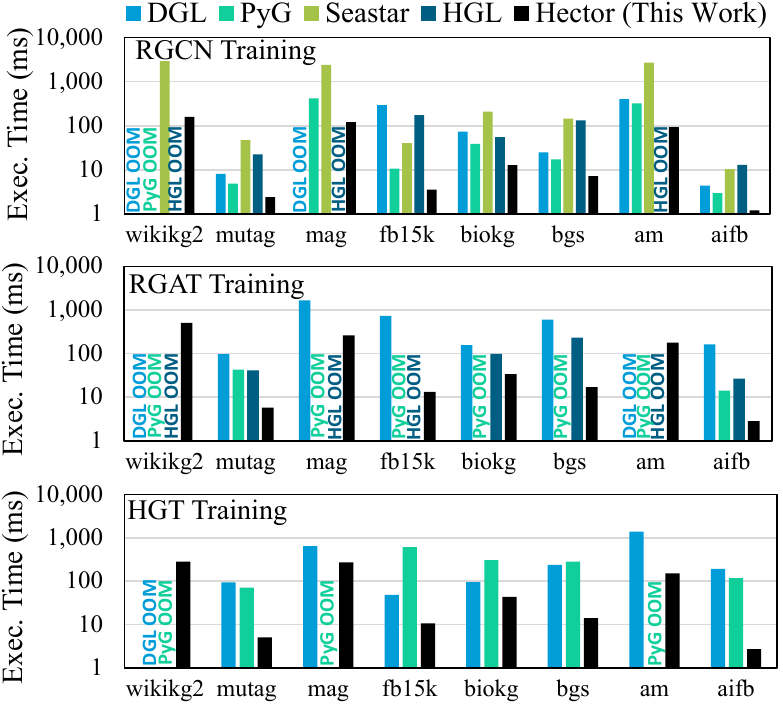}}
\subcaptionbox{Inference time}
[\linewidth]{\includegraphics[scale=0.8]{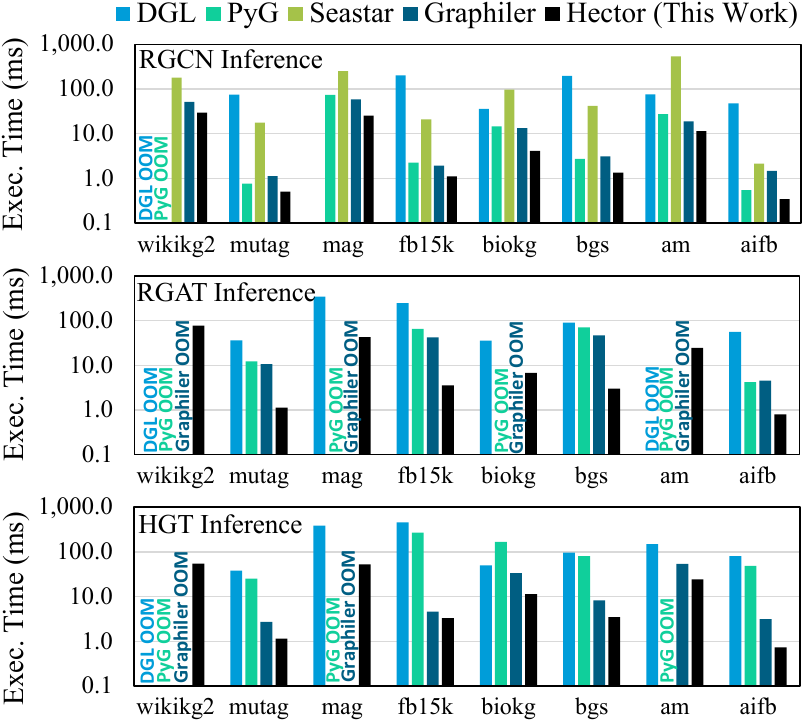}}
\caption{\label{fig:inference_results} Comparing the execution~(Exec.) time of Hector best optimized code with previous work. Table~\ref{tab:datasets} shows the datasets used. }
\end{figure}

Figure~\ref{fig:inference_results} shows that Hector's best-optimized code consistently outperforms state-of-the-art systems. It achieves up to 9.9$\times$ speed-up in inference and up to 43.7$\times$ speed-up in training against the best state-of-the-art systems. On geometric average, Hector gets 1.79$\times$, 8.56$\times$, 2.87$\times$ speed-up in inference via RGCN, RGAT, and HGT, respectively, and 2.59$\times$, 11.34$\times$, 8.02$\times$ speed-up in training RGCN, RGAT, and HGT, respectively. The performance advantage is larger in small graphs, demonstrating that \textbf{generating a single kernel that performs the computation across multiple edge types boosts the performance on small graphs %
compared to existing systems that run many small kernels}.

We see close performance achieved by Graphiler in RGCN and HGT inference. Graphiler leverages PyTorch utilities to produce TorchScript binaries before execution and utilizes edgewise parallelism for edgewise computation. Similarly to \texttt{RGCNConv}, it places node features into segments of the same type but runs separate kernels to perform a typed linear transformation. DGL and PyG, under similar configurations, achieve competitive performance. However, when it comes to RGAT, Graphiler suffers from performance degradation. Because Graphiler relies on pre-programmed fused kernels to deliver a significant portion of the performance boost~\cite{xieGraphilerCompilerGraph}, we postulate that the degradation is due to the non-exhaustiveness of these pre-programmed kernels~\cite{xieGraphilerRepositoryGithub2023}.
This reflects the drawbacks of compiler design without a code generation mechanism. By contrast, with two-level IR and a code generator, Hector achieves better performance, showing that \textbf{generating kernels with flexible access scheme that gather and scatter data on the fly eliminates redundant data movement and outperforms indexing/copying followed by hand-optimized GEMM and sparse kernels}. Besides, it is challenging to extend Graphiler's approach to training due to TorchScript's limited auto-differentiation support. For example, \texttt{dict} object creation is not supported, but it is a common way to express nodewise and edgewise data. 

By comparing Hector with Seastar, which lowers all logic to sparse kernels, we realize that \textbf{sparse kernel code generation alone is not efficient in RGNNs: it is better to lower to GEMM kernels as much as possible}.

There are two reasons why Hector is more efficient in device memory usage. First, Hector only keeps a single copy of weights, as discussed in Section~\ref{sec:materialization}. Replicating weights also affects backward propagation because the gradient of each copy will be derived, occupying extra memory. Second, our compact materialization reduces memory and redundant computation, as explained in Section~\ref{sec:dse_eval}.

Notably, even without compact materialization or linear operator reordering, Hector still consistently outperforms existing systems, as Table~\ref{tab:base_vs_baseline} shows. In addition, the unoptimized Hector code triggers fewer OOMs than existing systems, with the only exception where the RGAT inference is run on mag and wikikg2. 
For comparison, we also show the statistics of the best optimized Hector code in Table~\ref{tab:base_vs_baseline}.

\begin{table}[!htbp]
\centering
\begin{tabular}{lllllc|lllc}
 \toprule
&     & \multicolumn{4}{c}{\textbf{Training}}               & \multicolumn{4}{c}{\textbf{Inference}}              \\
&     & \multicolumn{1}{l}{\textbf{W}} & \textbf{M} & \textbf{B} & \multicolumn{1}{l}{\textbf{\#E}}  & \textbf{W} & \textbf{M} & \textbf{B} & \textbf{\#E}\\ \midrule
\multirow{3}{*}{\rotatebox[origin=c]{90}{\textbf{unopt.}}}& RGCN & 2.02 & 2.59 & 3.47  & \underline{0} & 1.51 & 1.79 & 2.19 & \underline{0} \\
& RGAT & 1.72 & 9.14 & 43.7 & \underline{2} & 1.41 & 5.02 & 9.89 & \underline{2} \\
& HGT  & 1.53 & 6.62 & 28.3 & \underline{0} & 1.20 & 1.90 & 4.31 & \underline{0} \\\hline
\multirow{3}{*}{\rotatebox[origin=c]{90}{\textbf{b.\ opt.}}}& RGCN & 2.02 & 2.76 & 3.48  & \underline{0} & 1.51 & 1.91 & 3.20 & \underline{0} \\
& RGAT & 4.61 & 11.3 & 55.4 & \underline{0} & 5.29 & 8.56 & 15.5 & \underline{0} \\
& HGT  & 2.17 & 8.02 & 43.1 & \underline{0} & 1.40 & 2.87 & 7.42 & \underline{0} \\
\bottomrule
\end{tabular}
\caption{Comparing to the best in state-of-the-art systems, speed-ups of Hector unoptimized~(unopt.) code and that of Hector best optimized~(b.\ opt.) code. Worst~(W), average~(M), and best~(B) cases. Numbers of OOMs Hector triggers~(\#E) are shown. \label{tab:base_vs_baseline}}
\end{table}

\subsection{Effects of Compact Materialization and Linear Operator Reordering}
\label{sec:dse_eval}
Now, we study the effects of compact materialization and linear operator reordering. They are detailed in \cref{sec:materialization,sec:inter_op_opt}. 
We investigate their effects on RGAT and HGT.

Table~\ref{tab:optimizations} shows the speed-up on top of Hector unoptimized code by these two optimizations. Due to compact materialization, Hector no longer triggers OOM errors when running RGAT on mag and wikikg2. In addition, in some cases, the layout speeds up the execution due to the common subexpression elimination brought forth by the layout. Compact materialization is hardly possible without a code generation scheme or an IR design that decouples the model semantics, data layout, and operator-specific schedule.
Besides, \textbf{data layout choice, compact materialization, in particular, allows further performance enhancement} while prior work usually focuses on improving the schedule given a specific sparse matrix format. This is shown by the significant speed-ups in the ``C[ompact]'' columns in Table~\ref{tab:optimizations}.

\begin{table}[htbp!]
\centering
\begin{tabular}{clccc|ccc} 
\toprule
\multicolumn{2}{l}{\multirow{2}{*}{\textbf{}}} & \multicolumn{3}{c}{\textbf{Training}}                                                                                          & \multicolumn{3}{c}{\textbf{Inference}}                                                                                          \\
\multicolumn{2}{l}{}                           & \multicolumn{1}{l}{\textbf{C}}           & \textbf{R}                               & \multicolumn{1}{c}{\textbf{C+R}}         & \textbf{C}                               & \textbf{R}                               & \textbf{C+R}                              \\ 
\midrule
\multirow{9}{*}{\rotatebox[origin=c]{90}{RGAT}}  & aifb    & \cellcolor[HTML]{D97460}0.80 & \cellcolor[HTML]{EEF4EC}\textbf{1.14}           & \cellcolor[HTML]{E08E7D}0.84          & \cellcolor[HTML]{FEFFFE}1.01          & \cellcolor[HTML]{E9F0E6}\textbf{1.19}           & \cellcolor[HTML]{F3F7F1}1.10          \\
 & am      & \cellcolor[HTML]{F2D1CB}0.94 & \cellcolor[HTML]{F2F6F0}\textbf{1.12}                    & \cellcolor[HTML]{FDF8F7}{0.99} & \cellcolor[HTML]{DAE6D5}1.31          & \cellcolor[HTML]{DEE8D9}1.28                    & \cellcolor[HTML]{C0D4B7}\textbf{1.54} \\
 & bgs     & \cellcolor[HTML]{F2D0C9}0.93 & \cellcolor[HTML]{EAF0E7}\textbf{1.18}           & \cellcolor[HTML]{FBFCFA}{1.04} & \cellcolor[HTML]{DDE8D9}1.29          & \cellcolor[HTML]{D7E4D2}{1.34}           & \cellcolor[HTML]{BCD1B3}\textbf{1.57} \\
 & biokg   & \cellcolor[HTML]{39771E}2.67 & \cellcolor[HTML]{E1EADD}1.26                    & \cellcolor[HTML]{38761D}\textbf{2.68} & \cellcolor[HTML]{38761D}\textbf{3.76} & \cellcolor[HTML]{D0DFC9}1.40                    & \cellcolor[HTML]{38761D}{3.74} \\
& fb15k & \cellcolor[HTML]{E8EFE5}1.20 & \cellcolor[HTML]{E7EFE4}1.20 & \cellcolor[HTML]{E0EADB}\textbf{1.27} & \cellcolor[HTML]{C5D7BD}1.50 & \cellcolor[HTML]{E1EADD}1.26 & \cellcolor[HTML]{B6CDAC}\textbf{1.62} \\
 & mag     & \cellcolor[HTML]{C3D6BB}1.51 & \multicolumn{1}{l}{\cellcolor[HTML]{F9FBF8}OOM} & \cellcolor[HTML]{BCD1B3}\textbf{1.57} & \cellcolor[HTML]{FFFFFF}1.00*          & \multicolumn{1}{l}{\cellcolor[HTML]{F3F7F2}OOM} & \cellcolor[HTML]{F8FAF7}\textbf{1.07} \\
 & mutag   & \cellcolor[HTML]{CC4125}0.70 & \cellcolor[HTML]{EFF4ED}\textbf{1.14}           & \cellcolor[HTML]{CC4125}0.73          & \cellcolor[HTML]{E4EDE0}1.23          & \cellcolor[HTML]{E3ECE0}{1.24}           & \cellcolor[HTML]{D5E2CF}\textbf{1.36} \\
 & wikikg2 & \cellcolor[HTML]{F4F8F3}1.09 & \multicolumn{1}{l}{\cellcolor[HTML]{F7FAF6}OOM} & \cellcolor[HTML]{F1F6EF}\textbf{1.12} & \cellcolor[HTML]{FFFFFF}1.00*          & \multicolumn{1}{l}{\cellcolor[HTML]{EDF3EB}OOM} & \cellcolor[HTML]{FEFEFD}\textbf{1.02} \\
 & AVERAGE & \cellcolor[HTML]{F0F5EE}1.13 & \cellcolor[HTML]{EBF1E8}1.17                    & \cellcolor[HTML]{EAF1E7}\textbf{1.18} & \cellcolor[HTML]{D5E2CF}1.36          & \cellcolor[HTML]{DEE8D9}1.28                    & \cellcolor[HTML]{C6D8BE}\textbf{1.49}
                  \\ 
\hline
\multirow{9}{*}{\rotatebox[origin=c]{90}{HGT}} & aifb    & \cellcolor[HTML]{F9E8E5}0.97 & \cellcolor[HTML]{C2D5B9}\textbf{1.52} & \cellcolor[HTML]{D1DFCA}1.40          & \cellcolor[HTML]{F0C9C1}0.92 & \cellcolor[HTML]{90B381}\textbf{1.94} & \cellcolor[HTML]{BBD0B1}1.58          \\
& am      & \cellcolor[HTML]{FAFCF9}1.05 & \cellcolor[HTML]{F1F6EF}1.12          & \cellcolor[HTML]{E9F0E6}\textbf{1.19} & \cellcolor[HTML]{F9FBF8}1.06 & \cellcolor[HTML]{DAE6D5}1.32          & \cellcolor[HTML]{CEDDC7}\textbf{1.42} \\
& bgs     & \cellcolor[HTML]{FEFCFB}1.00* & \cellcolor[HTML]{F2F6F0}1.11          & \cellcolor[HTML]{EAF1E8}\textbf{1.18} & \cellcolor[HTML]{F3D5CF}0.94 & \cellcolor[HTML]{E1EBDD}\textbf{1.25} & \cellcolor[HTML]{E3ECDF}1.24          \\
& biokg   & \cellcolor[HTML]{D6E3D1}1.35 & \cellcolor[HTML]{FCFDFB}1.03          & \cellcolor[HTML]{CFDEC8}\textbf{1.41} & \cellcolor[HTML]{CBDBC4}1.45 & \cellcolor[HTML]{F8FAF6}1.07          & \cellcolor[HTML]{BBD0B2}\textbf{1.58} \\
& fb15k   & \cellcolor[HTML]{E7A79A}0.88 & \cellcolor[HTML]{F2F6F0}\textbf{1.11} & \cellcolor[HTML]{F8E5E2}0.96          & \cellcolor[HTML]{D35E46}0.77 & \cellcolor[HTML]{ECF2EA}\textbf{1.16} & \cellcolor[HTML]{E59F91}0.86          \\
& mag     & \cellcolor[HTML]{E3ECDF}1.24 & \cellcolor[HTML]{F9FBF8}1.06          & \cellcolor[HTML]{D7E4D2}\textbf{1.34} & \cellcolor[HTML]{C9DAC1}1.46 & \cellcolor[HTML]{F3F7F2}1.10          & \cellcolor[HTML]{AAC59F}\textbf{1.72} \\
& mutag   & \cellcolor[HTML]{FEFDFD}1.00 & \cellcolor[HTML]{D9E5D4}\textbf{1.32} & \cellcolor[HTML]{DAE6D5}1.32          & \cellcolor[HTML]{F4D7D1}0.94 & \cellcolor[HTML]{AFC8A4}\textbf{1.68} & \cellcolor[HTML]{C4D6BC}1.50          \\
& wikikg2 & \cellcolor[HTML]{E6EEE2}1.22 & \cellcolor[HTML]{F7FAF6}1.07          & \cellcolor[HTML]{D8E5D3}\textbf{1.33} & \cellcolor[HTML]{E1EBDD}1.26 & \cellcolor[HTML]{EDF3EB}1.15          & \cellcolor[HTML]{C3D6BB}\textbf{1.51} \\
& AVERAGE & \cellcolor[HTML]{F6F9F5}1.08 & \cellcolor[HTML]{ECF2EA}1.16          & \cellcolor[HTML]{E1EADD}\textbf{1.26} & \cellcolor[HTML]{F7F9F5}1.07 & \cellcolor[HTML]{DBE6D6}1.31          & \cellcolor[HTML]{D0DFCA}\textbf{1.40}
        \\
\bottomrule
\end{tabular}
\begin{flushleft} \footnotesize *Normalized by the performance with compact materialization~(C) because the unoptimized version triggers OOM errors. 
\end{flushleft}
\caption{Speed-up on top of Hector unoptimized code due to compaction~(C) and linear operator reordering~(R). Input and output dimensions are both 64. The highest speed-ups per task are in bold. }
\label{tab:optimizations}
\end{table}

\begin{figure}[!htbp]
\centering
\includegraphics[width=\linewidth]{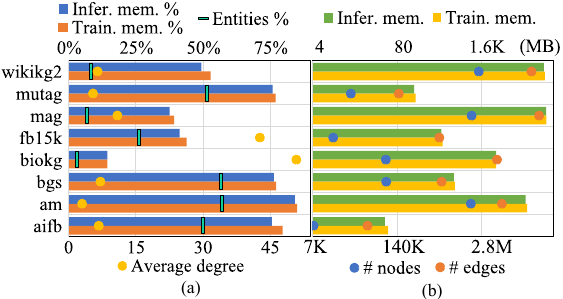}
\caption{\label{fig:dram_analysis} Memory usage when Hector runs training and inference on HGT. (b) shows the inference memory use~(Infer.\ mem.) and training memory use~(Train.\ mem.) of the unoptimized Hector code in MBs. (a) shows the portion of the memory use after applying compact materialization vs. the unoptimized Hector code. For comparison, the number of nodes~(\#\ nodes), number of edges~(\#\ edges), and average degree of datasets are shown as dot scatters. The entity compaction ratio of each dataset is also shown. Legend entries of each data series are placed next to the series' axis.}
\end{figure}

To study how compact materialization reduces the memory footprint, we illustrate the Hector DRAM usage without compact materialization in Figure~\ref{fig:dram_analysis}(b) and the portion of DRAM usage with compact materialization in Figure~\ref{fig:dram_analysis}(a). For simplicity, we define the entity compaction ratio as the number of unique $(\text{source node}, \text{edge type})$ pairs divided by the number of edges. Figure~\ref{fig:dram_analysis}(b) shows that the memory use of inference and training is highly proportional to the number of edges of the datasets. Figure~\ref{fig:dram_analysis}(a) shows that compact materialization significantly reduces DRAM usage in all datasets. The memory footprint ratio of compact materialization compared with the memory footprint of the unoptimized code correlates with the entity compaction ratio. The memory footprint ratio is higher than the entity compaction ratio, as the memory footprint consists of edgewise data, nodewise data, and weights, whereas the compaction applies to edgewise data only. Besides, in case the average degrees are larger, the memory footprint ratio reduces more significantly, getting closer to the entity compaction ratio.

\begin{figure}[!htbp]
\centering
\includegraphics[width=
\linewidth]{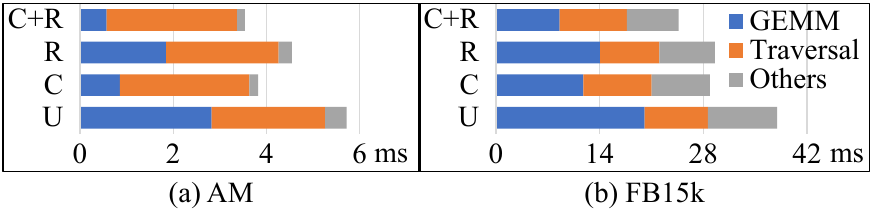}
\caption{\label{fig:perf_analysis} Breakdown of Hector RGAT inference on two datasets. Input and output dimensions are 64. Cases with compaction~(C), linear operator reordering~(R), and no optimization~(U) are presented.}
\end{figure}

To better understand the performance benefits of optimizations, Figure~\ref{fig:perf_analysis} studies two cases. 
The entity compaction ratio of AM and FB15k are 57\% and 26\%, respectively. On AM, the time GEMM instances take is greatly reduced. By comparison, in FB15k, compaction brings less performance improvement due to the less significant GEMM reduction.

In short, \textbf{due to the data-dependent nature of computation in RGNNs, there is no one-size-fits-all optimization strategy}. However, as shown in Table~\ref{tab:optimizations}, enabling compaction and reordering obtains fairly good performance consistently and is the best fixed strategy on average in all four scenarios, i.e., $\{\text{RGAT}, \text{HGT}\}\times\{\text{training},\text{inference}\}$. If Hector presumably chooses the best configuration in every run, it could further get 1.06$\times$, 1.33$\times$, 1.02$\times$, and 1.08$\times$ speed-up in the four scenarios above, respectively. We leave autotuning to future work.

\subsection{Analyzing the Architectural Characteristics}
\label{sec:arch_analysis}

\begin{figure}[!htbp]
\centering
\includegraphics[width=\linewidth]{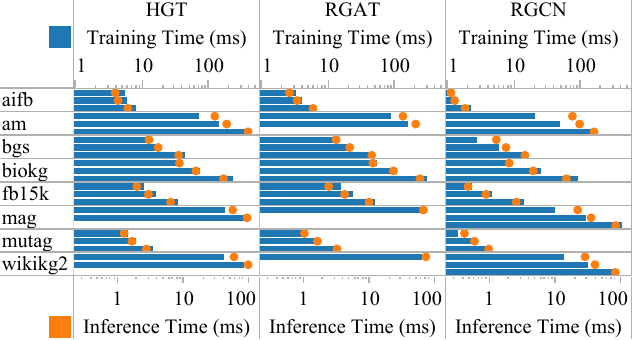}
\caption{\label{fig:perf_curve}Hector unoptimized performance. Each cell corresponds to one pair of dataset and model, where it is shown the time of (input dimension, output dimension) as (32, 32), (64, 64), and (128, 128) from the top to the bottom. Vacancy indicates OOM errors.}
\end{figure}

\begin{figure}[!htbp]
\centering
\includegraphics[width=0.8\linewidth]{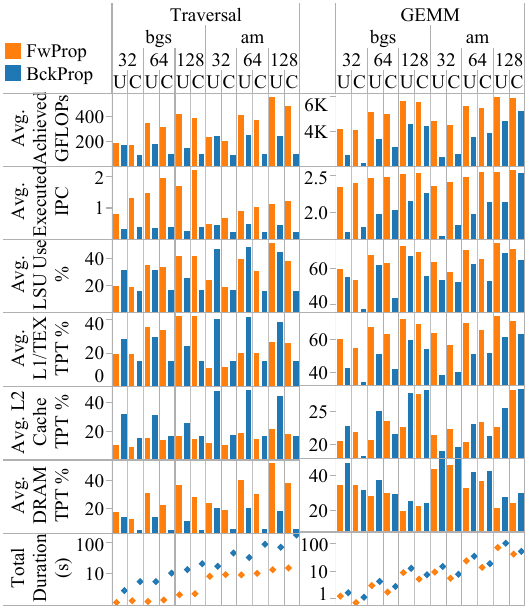}
\caption{\label{fig:arch_number}Architectural metrics of Hector kernels in the forward~(Fw) and backward~(Bck) propagation when running Hector on RGAT with compaction~(C) and without~(U). For each kernel category, aggregated duration and average~(Avg.) metrics, e.g., instructions per cycle~(IPC) and various throughputs~(TPT), are reported.}
\end{figure}

We show the average time of unoptimized Hector in Figure~\ref{fig:perf_curve}. We also profile generated kernels when running Hector on RGAT on bgs and am, as shown in Figure~\ref{fig:arch_number}.

One thing to note is the sublinear time increase in Figure~\ref{fig:perf_curve}: when the input and output dimension doubles, the amount of computation and memory accesses becomes close to 4$\times$ those of the original, but the time increase is typically lower than 2$\times$ of the original. The reason is increased computation throughput when the size increases, as corroborated by Figure~\ref{fig:arch_number}. Moreover, we observed higher throughput when the graph scale increases, e.g., from bgs to am in Figure~\ref{fig:arch_number}. Similarly, we witnessed the cuBLAS throughput increases steadily when we keep the right matrix size as (64, 64) and increase the number of rows of the left matrix from 1M~(2\textsuperscript{17}) to 8M~(2\textsuperscript{20}). These suggest that \textbf{an RGNN system should be memory-efficient in order to accommodate larger models and datasets to fully utilize the massive resources on GPUs}. By eliminating unnecessary data copies, Hector achieves better memory efficiency than state-of-the-art systems.

The instruction per cycle~(IPC) charts in Figure~\ref{fig:arch_number} indicate the traversal kernels are generally latency-bound: on RTX 3090, IPC is ideally four as each SM has four schedulers. Backward propagation kernels have lower throughput due to worsened latency and increased memory bandwidth consumption by doubled memory accesses compared to forward propagation. In backward propagation, backward traversal kernels compute gradients using atomic updates, therefore hindering the throughput; GEMM kernels also, on average, have lower performance due to outer products that compute the delta of weights.

\section{Related Work}
\label{sec:related_work}

\noindent\textbf{General GPU-accelerated GNN libraries.} DGL~\cite{wang2019deep} and PyG~\cite{fey2019fast} are among the most popular GNN Python packages that enable easy development and evaluation of GNN models.  DGL~\cite{wang2019deep} proposes to implement GNN as SpMM/SDDMM operations. PyG's key scheme is scatter and gather operations that switch between edge-parallel regions and node-parallel regions. Hector instead built upon GEMM and traversal templates. By lowering the operators to GEMM as much as possible, Hector obtains better RGNN performance.
Besides,  DGL, PyG, and work based on them do not currently provide inter-operator level IR. Hector shows the benefit of capturing
inter-operator and inter-relation opportunities, e.g., linear-operator reordering, by operator rewrite at the inter-operator level IR. Systems without IR at this level eagerly execute operators without support for such optimizations.

\noindent
\textbf{GNN end-to-end compilers.} Seastar~\cite{wuSeastarVertexcentricProgramming2021} proposes a vertex-centric compiler stack to generate performant kernels throughout the model's training and/or inference. Graphiler~\cite{xieGraphilerCompilerGraph} proposes to program the message passing data flow graph and devises several TorchScript transforms to emit highly optimized inference code. Similarly, HGL~\cite{guiHGLAcceleratingHeterogeneous} is an RGNN compiler. These prior arts 1)~expose PyTorch tensors as operands of all operations to users and 2)~replicate weight to unleash parallelism due to a lack of support for flexible data access schemes and/or code generation. Thus, they suffer more or less from memory inefficiency and performance degradation.
Although the general concept of multi-level IR is not new, Hector proposes new optimizations appropriate for each level and effective in reducing data movement and code bloat in the current state of practice:
Linear operator reordering and compact materialization are two key and novel features to capture and eliminate repetitive computation across edge types. Section~\ref{sec:gnn_applicability} discussed the generalizability of Hector.

\noindent
\textbf{Kernel code optimization.} FeatGraph~\cite{huFeatGraphFlexibleEfficient2020a} proposes a code optimization framework on top of TVM~\cite{chenTVMAutomatedEndtoEnd2018} for user-defined-function-enabled SpMM and SDDMM. Some work proposed optimizations for specific GNN kernels. GE-SpMM~\cite{huangEfficientSparseMatrix2021, huangGESpMMGeneralPurposeSparse2020}, and work~\cite{hidayetogluAtScaleSparseDeep2020} propose optimized schedules for SpMM. Others involve Seastar~\cite{wuSeastarVertexcentricProgramming2021}, PyTorch-Direct~\cite{minLargeGraphConvolutional2021}, and TLPGNN~\cite{fuTLPGNNLightweightTwoLevel2022}. As Hector shows, SpMM/SDDMM is not the only essential kernel in end-to-end RGNN execution. Hector is orthogonal to these prior arts as they can be incorporated into Hector as operator-specific schedules or new templates.

\noindent
\textbf{Code generation.} 
SparseTIR~\cite{yeSparseTIRComposableAbstractions2022} and TACO~\cite{kjolstadTensorAlgebraCompiler2017} propose IR and code generator for sparse tensor operations.
MLIR~\cite{lattnerMLIRScalingCompiler2021} proposes multi-level IR design for deep learning.
Aligned with this direction, FusedMM~\cite{rahmanFusedMMUnifiedSDDMMSpMM2021} unifies the SpMM and SDDMM CPU kernels. 
Hector is different as a higher-level compiler that optimizes the type of operators and generates efficient kernels to handle multiple edge/node types in the RGNN execution. SparseTIR and TACO are tensor-level compilers for sparse operators that may or may not specialize in deep learning.
While we do not intend to reinvent the general-purpose sparse tensor code generator for completeness or performance, some of these works inspire us. They may be incorporated to enhance the Hector code generator.

\section{Discussion on Extensibility}\label{sec:future_work}

\subsection{Support for New Optimizations}
Hector is designed as an extensible framework to prototype and evaluate new techniques. First, inter-operator optimizations can be prototyped as inter-operator level passes. Second, data layout optimizations can be supported by adding the corresponding intermediate data and adjacency access schemes discussed in Section~\ref{sec:inter_op_ir}.
Third, kernel optimizations can be prototyped as a kernel template and operator instances based on it. Alternatively, they can be implemented as operator-specific schedules.

Table~\ref{tab:feature_examples} shows how the proposed compiler could be extended to support common kernel optimizations from the high-performance computing community. Each row in the table shows an example of the new feature to support in bold, followed by an approach to add support to it in our system. For example, to enable row reordering to balance the load, we can add a schedule option at the intra-operator level such that, when enabled, the compiler remaps the row loop index to the row index.

\begin{table}[]
\centering
\begin{tabular}{p{13cm}}
\toprule
\textbf{Features}: How to support them in our system?    \\\midrule
\textbf{Different data reuse strategies \cite{yesilDenseDynamicBlocks2022}}: We can support it by adding primitives in the operator schedule at the intra-operator level IR including tile, tile\_edges, reuse, etc.                                               \\
\textbf{Row reorder to balance load \cite{hegdeExTensorAcceleratorSparse2019}}: We may add a schedule option such that, when it is enabled, the compiler uses a custom remapping function to get the row index from the loop index.                   \\
\textbf{Parameter tuning \cite{vuducAutomaticPerformanceTuning2003}}: We can specify the parameters and set their values in the operator schedule at the intra-operator level IR.                                                                           \\
\textbf{Occupancy and warp efficiency \cite{yangDesignPrinciplesSparse2018}}: At the intra-operator level IR, operator schedule supports GPU-model-specific specification of launch configurations involving threading block size, register number limits, etc. \\
\textbf{Novel sparse matrix format \cite{liuCSR5EfficientStorage2015,yanYaSpMVAnotherSpMV2014}}: Introduce new sparse format access logic into our system.\\
\textbf{Different intermediate data layout~\cite{STRZODKA2012429, HOMANN2018325}:} Introduce new intermediate data access logic into our system. \\
\textbf{Different parallel strategies \cite{yangDesignPrinciplesSparse2018}}: We may achieve this by conducting loop transform and changing the assignment of for-loop levels to architecture levels. We may further specify a custom mapping function to obtain a graph element, e.g., row index, from the loop index. \\\bottomrule              
\end{tabular}
\caption{Examples of new features and ways to incorporate kernel optimization techniques into the proposed compiler.}\label{tab:feature_examples}
\end{table}

\subsection{Use in Distributed Systems}
We focused Hector on single-GPU performance. The kernels Hector generated could serve distributed systems, e.g., DistDGL~\cite{zhengDistDGLDistributedGraph2020}. Since performance improvement results from the reduction of data movements and memory footprints, it also applies to distributed systems.

\subsection{Incorporating TACO} In Hector, we craft the code generators on our own for quick prototyping and focus on high-level optimizations. As Hector establishes our understanding of what constructs are needed in the code generators for the traversal kernels, we think it viable to incorporate TACO for the code generation in the future because TACO provides a mature compiler infrastructure that enables the expression and application of optimizations~\cite{kjolstadTensorAlgebraCompilation2019} for sparse tensor operations in a principled manner, e.g., loop transformations. However, RGNN scenarios still pose several open challenges to TACO, especially in edge-centric operations. Take the edgewise dot product when computing $att_{HGT}$ in Figure~\ref{fig:rgat_layer} as an example. First, to balance the workload, we evenly split the edgewise loop and assign them to threading blocks. If we specify the source-node-wise loop and destination-node-wise loop as two dimensions in the TACO iteration space, we need to fuse these two loop levels to form the edgewise loop to split, but such loop fusion between two loop levels of iteration variables is not supported by TACO yet. Alternatively, we can specify the edgewise loop index as one dimension in the iteration space. In this case, we need indirect addressing to retrieve node data: We need to retrieve \textcircled{1} the source/destination node index by edgewise loop index and then \textcircled{2} the node data. However, indirect addressing is not natively supported in TACO and thus poses the second challenge.

\section{Conclusion}
RGNNs are graph neural networks with dedicated structures for modeling the different types of nodes and edges in heterogeneous graphs. While RGNNs have been increasingly adopted in many real-world applications due to their versatility and accuracy, they pose performance and system design challenges: inherent memory-intensive computation patterns, the gap between the programming interface and kernel APIs, and heavy programming effort required to optimize kernels caused by their coupling with data layout and heterogeneity. To systematically address these challenges, we propose Hector, a novel two-level intermediate representation and its code generator framework that 
(a)~\textit{captures} the key properties of RGNN models, and opportunities to reduce memory accesses in inter-operator scheduling and materialization,
(b)~\textit{generates} code with flexible data access schemes to eliminate redundant data copies, and
(c)~\textit{decouples} model semantics, data layout, and operators-specific optimizations from each other to reduce programming effort. 
By building on one GEMM template and a node/edge traversal template, Hector achieves up to 9.9$\times$ speed-up in inference and 43.7$\times$ speed-up in training compared with the state-of-the-art public systems on select models, RGCN, RGAT, and HGT, when running heterogeneous graphs provided by DGL and OGB. 
In addition, Hector does not trigger any OOM exceptions in these tests. 
We also propose linear operator reordering and compact materialization to further accelerate the system by up to 3.8$\times$. 
As an indicator of the reduction of programming effort, Hector takes in 51 lines of code expressing the three models and generates 8K lines of CUDA and C++ code.
Through profiling, we found that higher memory efficiency allows Hector to accommodate larger input and, therefore, attain higher throughput in forward propagation. In contrast, backward propagation is bound by latency introduced by atomic updates and outer products.

\chapter{PyTorch-Direct: Enabling GPU-Centric Data Access for Very Large Graph Neural Network Training}
\label{ch:pytorch_direct}
\section{Introduction}

\kwa{(Paragraph removed.)}

Compared with traditional neural networks, GPU-accelerated systems in large-scale GNNs suffer from performance penalties caused by low effective PCIe bandwidth.
The scale of graphs in real \kwc{world} is way larger than the tens of gigabytes of capacity the GPU device memory offers;
Therefore, raw data of the graph is stored in the host memory, and during each mini-batch, the input to the model is transferred to the GPU.
\hide{\hl{This introduces performance overhead because scatter in memory and transfer to GPU are expensive.}}
Figure~\ref{fig:GrpahSAGE} illustrates the data layout and transfer during the training of a GNN model.
The features of all nodes in the graph are stored in a two-dimensional array, as shown on the left in Figure~\ref{fig:GrpahSAGE}.
They encode prior knowledge and stay constant during training.
In this example, the vector for node 9 is to be output.
Its neighbors and two-hop neighbors are sampled and required as the input for the graph.
These sampled neighbors are scattered in the node features array.
Unfortunately, transferring the scattered data to GPUs with the existing deep neural network (DNN) libraries is not straightforward.
Initiating a DMA call on each data fragment is too expensive; therefore, the CPUs must first gather the scattered data before the transfer.
For small graphs, this inefficiency can be bypassed by simply loading the whole features into GPU memory, but real-world graphs can go beyond billions of nodes~\cite{Pinterest} and thus far exceed the GPU memory capacity.

Conventional wisdom would argue that since the graph feature data is in host memory, the CPU should have a significant latency and bandwidth advantage over GPUs in performing the gather operations on these features. However, with their ability to issue a massive number of concurrent memory accesses to tolerate latency, GPUs have recently been shown to be effective in accessing data with irregular structures like graphs in the host memory~\cite{minEMOGIEfficientMemoryaccess2020}. If successful, having the GPUs perform gather operations also eliminates the need to perform a data copy from the CPU to the GPU after the feature data has been gathered.
It is, therefore, desirable to explore the use of GPUs to perform feature gather accesses to significantly reduce end-to-end GNN training time. %
This chapter presents PyTorch-Direct, a GPU-centric data access design for GNN training.

PyTorch-Direct adopts zero-copy, \kwc{in which} the node features array is stored in host-pinned memory and can be accessed by GPU \kwc{kernels directly}.
In a zero-copy access, the GPU sends a PCIe read request to the host memory at the time the GPU core dereferences the pointer.
Contrary to the usual belief, after careful optimization on access pattern, zero-copy access yields close to peak PCIe bandwidth~\cite{minEMOGIEfficientMemoryaccess2020}.
Moreover, it removes the redundant data copy in the host memory incurred during a block transfer.
Figure~\ref{fig:design_comparison}(b) shows the transfer procedure after adopting zero-copy access.
Comparing it with the original procedure in Figure~\ref{fig:design_comparison}(a) shows that 1) redundant data copy is eliminated, and 2) finer-granularity zero-copy access replaces block transfer.

\kwc{Nevertheless, i}ncorporating zero-copy into PyTorch is non-trivial.
PyTorch does not support zero-copy.
Nor did PyTorch take cross-device access into consideration in its tensor abstraction.
Specifically, every tensor in PyTorch is bound to a specific device, as illustrated in Figure~\ref{fig:design_comparison}(b).
Such device binding governs the computation device and the physical location of the result tensor.
\hide{\hl{PyTorch is compute-oriented not data-oriented}} PyTorch-Direct devises and implements a full-fledged new tensor type, the unified tensor, accessible by both CPU and GPU.
It is underlain by zero-copy access, enabling the scheme in Figure~\ref{fig:design_comparison}(b), and is seamlessly integrated into the PyTorch APIs and runtime.
Changes needed to adopt it in GNN scripts are minimal.

The contributions of PyTorch-Direct are as follows:

\begin{enumerate}
    \item We identify inefficient host-to-GPU data transfer patterns in existing GNN training schemes that cause high CPU utilization and increase end-to-end training time.
    \item We propose
          a GPU-centric data access paradigm with a novel circular shift indexing optimization for GNN training to reduce training time and CPU utilization.
    \item We seamlessly incorporate the proposed system level changes into a popular DNN library, PyTorch, with a comprehensive implementation to benefit a broad range of GNN architectures.
\end{enumerate}

\section{Background and Related Work}
\subsection{Neighbor Sampling for GNNs}
One shortcoming of early GNN models is the large memory footprint.
As inspired by the Laplacian filter, they usually involve an adjacency matrix in each of their hidden layers, which scales up as the size of the graph increases.

\begin{figure}[!htbp]
    \centering
    \includegraphics[width=0.5\linewidth]{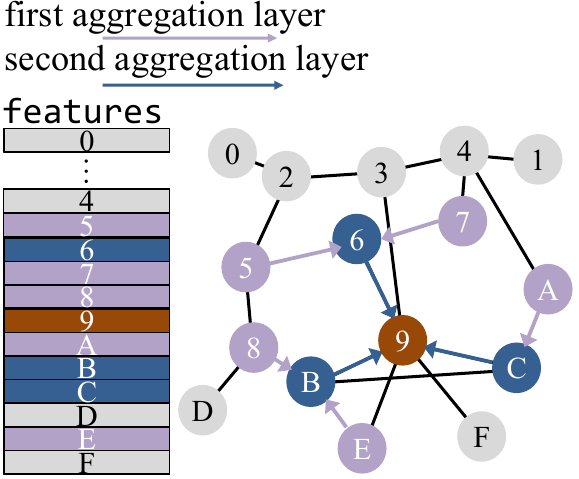}
    \caption{An example demonstrating the GraphSAGE neighbor sampling approach for output node 9. There are two aggregation layers in the model. The left shows the data layout of node features of this graph.}\label{fig:GrpahSAGE}
\end{figure}

To mitigate this, GraphSAGE~\cite{hamilton2017inductive} proposes neighbor sampling along with mini-batch.
It takes in the node pairs chosen in the mini-batch, their sampled neighbors, and multi-hop neighbors rather than the nodes in the whole graph.
This dramatically reduces the memory footprint.
A GraphSAGE model includes two to three aggregation layers, which can be mean, pooling, LSTM, etc.
Figure~\ref{fig:GrpahSAGE} shows an example of neighbor sampling on node 9.
Node indices are represented in hexadecimal.
Neighbors of node 9 are sampled, constituting the input of the second aggregation layer.
Similarly, the neighbors of these neighbors are sampled as the input for the first aggregation layer.
The input of the first aggregation layer is node features from the graph.
Consequently, the sampled node features are scattered in the node features tensor, as the illustration on the left shows.
Gathering needs to be done before DMA block transfer in the original mini-batch input transfer scheme, as Figure~\ref{fig:design_comparison}(a) shows.

\subsection{GPU Out-Of-Memory Solution for GNN Training}
In GNN training, the input features are located in a two-dimensional array where the row indices are the \kwc{identifiers} of nodes and the columns are the features of each node.
In Figure~\ref{fig:design_comparison}, we show a case of retrieving the node features of the neighboring nodes during the GNN training.
Due to the structural discrepancy between the graph and the array, accessing the features of neighboring nodes in the graph results in accessing rather unpredictable and non-sequential rows of the Feature Array.

A straightforward approach to sending these non-consecutive rows to the GPU is to call data copying functions like \texttt{cudaMemcpy()} multiple times, once for each row. 
Unfortunately, making multiple calls to data copying functions incurs significant overhead and can be highly inefficient. 
When the input graphs are small, one can bypass this issue by simply placing
the entire feature array into the GPU memory at the beginning of GNN training.
However, in reality, it is not reasonable to assume that the entire feature array can always fit into the GPU memory.

Currently, the solutions for training GNNs on huge graphs can be divided mainly into two categories:
1) Only the immediately necessary features for the current mini-batch are gathered by the CPU and then sent to the GPU memory~\cite{Pinterest}.
2) Before training, partition the input graphs into multiple smaller subgraphs that can be fit into the GPU memory and then train on them one by one~\cite{Chiang2019ClusterGCNAE,graphsaint-iclr20}.
In the former category, the CPU can become a bottleneck, slowing down the training pipeline. In the latter category, the subgraphs inevitably lose some of the distinct structural patterns of the original graphs~\cite{zu2019survey}.
PyTorch-Direct addresses these deficiencies by enabling the GPU to directly gather all the needed features from the host memory on demand.

\subsection{GNN Frameworks with Python DNN Libraries}
\kwc{To facilitate GNN development, efforts are made to create frameworks that incorporate commonly required functionalities in GNN training based on popular Python-based DNN libraries such as PyTorch and TensorFlow.}
DGL~\cite{wang2019deep} is developed based on MXNet, PyTorch, and TensorFlow. 
PyTorch-Geometric~\cite{fey2019fast} is a PyTorch-based GNN framework.
StellarGraph~\cite{StellarGraph} and Spektral~\cite{grattarola2020graph} are based on TensorFlow and Keras API. In PyTorch-Direct, we demonstrate the benefit of our approach by extending PyTorch.

\subsection{Large Scale GNN Systems}
There is rich literature \kwc{addressing the challenges of} large-scale GNNs.
While \kwc{this body of work highlights} the demands and issues with large-scale GNNs, the novelty of PyTorch-Direct is unique: it mitigates the PCIe bottleneck for these applications by proving the close-to-peak effective PCIe bandwidth of zero-copy in node features gathering and transfer and devising the new APIs and runtime modifications to integrate into PyTorch.
Works~\cite{sign_icml_grl2020,graphsaint-ipdps19,graphsaint-iclr20,Chiang2019ClusterGCNAE, Jia2020ImprovingTA} propose new models to mitigate the memory footprint, such as graph partitioning, layer sampling, etc. They change the algorithm and empirically may worsen the accuracy~\cite{ogbLeaderboard}.
In comparison, PyTorch-Direct applies to all GNN models using neighbor sampling.
Work~\cite{SAGA} proposes a general GNN processing framework able to utilize multiple GPUs and conduct high-level optimizations, including kernel fusions and dataflow optimizations. Still, it does not account for the PCIe transfer efficiency.
Work~\cite{zhengDistDGLDistributedGraph2020} devises a distributed CPU-only GNN processing system, which does not exploit the massive parallelism of GPU\kwc{s}.

Besides, there is much research on large-scale graph processing systems.
Work~\cite{10.14778/3384345.3384358} utilizes unified memory and static graph ordering to mitigate irregular data access, but our work \kwc{also} applies to dynamic graphs. Work~\cite{10.1145/3342195.3387537} proposes a subgraph generation algorithm and uses DMA engine\kwc{s} to perform host-to-device transfer.

\subsection{Ways of Data Transfer among CPU and GPUs}
There are three ways to transfer data among the CPU and GPUs, i.e., API calls for DMA transfers, on-demand paging by Unified Virtual Memory (UVM)~\cite{P100Whitepaper, V100Whitepaper, A100Whitepaper, UVMPrimer, pearson19}, and zero-copy access~\cite{UVMPrimer, minEMOGIEfficientMemoryaccess2020,relTransfer1}.

\kwc{The first way is through} explicit API calls.
Host logic in the program invokes the corresponding APIs to perform data transfer.
The two most commonly used APIs are \texttt{cudaMemcpy()} and \texttt{cudaMemcpyAsync()}, which perform synchronous and asynchronous data copy, respectively.
The programmer must also specify data movement direction, e.g., host to device, device to device, etc.
When the API is invoked, the driver uses the DMA engine to perform the transfer.
If the source data are in the host memory, \kwc{they} will be first copied to a pinned region in the host memory by the CPU, causing extra data movement~\cite{pearson19, gpudirectrdma}.
As Pearson et al.~\cite{pearson19}\ measured, the effective bandwidth is very low when the transfer size is a few KBs and reaches 50\% of the peak bandwidth only when the transfer size is at least 2\textsuperscript{17} to 2\textsuperscript{19} bytes.
Given that each node feature typically takes around 1 KB, the host must first gather the node features into a temporary array in host memory before DMA transfer to well utilize PCIe bandwidth.

On-demand paging is the second way. CUDA provides UVM~\cite{UVMPrimer} to reduce the burden of memory management.
\texttt{cudaMallocManaged()} \kwc{calls} allocate UVM-managed memory regions, which can be accessed by either the host or GPUs in the system.
During a miss, the driver transparently migrates the page from the remote to the local memory.
The migration granularity is between 4 KB and 2 MB.
\kwc{Since} Pascal architecture\kwc{, Nvidia GPUs use the page faulting mechanism to handle missing pages} when access\kwc{ing} a location not in its device memory~\cite{UVMPrimer2, P100Whitepaper}.
UVM provides the programmers with convenience.
Especially, they do not need to explicitly perform deep copies for every referenced memory region.
But \kwc{UVM} is not designed to maximize performance.
As Chien et al.~\cite{Chien_2019} have measured, page faults by unified virtual memory cause non-negligible negative impacts on bandwidth.
Besides, in GNN, in particular, only a few node features may be accessed per page migration, reducing the effective bandwidth.
Furthermore, since the \kwc{total} size of all node features is way larger than the device memory, it may cause excessive eviction, further aggravating the originally severe PCIe bottleneck.

The third method is zero-copy access.
GPU can access any data in the system as long as it is pinned and memory-mapped into the GPU's address space.
In zero-copy access, GPU sends the request through the interconnect to get data, without explicit copying or migration in the previously mentioned two mechanisms.
When accessing host memory, the GPU issues at most cacheline-sized, i.e., 128 bytes, data requests onto PCIe~\cite{minEMOGIEfficientMemoryaccess2020}.
There are three APIs or combinations that enable zero-copy in a memory region~\cite{min2021pytorchdirect}, but for simplicity, we choose \texttt{cudaMallocHost()} in PyTorch-Direct.

\section{Motivation}

\begin{figure}[!htbp]\captionsetup[subfigure]{font=small}
\centering
\subcaptionbox{Baseline PyTorch Approach}
[\linewidth]{\includegraphics[scale=0.55]{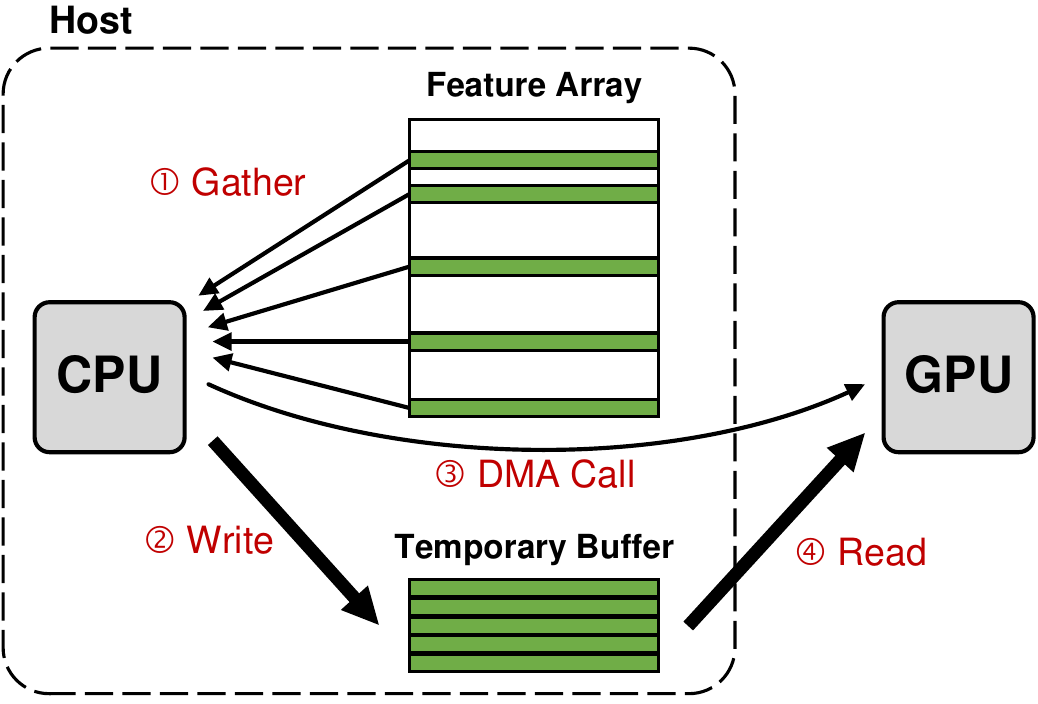}}
\subcaptionbox{PyTorch-Direct Approach}
[\linewidth]{\includegraphics[scale=0.55]{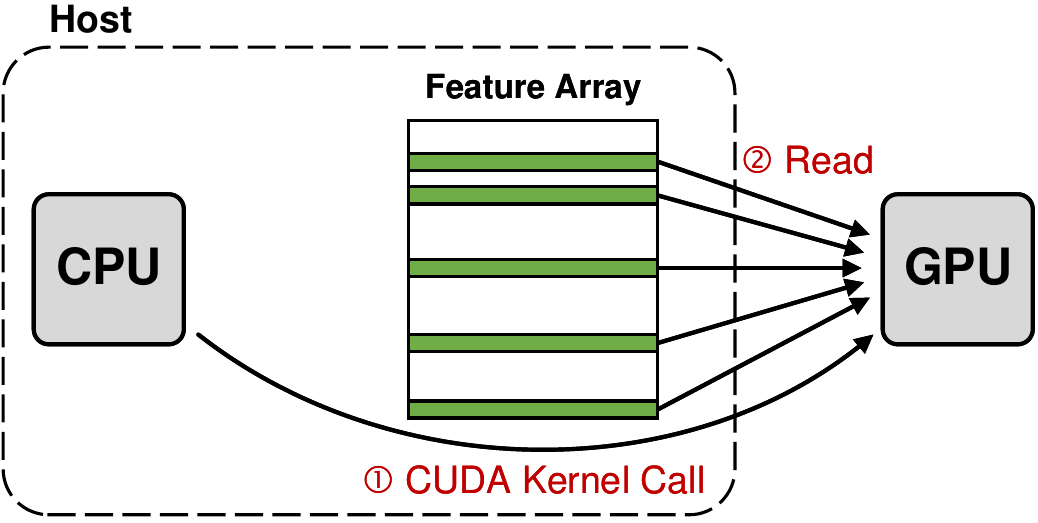}}
\caption{\label{fig:design_comparison} (a) High-level depiction of data transfer mechanism in current PyTorch implementation. (b) Simplified data transfer mechanism in PyTorch-Direct with direct access.}
\end{figure}

In current implementations of deep learning frameworks, the host-to-GPU data loading process is CPU-centric.
When data that needs to be processed by the GPU is scattered in host memory, it is the CPU's responsibility to gather the data fragments before calling a DMA.  Figure~\ref{fig:design_comparison}(a) shows the four main steps of this CPU-centric approach. The CPU first reads (gathers) the features, i.e., relevant rows of the Feature Array in this example, into its cache (\textcircled{1}),
it then writes them into consecutive locations in a temporary buffer (\textcircled{2}) 
before it calls a data copy function to set up a DMA operation (\textcircled{3}) 
and finally, the DMA hardware on the GPU reads the data from the temporary buffer in host memory into a corresponding buffer in the GPU memory (\textcircled{4}).

\begin{figure}[!htbp]
    \centering
    \includegraphics[width=\linewidth]{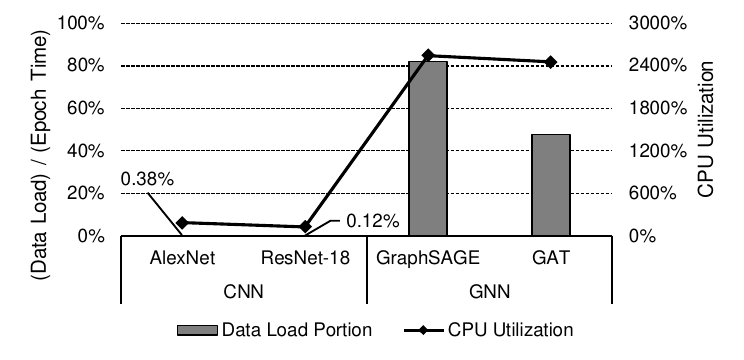}
    \caption{CPU utilization and data loader time comparison between CNN and GNN training. CPU utilization can go beyond 100\% as it is multithreaded.}
    \label{fig:pyd_motivation}
\end{figure}

In Figure~\ref{fig:pyd_motivation}, we show the impact of this CPU-centric data loading approach on GNN training.
As a comparison, we use AlexNet~\cite{krizhevsky2012imagenet} and ResNet-18~\cite{He2015} as CNN examples and GraphSAGE~\cite{hamilton2017inductive} and graph attention network (GAT)~\cite{attention2018graph} as GNN examples.
We use Torchvision~\cite{torchvision} for CNN training and DGL backed by PyTorch for GNN training.
While the time spent for data loading is less than 1\% of the CNN training time, it consumes 47\% and 82\% of the GNN training time for GrapSAGE and GAT, respectively.
As the vertical axis on the right of Figure~\ref{fig:pyd_motivation} shows, CPU utilization is also much higher in GNN training.
This happens partly because the data gathering part of the code is multithreaded and tries to maximize the throughput and thus minimize latency.
Additionally, multi-threading is also used to maximize the performance of graph traversal and subgraph generation during data loading.

In short, in GNN training, unlike CNN training, data loading incurs significant time and resource overheads. In PyTorch-Direct, we aim to reduce this overhead from inefficient use of CPU resources in gather operations.
We propose a GPU-centric approach to accessing data for GNN training based on the direct host-memory-access capability of modern GPUs (Figure~\ref{fig:design_comparison} (b)).
Modern GPUs have their own address translation units and can access host memory directly.
If GPUs are connected over PCIe, they can simply generate PCIe read/write I/O
requests to the host.
From the programmer's point of view, accessing host memory can be simply done by dereferencing unified memory pointers, just like dereferencing device memory pointers.

This direct access feature is different from the conventional unified virtual memory (UVM) method, which is based on page migration.
In UVM, the data transfer between the host and GPU is done in page granularity, which is at least 4 KB per page in modern computing systems.
Whenever a required page is missing from the GPU, the CPU needs to handle the page fault through a hardware interrupt service.
Since the minimum data transfer granularity is a page and the hardware interrupt service process is costly, the performance of the UVM method depends on the applications' spatial and temporal localities~\cite{uvmguide}.
When dealing with highly irregular data structures such as a graph, using UVM incurs excessive page faults and I/O amplification~\cite{10.14778/3384345.3384358,minEMOGIEfficientMemoryaccess2020,10.1145/3342195.3387537}.

In the following section, we describe our implementation of PyTorch-Direct, which enables GPU-centric data accesses for the PyTorch DNN library.
We mainly focus on PyTorch in PyTorch-Direct due to its straightforward and intuitive way of binding data to a certain physical location from the user's perspective.
However, the main idea of the GPU-centric data accessing mechanism can still be applied to other DNN frameworks, such as TensorFlow.

\section{Design and Implementation}

\kwa{(The unused Listing 3.3 is removed.)}

 This section describes the design and implementation of PyTorch-Direct.
First, we provide an overview of design goals and introduce a new type of tensor, i.e., \textit{the unified tensor}, which incorporates new concepts in need.
We then discuss the unified tensor API and its advanced configurations.
Finally, we describe our implementation and optimizations.

\subsection{Overview}
\label{sec.PyTorch_direct.Overview}

PyTorch-Direct aims to enable GPU out-of-memory training and inference for GNN while incorporating the direct access feature to improve data access performance.
To achieve this, PyTorch-Direct presents to the developers several API features centered around a new type of tensor called ``unified tensor''.
It is a new, independent type parallel to PyTorch native GPU or CPU tensors from both the user interface perspective and its implementation in the runtime system. We have developed all the supporting code that allows unified tensors to be used as a full-fledged tensor type in all PyTorch runtime activities such as memory allocator, \texttt{torch.device} class, dispatch, etc. This makes it extremely easy for the application developers to adapt their PyTorch code to use unified tensors.

Unified tensors are at the core of the PyTorch-Direct design, which enables GPUs to directly operate on the host memory.
\kwc{All} CUDA and CPU C++ \kwc{kernels in} PyTorch runtime can directly access unified tensors by simply dereferencing their memory pointers.
In comparison, PyTorch native CPU tensors can only be accessed by CPU, and CUDA tensors can only be accessed by GPU, thus limiting the type of computation devices that can participate in processing these tensors.
Unified tensors eliminate these limitations.
\hide{\hl{Compute-oriented, not data oriented}}

By default, PyTorch-Direct allocates the unified tensors in the host memory and allows GPUs to directly access them over the PCIe.
Since the unified tensors are located in the host memory, their sizes can grow beyond the GPU memory size.
From the CPU's perspective, accessing the unified tensors is identical to accessing CPU tensors.

\begin{lstlisting}[label=unified-example-original,caption=An example of GNN training in PyTorch.,frame=tb,float=]
 # Load features into regular CPU tensor
 features = dataload()
  
 for epoch in range(num_epochs):
   for (neighbor_id, <@\ldots@>) \
     in enumerate(neighbor_sampler):
      
     # Gather features using neighbor_id
     # and then copy to GPU
     input_features = \
       features[neighbor_id].to("cuda")
          
     train(input_features, <@\ldots@>)<@\lstsetnumber{\ldots}@>
     <@\ldots@><@\lstsetnumber{}@><@\lstresetnumber\setcounter{lstnumber}{299}@>

\end{lstlisting}

Application developers can adapt their PyTorch code to use unified tensors with minimal changes to their code.
In Listing \ref{unified-example-original}, we show a simplified example of GNN training in PyTorch.
After loading all the features into host memory, in every training step, it sends the features in the mini-batch to the GPU by calling \texttt{to("cuda")} before invoking the \texttt{train} function (lines 10--13).

The procedure with the unified \kwc{tensor} is shown in Listing \ref{unified-example}.
In this example, to migrate to the unified tensor scheme, the developer only needs to remove the \texttt{to("cuda")} invocation on \texttt{features[neighbor\_id]} and instead invoke \texttt{to("unified")} on \texttt{features} at the beginning.
The features of the whole graph are now stored in a unified tensor that can hold data beyond the GPU memory capacity.
After that, GPU kernels that are launched by the \texttt{train()} function can directly access \texttt{features} since it can access \kwc{a} unified tensor and its derived tensors.
Therefore, \texttt{to()} \kwc{calls are} not needed anymore.
Section \ref{sec.PyTorch_direct.API} describes more about the API design, including advanced configurations.

\kwc{As a full-fledged tensor type,} unified tensor \kwc{facilitates} a clean implementation of complicated rules in runtime systems and easy future extensions.
For example, PyTorch-Direct clearly defines the whole set of rules to resolve computation placement and output tensor placement for computation that involves unified tensors, as detailed in Section \ref{sec:placement_rules}. Thanks to the completeness of the unified tensor, this \kwc{ruleset} is well integrated into the PyTorch runtime system. The implementation details are discussed in Section \ref{sec:impl},

\begin{lstlisting}[label=unified-example,caption=GNN training in PyTorch-Direct with unified tensor. Only two lines (2 and 11) from Listing \ref{unified-example-original} are changed to incorporate unified tensor., frame=tb,float=]
 # Load features into unified tensor
 features = dataload().to("unified")
  
 for epoch in range(num_epochs):
   for (neighbor_id, <@\ldots@>) \
     in enumerate(neighbor_sampler):
      
     # GPU directly fetches required
     # features from unified tensor
     input_features = \
       features[neighbor_id]
          
     train(input_features, <@\ldots@>)<@\lstsetnumber{\ldots}@>
     <@\ldots@><@\lstsetnumber{}@><@\lstresetnumber\setcounter{lstnumber}{299}@>
\end{lstlisting}

\subsection{API Design}
\label{sec.PyTorch_direct.API}

\begin{table}[]
\centering
\begin{tabular}{ll}
\toprule
Example                           & Description                                                                                      \\\midrule
\texttt{t.to("unified")}                   & \begin{tabular}[c]{@{}l@{}}Copy the tensor \texttt{t} to unified\\  device.\end{tabular}                   \\
\texttt{torch.ones(16, device="unified")} & \begin{tabular}[c]{@{}l@{}}Specify unified device in \\ PyTorch native APIs.\end{tabular}        \\
\texttt{t.is\_unified}                     & \begin{tabular}[c]{@{}l@{}}Return true if the tensor \texttt{t} \\ is a unified tensor.\end{tabular}      \\
\texttt{unified\_tensor + cpu\_tensor}       & \begin{tabular}[c]{@{}l@{}}Compute with hybrid tensors \\ of unified and CPU types.\end{tabular} \\
\texttt{unified\_tensor[gpu\_tensor]}   & \begin{tabular}[c]{@{}l@{}}Subscript unified tensor with \\  CUDA tensor.\end{tabular}     \\
\bottomrule
\end{tabular}
    \caption{Typical usage of APIs with unified tensor. Unified tensors are allowed for easy creation and flexible computation.}
    \label{tab:APIUnified}
\end{table}

PyTorch-Direct APIs are designed to provide an interface to unified tensors in the idiomatic PyTorch manner.
Table \ref{tab:APIUnified} demonstrates the typical use of unified tensor APIs.
Developers can create a unified tensor by copying from another tensor via PyTorch built-in \texttt{to()} method of \texttt{torch.Tensor}.
It can also be created from scratch by specifying the \texttt{device} argument as the unified device in PyTorch APIs, such as \texttt{torch.ones}.
The user can check if a tensor is of unified type by \kwc{checking} the \texttt{is\_unified} \kwc{attribute}.

Unified tensors can be computed with CPU or CUDA tensors, providing great flexibility.
Meanwhile, they are free from redundant data movements since the CPU and GPU can directly access their underlying memory without creating temporary copies.
By contrast, in the native PyTorch API, CPU tensors typically cannot work with CUDA tensors because of the device binding unless additional routines to handle them have been implemented in the PyTorch runtime system.
For example, the subscript operator allows a CUDA tensor to be indexed by a CPU tensor, and binary and comparison operators accept GPU scalar and CPU scalar as the two operands.

\subsection{Computation and Storage Placements}
\label{sec:placement_rules}
\begin{table}[!htbp]
\begin{subtable}[]{\textwidth}\centering
\begin{tabular}{lll}
\toprule
\multicolumn{3}{c}{All unified tensors are GPU-affinitive.}        \\
\midrule
\multirow{2}{6.2cm}{No less than one operand is non-scalar CPU tensor.}   & \multicolumn{1}{l|}{compute on} & GPU  \\
 & \multicolumn{1}{l|}{output type}& host-affinitive unified \\
\hline
\multirow{2}{6.2cm}{Previous row is false. And no less than one operand is CUDA type.}  & \multicolumn{1}{l|}{compute on} & GPU  \\
 & \multicolumn{1}{l|}{output type}& GPU   \\
\hline
\multirow{2}{6.2cm}{Operands are either unified tensors or CPU scalars.}   & \multicolumn{1}{l|}{compute on} & GPU  \\
 & \multicolumn{1}{l|}{output type} & GPU   \\
\bottomrule
\vspace{0.05em}
\end{tabular}
\end{subtable}

\begin{subtable}[]{\textwidth}\centering
\begin{tabular}{lll}
\toprule
\multicolumn{3}{c}{At least one unified tensor is host-affinitive.}      \\
\midrule
\multirow{2}{5cm}{No less than one operand is non-scalar CPU tensor.}   & \multicolumn{1}{l|}{compute on}  & \begin{tabular}[c]{@{}l@{}}CPU if no operand is \\ GPU-affinitive, else GPU\end{tabular} \\
& \multicolumn{1}{l|}{output type} & host-affinitive unified  \\
\hline
\multirow{2}{5cm}{Previous row is false. And, no less than one operand is CUDA tensor.} & \multicolumn{1}{l|}{\multirow{2}{*}{compute on}} & \multirow{2}{*}{GPU}\\
& \multicolumn{1}{l|}{}                            &  \\
& \multicolumn{1}{l|}{output type}                 & host-affinitive unified     \\
\hline
\multirow{2}{5cm}{Operands are either unified tensors or CPU scalars.}   & \multicolumn{1}{l|}{compute on}  & \begin{tabular}[c]{@{}l@{}}CPU if no operand is \\ GPU-affinitive, else GPU\end{tabular} \\
& \multicolumn{1}{l|}{output type} & host-affinitive unified  \\
\bottomrule
\end{tabular}
\end{subtable}
\caption{Rules of placements for operators involving unified tensors as operands.}
\label{tab:place_rule}
\end{table}
Though unified tensors can be accessed by both CPU and GPUs, \kwc{we need to define} scheme to determine the computation device and the location of result tensors.
Especially, this \kwc{scheme} may be \kwc{complicated} in scenarios where the operator involves more than two tensors or a hybrid of native tensors and unified tensors.

In the original PyTorch, the dispatch mechanism determines the computation device and result tensor type based on input tensor metadata before executing the operator.
We followed the same idea and integrated a set of lightweight rules into the existing dispatch mechanism.
This allows it to be better integrated into the PyTorch runtime and leads to low overhead in performance and programmer effort to adopt the unified tensor.
There might be more sophisticated ideas, such as computational graphs, but they may drastically change the APIs or cause a bigger performance overhead.

Each unified tensor is designated an affinity mode, either host-affinity or GPU-affinity.
In the simplest scenario where an operator is applied to a unified tensor in host-affinity mode, the computation and results tensors are placed on the host during execution.
Similarly, if this happens to a unified tensor in GPU-affinity mode, the computation and result are placed on the GPU.
The reasoning behind the two modes is simple.
In GNN mini-batch input transfer, we want the results to stay on the GPU as they are consumed by kernels executing the GNN on the GPU, thus avoiding unnecessary data transfers over PCIe.
Therefore, the output tensor should be of CUDA type.
This is what the GPU-affinity mode is for.
On the other hand, the host-affinity mode allows the result tensor to stick to the unified tensor type and allows for more preprocessing.
One can switch \kwc{a host-affinitive unified tensor} to GPU-affinity mode once the preprocessing is done.

Switching the affinity mode of a unified tensor can be done easily by a new tensor method.
It does not incur movements.

Table~\ref{tab:place_rule} shows the complete set of rules.
The number of scalar CPU tensors influences the placement to stay consistent with the existing PyTorch dispatch logic.
The other factor is if CUDA tensors are participating in the operator because, in that case, the only feasible computation devices are GPUs.

\subsection{Implementation}
\label{sec:impl}
While offering seamless API integration into the existing PyTorch design, this project also integrates it into the PyTorch runtime C++ code in a neat, modular, and extensible way.

The goal of implementation is to realize the flexibility and performance benefits of the unified tensor while keeping modifications to existing logic as minimal as possible, especially with the large number of operator definitions.

The core object in the PyTorch runtime system is \texttt{at::Tensor}.
Every PyTorch tensor (\texttt{torch.Tensor} object) is a \texttt{THPVariable}\footnote{\kwc{``\texttt{THP}'' stands for TorcH Python}~\cite{edwardz.yangPytorchTorchCsrc2017}.} object in C++ runtime code, which is the wrapper class combining an \texttt{at::Tensor} object with Python metadata.
The PyTorch runtime dispatches each method call to the proper definition according to the device and data types of the tensor arguments.
A PyTorch method operating on tensors eventually goes into a function of \texttt{at::Tensor}\footnote{\kwc{``at'' stands for the ``A TENsor'' library}~\cite{edwardz.yangATenAtenSrc2018}.}.

PyTorch-Direct implements the unified tensor mechanisms in the PyTorch runtime as a complete type of tensor.
This makes the design modular, extensible, and well-integrated into the PyTorch runtime code.
A new memory allocator is implemented to govern the memory allocation for all unified tensors.
It adapts the allocation \kwc{pool} mechanism from the PyTorch CUDA allocator to reduce the number of CUDA API invocations.

Two dispatch keys are added, corresponding to the two affinity modes mentioned in Section~\ref{sec:placement_rules}.
Dispatch keys specified by each tensor inform the dispatcher to dispatch the operator to the correct backend to get executed.
The introduced zero-copy memory allocator uses PyTorch's pooling idea in PyTorch's original CUDA to reduce API invocations.
Besides, auxiliary logic in the build system and runtime is modified to incorporate the changes.
Only the device-checking logic needs to be changed for most operator definitions, as it now needs to recognize the new unified tensor type.

This project is first developed on top of PyTorch 1.6.
Around 2.6K lines of code are added or modified to incorporate the complete mechanism detailed in this section.
To support the latest CUDA microarchitecture in Section~\ref{sec:pytorch_direct_evaluation}, we then migrated the minimal functional part to nightly PyTorch 1.8.

\subsection{Memory Alignment Optimization}
\label{sec.PyTorch_direct.implementation.alignment}

\begin{figure}[]
    \centering
    \includegraphics[width=0.7\linewidth]{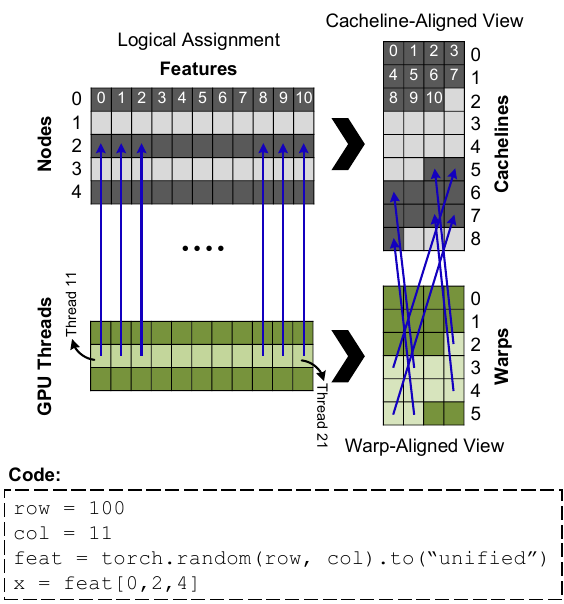}
    \caption{Data access misalignment occurring in PyTorch-Direct when using unmodified PyTorch indexing scheme. Based on the code, thread 0--10 access \texttt{feat[0]}, thread 11--21 access \texttt{feat[2]}, and thread 22--32 access \texttt{feat[4]}. For the case accessing \texttt{feat[2]} (blue arrows), we can easily identify the accesses are fragmented into multiple warps and cachelines.}
    \label{fig:alignment_problem}
\end{figure}

To achieve efficient PCIe data transfer, memory requests from the GPU threads in the same warp should be aligned and merged to the GPU cacheline (128-byte) granularity~\cite{minEMOGIEfficientMemoryaccess2020}.
However, the default PyTorch GPU indexing function does not guarantee memory alignment unless the input feature tensors are naturally aligned with the GPU cacheline size.
In Figure~\ref{fig:alignment_problem}, we depict a simplified working mechanism of the default PyTorch GPU indexing function.
In this specific example, we scale down the warp size (32 threads in real) and the GPU cacheline size (128 bytes in real) by a factor of eight.
We assume each feature is 4 bytes, and each node has 11 features.
Now, due to the size mismatch between the cacheline (16-byte) and the node feature (44-byte), misaligned accesses can occur.

In the example of Figure~\ref{fig:alignment_problem}, assume that the GPU needs to access nodes 0, 2, and 4.
To achieve this, each thread accesses a single feature.
For example, the first 11 threads access the 11 features of node 0; the following 11 threads access the 11 features of node 2, and so on.
This looks simple in a logical view on the left side of Figure~\ref{fig:alignment_problem}, where we highlight the accesses of threads 11--21 to features of node 2.
However, when we redraw the access patterns based on cacheline and warp alignments on the right side of Figure~\ref{fig:alignment_problem}, we see that the accesses are fragmented into multiple cachelines and warps.

\begin{figure}[]
    \centering
    \includegraphics[width=0.7\linewidth]{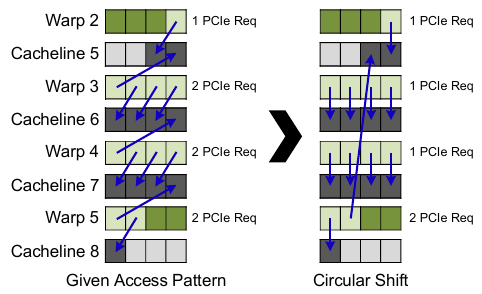}
    \caption{Memory alignment optimization with a circular shift. The example is identical to the case in Figure~\ref{fig:alignment_problem}. Alignment reduces the total number of PCIe requests (req) from seven to five in this case.
    }
    \label{fig:alignment_fix}
\end{figure}

To solve the problem of misaligned access patterns, we use a circular shift method as described in Figure~\ref{fig:alignment_fix}.
In this method, all threads calculate the required index offset values to make aligned accesses.
In the case of Figure~\ref{fig:alignment_fix}, the threads need to do a right shift by an offset of one.
The threads on the edges check the boundary conditions and make additional adjustments by adding or subtracting the length of the node feature so that they do not accidentally access the other node features.
When the indexed values are written to the output, the output indices are also identically adjusted to maintain the ordering.
With the optimization, PyTorch-Direct reduces the number of total PCIe requests from seven to five in this case.
Inside the PyTorch GPU indexing kernel, we check the input tensors and apply this optimization only when the input tensors are unified tensors and the feature widths are not naturally aligned to 128-byte granularity.
All these adjustments are automatically made due to our modifications to PyTorch source code. As such, no programmer effort is required to solve the memory alignment problem.

\section{Evaluation}
\label{sec:pytorch_direct_evaluation}
This section evaluates PyTorch-Direct performance using a well-defined microbenchmark and end-to-end GNN training.
Using the microbenchmark, we demonstrate that (1) PyTorch-Direct is faster than the baseline PyTorch approach in accessing features from the GPU under different combinations of data sizes and systems and (2) the effectiveness of our optimized memory alignment mechanism.
In GNN training, we show the benefit of using PyTorch-Direct for faster training.

\subsection{Evaluation Setup}
\begin{table}[]
\centering
                \begin{tabular}{llcccc}
                    \toprule
                    Abbv.   & Dataset             & \#Features & Size  & \#Node & \#Edge \\
                    \midrule
                    reddit  & reddit              & 602     & 561MB & 233.0K & 11.6M  \\
                    product & ogbn-products       & 100     & 960MB & 2.4M   & 61.9M  \\
                    twit    & twitter7            & 343     & 57 GB  & 41.7M  & 1.5B   \\
                    sk      & sk-2005             & 293     & 59 GB  & 50.6M  & 1.9B   \\
                    paper   & ogbn-papers100M     & 128     & 57 GB  & 111.1M & 1.6B   \\
                    wiki    & wikipedia\_link\_en & 800     & 44 GB  & 13.6M  & 437.2M \\
                    \bottomrule
                \end{tabular}
    \caption{Datasets for GNN training, their characteristics, and abbreviations (abbv.) used in the text.}
    \label{tab:pydarxiv_datasets}
\end{table}

\noindent\textbf{Datasets.} The datasets we use for the GNN training evaluation are shown in Table~\ref{tab:pydarxiv_datasets}.
For the sk-2005~\cite{BoVWFI}, twitter7~\cite{Kwak10www}, and wikipedia\_link\_en~\cite{konect} datasets, we have created them from existing real-world graphs but with synthetic feature values just for the purpose of training time evaluation.
Datasets reddit~\cite{hamilton2017inductive}, ogbn-products, and ogbn-papers100M~\cite{huOpenGraphBenchmark2021} are commonly used datasets in the field for comparing the training accuracies between different GNN architectures.

\noindent\textbf{Test System.} The platforms we have used for the evaluation are described in Table~\ref{tab:hardware}.
We use NVIDIA 450.51.05 driver and CUDA 10.2 on the evaluation platforms.
System2 and System3 configurations are only used in Section~\ref{sec.evaluation.microbenchmarkI}.

\begin{table}[]
\centering
            \begin{tabular}{ccl}
                \toprule
                Config                   & Type & \multicolumn{1}{c}{Specifications} \\
                \midrule
                System1                  & CPU  & AMD Threadripper 3960X 24C/48T    \\
                (Primary)                & GPU  & NVIDIA TITAN Xp 12 GB              \\
                \midrule
                \multirow{2}{*}{System2} & CPU  & Dual Intel Xeon Gold 6230 40C/80T \\
                                         & GPU  & NVIDIA Tesla V100 16 GB            \\
                \midrule
                \multirow{2}{*}{System3} & CPU  & Intel i7-8700K 6C/12T             \\
                                         & GPU  & NVIDIA GTX 1660 6 GB               \\
                \bottomrule
            \end{tabular}
    \caption{Evaluation platforms. \kwc{The number of cores (C) and threads (T) of CPUs are listed in the specifications column.}}
    \label{tab:hardware}
\end{table}

\noindent\textbf{Microbenchmark.} We would like to answer the following questions with the microbenchmark:

\begin{itemize}
    \item How does increasing the feature size affect the PyTorch-Direct performance? The feature sizes vary greatly across datasets. For example, while a node of ogbn-products~\cite{huOpenGraphBenchmark2021} has 100 features, a node of reddit~\cite{hamilton2017inductive} has 602 features.
    \item How does increasing the number of features to be copied affect the PyTorch-Direct performance? Depending on factors such as the connectivity of the input graph and the batch size, the number of neighboring nodes that need to be fetched per batch can vary.
    \item How well does the alignment optimization as discussed in Section~\ref{sec.PyTorch_direct.implementation.alignment} work with misaligned input features?
    \item What is the performance impact of using PyTorch-Direct on different systems?
\end{itemize}
The microbenchmark is designed to mimic the behavior of the data gathering and copy processes in the GNN training.
The microbenchmark uses a random number generator (RNG) to generate random indices, which are used to index feature values.
The total number of items is fixed to 4M for all experiments.

\noindent\textbf{GNN Training.} In this evaluation, we use GraphSAGE~\cite{hamilton2017inductive} and GAT~\cite{attention2018graph} implementations from DGL.
Both implementations have all necessary \kwc{utilities} (e.g., subgraph generation) to perform GNN mini-batching, which makes it suitable to work even if the input graphs cannot fit into the GPU memory.
The features are located in host memory, and during training, only the immediately required features are transferred to the GPU memory.
In the baseline implementation with PyTorch, the required features are gathered by the CPU and then copied to the GPU memory through DMA.
In the PyTorch-Direct implementation, the entire features are located in the unified tensor and the GPU directly accesses only the immediately required features.
Besides the data movement parts, the core training algorithms of the DGL implementations are left unmodified.

\subsection{Microbenchmark - Size and System Dependency}
\label{sec.evaluation.microbenchmarkI}

\begin{figure}[]
    \centering
    \includegraphics[width=0.85\linewidth]{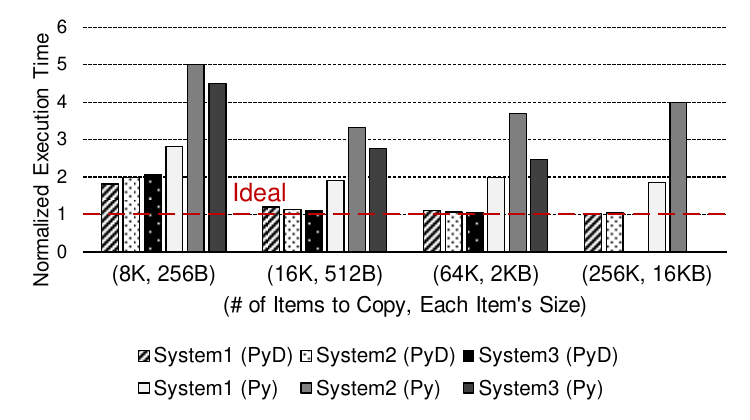}
    \caption{Irregular host data access pattern microbenchmark comparisons between PyTorch (Py) and PyTorch-Direct (PyD) on different systems. The ideal case shows only the pure data transfer time with a peak PCIe bandwidth.}
    \label{fig:micro_fig}
\end{figure}

The result of copying different numbers of features with various sizes is shown in Figure~\ref{fig:micro_fig}.
The ideal case only includes the pure data transfer time under the theoretical peak bandwidth of the interconnect.
Due to the lack of system memory, we do not run the \texttt{(256K, 16KB)} setup with System3.
With the baseline PyTorch approach, the performance varies greatly depending on the system configurations.
While the slowdowns in System2 are about 3.31$\times$ to 5.01$\times$, the slowdowns in System1 are about 1.85$\times$ to 2.82$\times$.
On the other hand, with PyTorch-Direct, we can consistently reach near the ideal performance regardless of the system configuration unless the data transfer volume is very small.
When the total data transfer volume is very small, the overall execution time is dominated by the CUDA API calls and kernel launch overheads.
Except for the \texttt{(8K, 256B)} case, the baseline PyTorch approach shows 1.85$\times$ to 3.98$\times$ slowdowns, while PyTorch-Direct shows only 1.03$\times$ to 1.20$\times$ slowdowns compared with the ideal case.
Overall, PyTorch-Direct shows about 2.39$\times$ of performance improvement on average compared to the baseline PyTorch approach.

\subsection{Microbenchmark - Memory Alignment}

\begin{figure}[]
    \centering
    \includegraphics[width=0.85\linewidth]{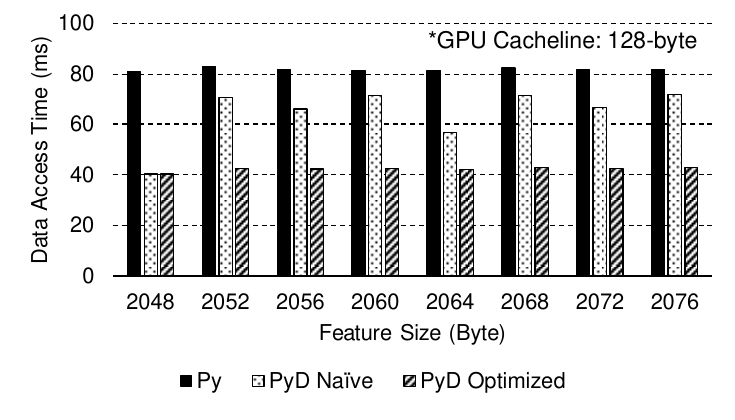}
    \caption{Memory access alignment and its impact on PyTorch-Direct (PyD) performance. PyTorch (Py) results were added for comparison.}
    \label{fig:alignment}
\end{figure}

To evaluate the impact of the memory alignment optimization in PyTorch-Direct, we measure data access times for various feature sizes from 2048-byte to 2076-byte in a 4-byte stride.
The result is shown in Figure~\ref{fig:alignment}.
For the \texttt{PyD Naïve} case, we use the unmodified GPU indexing kernel from PyTorch, and the kernel has no knowledge of memory alignment.
For the \texttt{PyD Optimized} case, the optimization from Section~\ref{sec.PyTorch_direct.implementation.alignment} is applied.

Figure~\ref{fig:alignment} shows that PyTorch-Direct reduces the data access time significantly compared to the PyTorch baseline.
However, the benefit is limited without the memory-alignment optimization.
For example, when the feature size is 2052 bytes, the \texttt{PyD Naïve}  provides only 1.17$\times$ of performance improvement over \texttt{Py}, while the \texttt{PyD Optimized} provides 1.95$\times$ of performance improvement.
Based on the results, we observe the optimization provides a consistent benefit over the PyTorch baseline (averagely 1.93$\times$) regardless of the data alignment.

\subsection{GNN Training Performance}

\begin{figure}[!htbp]\captionsetup[subfigure]{font=small}
\centering
\subcaptionbox{GraphSAGE}
[\linewidth]{\includegraphics[scale=1.2]{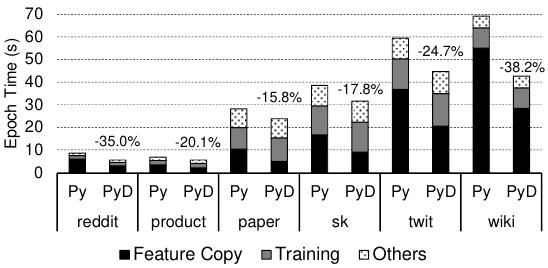}}
\subcaptionbox{GAT}
[\linewidth]{\includegraphics[scale=1.2]{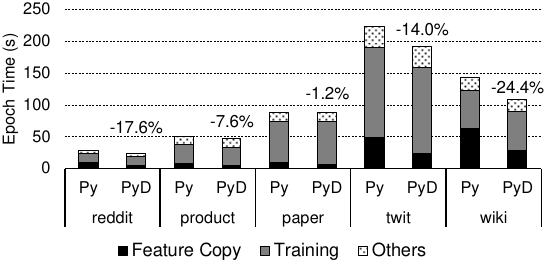}}
\caption{\label{fig:graphsage} Single epoch execution time breakdown for both PyTorch (Py) vs. PyTorch-Direct (PyD) when running (a) GraphSAGE and (b) GAT in different datasets. Training epoch time reductions are written on the bars.}
\end{figure}

In Figure~\ref{fig:graphsage}, we compare the breakdown of the training epoch time when using unmodified DGL implementations in PyTorch vs. PyTorch-Direct.
In the GAT training, we do not run \texttt{sk} dataset due to the DGL's out-of-host-memory error for both PyTorch and PyTorch-Direct cases.
Similar to the microbenchmark results in Section~\ref{sec.evaluation.microbenchmarkI}, we observe about 47.1\% reduction in the feature copy times.
The other portions of the training epoch times remain almost identical to the baseline case.
PyTorch-Direct gives less benefit for datasets with smaller feature sizes (e.g., \texttt{paper}) because the feature copy time is smaller in the end-to-end training time.
Similarly, GAT training is computationally heavier than GraphSAGE, and therefore, we observe less benefit of PyTorch-Direct.
Overall, we observe between 1.01$\times$ to 1.45$\times$ speedup when we use PyTorch-Direct in GNN training.

\section{Conclusion}
With the increasing adoption of GNNs in the machine learning community, GPUs have become essential to accelerate GNN training.
However, training GNNs on massive graphs that do not fit in GPU memory is still a challenging task.
Unlike conventional neural networks, mini-batching input samples
in GNNs requires complicated tasks such as traversing neighboring nodes and gathering their feature values.
While this process accounts for a significant portion of the training time, existing GNN implementations using popular deep neural network libraries such as PyTorch are limited to a CPU-centric approach for the entire data preparation step.
This ``all-in-CPU'' approach negatively impacts the overall GNN training performance as it over-utilizes CPU resources and hinders GPU acceleration of GNN training.
To overcome such limitations, we introduce PyTorch-Direct, which enables a GPU-centric data-accessing paradigm for GNN training.
In PyTorch-Direct, GPUs can efficiently access complicated data structures in host memory directly without CPU intervention.
Our microbenchmark and end-to-end GNN training results show that PyTorch-Direct reduces data transfer time by 47.1\% on average and speeds up GNN training by up to 1.6$\times$.
To minimize programmer effort, we introduce a new ``unified tensor'' type along with necessary changes to the PyTorch memory allocator, dispatch logic, and placement rules. 
As a result, users need to change at most two lines of their PyTorch GNN training code for each tensor object to take advantage of PyTorch-Direct.

\chapter{SSDTrain: Enhancing Large Language Model Training Throughput by Using SSDs to Keep Activations}
\label{ch:ssdtrain}
\section{Introduction}
\label{sec:intro}

\kwa{(Paragraph removed.)}

\begin{figure}[!t]
\centering
\includegraphics[width=\linewidth]{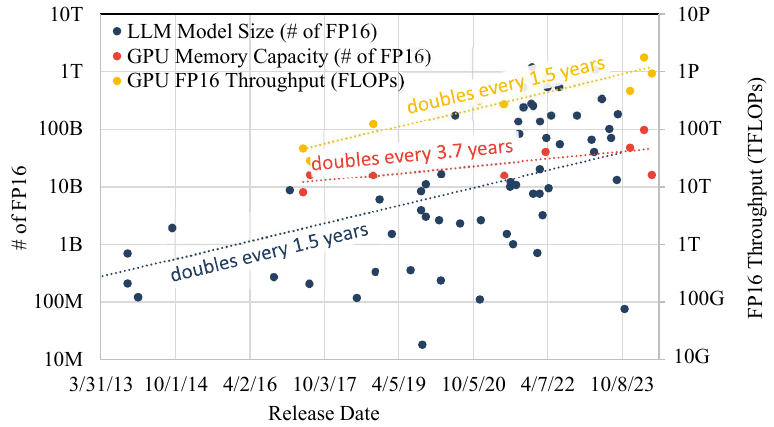}
\caption{\label{fig:trend_scale} The growth of FP16 throughput (right vertical axis) of GPUs for deep learning training is aligned with the model size of LLMs (left vertical axis), but GPU memory capacity (left vertical axis) falls behind~\cite{theepochaiAnnouncingEpochAI2023}. The horizontal axis shows the release date.  Points represent both Nvidia 100-level GPUs since K100 and Google TPUs. The growth rate of FP16 throughput (yellow dotted line) is more than 2$\times$ of that of the memory capacity growth rate (red dotted line).}
\end{figure}

GPU memory capacity has become a bottleneck for the continued growth of LLMs.
As Figure~\ref{fig:trend_scale} shows, the increase of GPU memory capacity is around 60\% slower than the LLM size scaling speed and the GPU FP16 throughput improvement. About 80\% of the GPU memory used to train recent LLMs consists of activations~\cite{liuWinnerTakeAllColumnRow2023,korthikantiReducingActivationRecomputation2022}, the intermediate tensors produced by forward propagation and reused in backward propagation.
Furthermore, the memory needed for activations is growing more rapidly than any other memory use, making GPU memory a more severe constraint for future LLM training (see Section~\ref{sec:llm_scaling} for details).

Common mitigations are to reduce batch size or through gradient accumulation.  With gradient accumulation, a batch is divided into micro-batches that are processed separately between gradient updates.  Although gradient accumulation has been adopted by many LLMs~\cite{jiangMegaScaleScalingLarge2024,shoeybiMegatronLMTrainingMultiBillion2020a,workshopBLOOM176BParameterOpenAccess2023}, the GPU computation stack is not designed for small inputs, and both mitigations lead to device under-utilization~\cite{DissectingBatchingEffects,anthonyCaseCoDesigningModel2024} and suboptimal math library performance~\cite{aminabadiDeepSpeedInferenceEnabling2022}. 
Intuitively, a smaller batch size might reduce total training computation through faster convergence. However, LLM trainers have identified a critical batch size for each model, below which convergence speed increases negligibly or even decreases~\cite{kaplanScalingLawsNeural2020,mccandlishEmpiricalModelLargeBatch2018}.  Notably, critical batch size grows during training as training loss is reduced.

Another common approach to reducing GPU memory use is activation checkpointing.  With this strategy, only some activations are kept in GPU memory, while others are flushed and then recomputed during backward propagation.  For a model with $L$~layers, activation checkpointing can reduce memory requirements from $O(L)$ to $O(\sqrt{L})$~\cite{chenTrainingDeepNets2016}.  However, as we show in Section~\ref{sec:llm_scaling}, even this reduction is insufficient to eliminate the bottleneck posed by the GPU memory limits for future LLMs.

This chapter proposes SSDTrain, a software framework that offloads activations to NVMe SSDs and reloads activations just before they are needed in backward propagation.  SSDTrain can fully overlap activation transfers with computation, reducing activation memory usage without incurring significant performance overhead. 

\begin{figure}[!tb]
\centering
\includegraphics[width=0.6\linewidth]{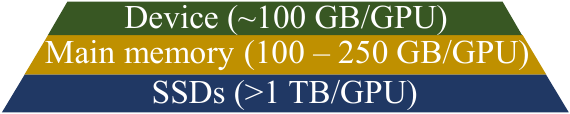}
\caption{\label{fig:current_sys} Current clusters and cloud instances usually have limited main memory~\cite{microsoftNDA100V4series2024,googleGPUMachineTypes,ncsaDeltaProjectProfile}.}
\end{figure}

SSDs are a more attractive target than main~(CPU) memory for several reasons. First, as illustrated in Figure~\ref{fig:current_sys}, clusters and cloud instances~\cite{microsoftNDA100V4series2024,googleGPUMachineTypes,ncsaDeltaProjectProfile} typically have limited host memory capacity (100--250 GB per GPU), while SSDs offer much greater capacity. Host memory capacity is further consumed by input data, checkpointing buffers, and other training management buffers, leaving even less capacity for activation offloading. \kwc{In contrast, as modeled in Section~\ref{sec:projected_life}, the activation size per GPU per training step in large LLM models can reach hundreds of GBs or even TBs, exceeding the capacity of host memory. Additionally, as Section~\ref{sec:llm_scaling} will detail, SSD capacity is increasing faster than the main memory, making SSD the more viable choice in the future.} Second, host memory bandwidth is shared across training management tasks and offloaded computation~\cite{kamahoriFiddlerCPUGPUOrchestration2024,renZeROOffloadDemocratizingBillionScale2021,songPowerInferFastLarge2023} running on the host CPU~\kwc{(Please see further elaboration on \textbf{Swapping and offloading} in Section~\ref{sec:ssdtrain_related})}. \kwc{This shared usage can make host memory bandwidth both limited and unpredictable}~\cite{baeFlashNeuronSSDEnabledLargeBatch2021} for saving and restoring activations. In contrast, the SSD bandwidth can be dedicated to activation offloading during training. 
Third, SSDs are more elastic, both by adding more SSDs and even PCIe switches if necessary---as well as through the use of optional remote high-throughput storage~\cite{googleGoogleCloudHyperdisk,lockwoodArchitecturePerformancePerlmutter2024}. Such elasticity allows the data centers to keep up with the fast-growing size of activations. In contrast, the memory capacity of GPU cloud instances and cluster nodes is much more challenging to extend.

SSDTrain makes the following main contributions:
\begin{enumerate}[1.]
\item To address the GPU memory capacity issue and the resulting GPU under-utilization during LLM model training, we design and implement the SSDTrain framework to offload activations in LLM training to NVMe SSDs. We demonstrate the viability of SSDTrain on large-scale systems by modeling the performance, estimated SSD lifespan, and the required per-GPU PCIe bandwidth. 
\item With all code in Python except for a tiny CUDA memory allocation API hooking library, SSDTrain works with the latest PyTorch and distributed frameworks, including Megatron~\cite{shoeybiMegatronLMTrainingMultiBillion2020a} and DeepSpeed~\cite{rasleyDeepSpeedSystemOptimizations2020}. We developed and tested SSDTrain with Megatron-DeepSpeed~\cite{microsoftMicrosoftMegatronDeepSpeedOngoing2019} on a two-GPU node with seven Intel Optane SSDs. 
\item Because SSDTrain overlaps the data transfer entirely with computation, it incurs almost no performance overhead. To achieve this, we introduce several optimization techniques, including tensor deduplication, tensor forwarding, and adaptive offloading algorithm. 
\item Evaluation shows SSDTrain achieves almost the same training time per step as the original system without SSDTrain while reducing the activations peak memory use by up to 47\%. We introduce the recompute-offload-keep (ROK) curve to compare the SSDTrain offloading with two other tensor placement strategies, keeping activations in memory and layerwise full recomputation. SSDTrain has the same performance as keeping activations in memory and a lower memory peak than activation checkpointing. 
\item \kwc{We further analyze how the reduced activation memory use may be leveraged to increase throughput by increasing micro-batch size and reducing pipeline parallelism bubbles.}
\end{enumerate}

\section{Background and Motivation}

\subsection{GPU Memory Capacity and Model Throughput}
\label{sec:llm_scaling}
As Figure~\ref{fig:eval_dse} of Section~\ref{sec:evaluation} will show, the GPU memory capacity limits the model throughput. By offloading the activations to SSDs, SSDTrain can alleviate this limitation and improve the per-GPU model throughput. An important question is whether the GPU memory capacity will continue to be the limiting factor of per-GPU model throughput according to the trend of LLM scaling. This section shows that the historical trend will make GPU memory capacity an even more critical limiting factor of the per-GPU model throughput.

Neural scaling laws~\cite{jordanhoffmannTrainingComputeOptimalLarge2022,kaplanScalingLawsNeural2020,mccandlishEmpiricalModelLargeBatch2018} guide LLM scaling as computing power increases. We follow these laws in our reasoning.
The whole-system GPU compute throughput $C\propto ND_{batch}$, where $N$ is the number of parameters and $D_{batch}$ is the number of tokens in a batch~\cite{brown2020languagemodelsfewshotlearners}. The Chinchilla scaling law~\cite{jordanhoffmannTrainingComputeOptimalLarge2022} concludes that the optimal model design follows $N\propto C^{0.5}$, which implies $D_{batch}\propto C^{0.5}$ to saturate the GPU throughput.  Whole-system GPU memory use consists of two parts: activations, which require $S_{activations}\propto \frac{N}{h}D_{batch}$, where $h$ is the hidden dimension in the layers and is a slow-growing function of $N$, e.g., $h\propto N^{1/3}$, and all other memory use, $S_{others}\propto N$, including parameters, gradients, and optimizer states. Comparing the factors, we can deduce that (1) $S_{activations}$ grows faster than $S_{others}$, and (2) whole-system memory use, which is dominated by the activations, grows %
slightly slower than the compute throughput $C$ (approximated $C^{5/6}$).
However, Figure~\ref{fig:trend_scale} shows that the historical growth rate of GPU memory capacity (red dotted line) is less than 50\% of that of the compute throughput (yellow dotted line). Therefore, \textbf{GPU memory capacity will become increasingly inadequate for saturating the compute throughput, and memory for activations will continue to dominate the GPU memory usage.}

What about activation checkpointing? Revisiting the prior equation, $S_{activations}\propto \frac{N}{h}D_{batch}\propto LhD_{batch}$ where $L$ is the number of layers. Activation checkpointing makes the new activations memory use $S_{activations}^\prime \propto \sqrt{L}hD_{batch}$. Since $L$ and $h$ grow when $N$ increases and $D_{batch}\propto C^{0.5}$, $S_{activations}^\prime$ still grows faster than $S_{others}$.

\begin{figure}[]
    \centering
    \includegraphics[width=0.8\linewidth]{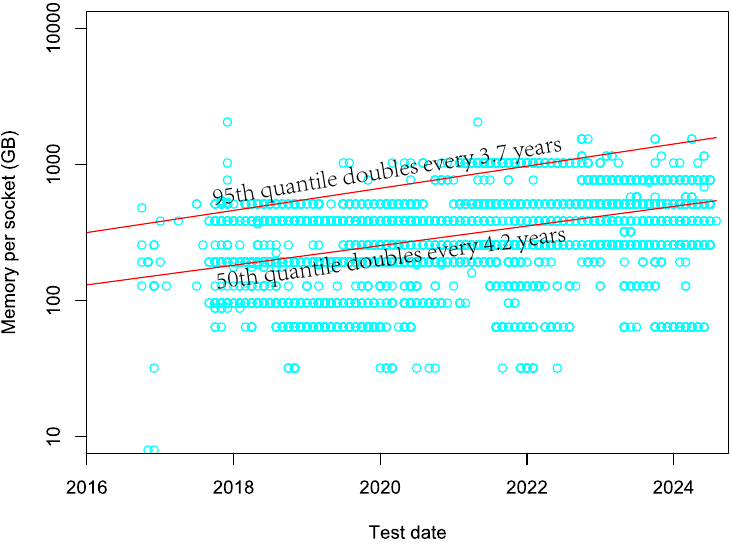}
    \caption{The trend of main memory capacity per CPU socket. Data are a dump of all submitted results of SPEC Open System Group (OSG) benchmarks and High-Performance Group (HPG) benchmarks~\cite{specAllSPECOSG2024}. Data points are deduplicated according to (system vendor, system name, CPU model, main memory capacity). Red lines show the growth rates predicted by quantile regression.  The visualization code is adapted from Derek Jones's work~\cite{derekjonesShapeCodeMemory2020}. }
    \label{fig:main_mem_trend}
\end{figure}

\begin{figure}[]
    \centering
    \includegraphics[width=0.8\linewidth]{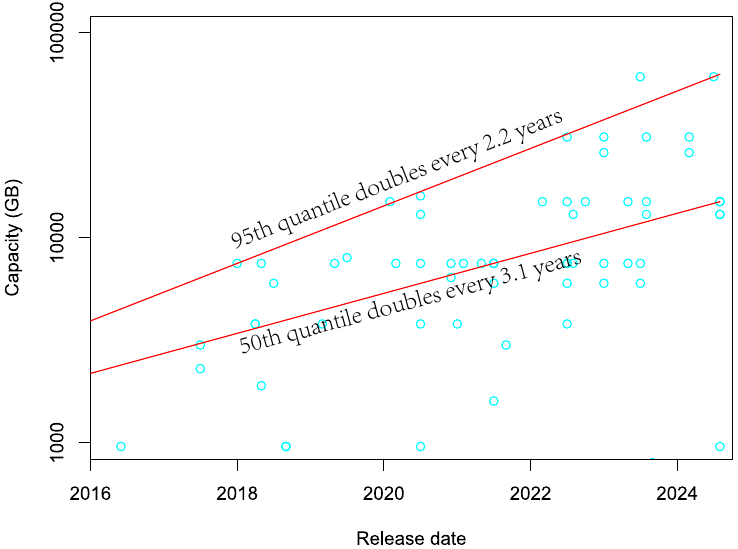}
    \caption{The trend of enterprise SSD capacity~\cite{techpowerupEnterpriseSSDDatabase2024}. For each model, only the data of the variant with maximal capacity is collected. Red lines show the growth rates predicted by quantile regression.  The visualization code is adapted from Derek Jones's work~\cite{derekjonesShapeCodeMemory2020}. }
    \label{fig:ssd_trend}
\end{figure}

Figure~\ref{fig:main_mem_trend} illustrates the trend of the main memory capacity and Figure~\ref{fig:ssd_trend} illustrates the SSD capacity's trend. As shown, the growth of the main memory capacity still falls behind the demand to sustain GPU throughput growth. On the contrary, the SSD capacity \kwc{better} keeps up with such demand.

\subsection{SSD Endurance}

Trends in price, latency, and bandwidth have led to the widespread adoption and integration of SSDs into cloud instances and clusters~\cite{microsoftNDA100V4series2024,googleGPUMachineTypes,ncsaDeltaProjectProfile}.
The random write latency of flash has been reduced to tens of microseconds~\cite{samsungUltraLowLatencySamsung2017}, and NVMe SSD data rates are now a few GB/s.

SSD endurance remains a concern: how long will SSDs last in a write-intensive scenario such as activation offloading?
SSD endurance is determined by the type and number of cells, write amplification factor (WAF), and over-provisioning.
SSD cells can be purposed to store one bit, i.e., single-level cells (SLCs), or multiple levels, e.g., triple-level cells (TLCs).
Generally, the more bits a cell stores, the shorter its lifetime in program/erase (P/E) cycles. WAF is the ratio of media write amount to host write amount---SSD writes pages at a time but erases blocks of pages, a coarser granularity. Erasing a partially empty block requires the remaining valid pages to be relocated, causing write amplification. 
In turn, vendors adopt over-provisioning to reserve some blocks for wear leveling, evening out the writes across blocks.

Table~\ref{tab:ssd} samples current SSD models. The \mbox{D7-P5620} represents a mainstream data center model with 144-layer~(L) TLC cells and a rating of three disk writes per day~(DWPD). The FL6 and \mbox{D7-P5810} SSDs are designed for write-intensive scenarios and have much higher endurance. Notably, SSD endurance rating uses the JESD testing method~\cite{jedecsolidstatetechnologyassociationJESD218BSolidStateDrive2016}, performing random writes after tough preconditioning. In our scenario, the writes are large and sequential, as each tensor being offloaded is easily hundreds of MBs. Such writes are more endurance-friendly than those used to determine the JESD rating. 
For example, \mbox{three-DWPD} SSDs generally allow about 2.5$\times$ as many sequential writes as expected from the JESD rating~\cite{lenovoWhatNeedKnow2023,qnapsystemsinc.QNAPNASSolution2108, smartmodulartechnologiesinc.WhySMARTOverProvisioning2024}. Vendor guidelines~\cite{solidigmSolidigmSSDEndurance,intelOverProvisioningNANDBasedIntel2018,samsungOverProvisioningBenefitsSamsung2019} and empirical data~\cite{maneasOperationalCharacteristicsSSDs2022} corroborate this difference.
Section~\ref{sec:projected_life} conducts modeling to demonstrate why mainstream data center SSDs similar to \mbox{D7-P5620} are viable options to support the deployment of SSDTrain in a large-scale LLM training system.

\begin{table}[!t]
\centering
\begin{tabular}{llll}
\toprule
& \begin{tabular}[c]{@{}l@{}}Kioxia\\ FL6\end{tabular} & \begin{tabular}[c]{@{}l@{}}Solidigm\\ D7-P5620\end{tabular} & \begin{tabular}[c]{@{}l@{}}Solidigm\\ D7-P5810\end{tabular}  \\\midrule
\textbf{3D NAND technology} & 96L SLC & 144L TLC & 144L SLC\\\cline{1-1}
\textbf{\begin{tabular}[c]{@{}l@{}}Endurance rating\\ (DWPD)\end{tabular}}   & 60 & 3 & \begin{tabular}[c]{@{}l@{}}65 (sequential)\\ 50 (random)\end{tabular} \\\cline{1-1}
\textbf{Max capacity} & 3.2 TB & 12.8 TB & 1.6 TB\\\cline{1-1}
\textbf{Max endurance} & 342 PBW  & 65.4 PBW & 146 PBW\\\cline{1-1}
\textbf{Price per PBW} & US\$13.9 & US\$43.8 & US\$11.1\\
\bottomrule
\end{tabular}
\caption{A sample of SSD models in mass production with high endurance in PB writes (PBW)~\cite{solidigmD7P5620MidEndurancePCIe2023,solidigmD7P58102023,neweggSolidigmSolidState2024,kioxiaFL6Series5inch2022,serverorbitKioxiaFL6XHUL1T606TB2024,dihuniSOLIDIGMSSDPF2SQ800GZ01D7P58102024}. \label{tab:ssd}}
\end{table}

\subsection{SSD Offloading Systems for LLM}
\kwc{To offload tensors to SSDs with high performance, SSDTrain utilizes} GPUDirect Storage (GDS)\kwc{, which} enables a direct data path between GPU and local or remote NVMe SSDs~\cite{inupakutikaQuantifyingPerformanceGains2022}. By eliminating the need to involve the CPU for the bounce buffer, \kwc{GDS} enhances bandwidth and reduces both latency and CPU load. 

\kwc{SSDTrain aims to mitigate the training overhead caused by the GPU memory capacity limit, e.g., device underutilization, large pipeline bubble time fraction, etc. In contrast, most existing projects incorporate the offloading mechanism to execute larger models than the original system can fit without offloading at the cost of performance. To this end, there are three differences between SSDTrain and existing work: SSDTrain offloads (a)~activations to (b)~the SSDs (c)~with negligible performance overhead. To the best of our knowledge, SSDTrain is the first work that leverages SSD to offload activations for LLM training.}

\kwc{To take a closer look at the uniqueness of SSDTrain, let us compare it with related work Stronghold}~\cite{sunSTRONGHOLDFastAffordable2022} \kwc{and ZeRO-Infinity}~\cite{rajbhandariZeROinfinityBreakingGPU2021}. \kwc{For difference~(a), SSDTrain offloads activations. Data in LLM training can be categorized into mutually exclusive types: parameters, optimizer states, gradients, and activations.  
In contrast, existing work offloads other data than activations. E.g., Stronghold offloads parameters and gradients. Although ZeRO-Infinity offloads many types of data, when it comes to activations, only a subset as defined in the activation checkpoints, is optionally offloaded.  Activations are the intermediate tensors produced in the forward propagation and kept for gradient computation. They are consumed in the backward propagation immediately after the forward propagation. Due to the high computing cost, gradient computation is best done by GPUs. In comparison, parameter updates associated with parameter and gradient offloading are also light and suitable for CPUs, which is why some work leverages the CPU computing power to update the gradients to improve overall throughput.

For difference~(c), neither ZeRO-Infinity nor Stronghold is designed to hide long data transfer latency. With activation checkpointing in the CPU memory enabled, at the beginning of the backward propagation of each layer, ZeRO-Infinity loads its checkpoint from the CPU memory and waits until it is done. The data transfer latency is in the critical path. Because Stronghold overlaps data transfer with computation, Stronghold’s evaluation performs significantly better than ZeRO-Infinity. Nevertheless, Stronghold  exhibits performance degradation compared with the no-offloading Megatron due to the long transfer latency when using NVMe, as Figures 11 and 14 of Stronghold’s publication show. In contrast, SSDTrain incurs no performance degradation as it overlaps the computation and data transfer well and uses GDS to reduce the SSD access latency.

In summary, SSDTrain must tackle unique challenges, including (i)~the micro-second level SSD latency and (ii)~the short interval between producing activations in the forward propagation and their consumption in the backward propagation. To achieve this, SSDTrain uses GDS to reduce SSD access latency and carefully schedules data movement so that the computation hides the latency.}

Table~\ref{tab:salesman} \kwc{compares the features of} earlier LLM systems supporting \kwc{activation} offloading and SSDTrain:

\noindent
\textbf{Direct GPU--SSD data path.} As Section~\ref{sec:intro} mentions, transfer via CPU interferes with CPU workloads, affecting efficiency. 

\noindent
\textbf{Async data transfer.} These systems either block the training computation when loading the offloaded data or synchronize at each layer. Consequently, the I/O latency is exposed in the critical path. SSDTrain hides the I/O latency by overlapping I/O with GPU computation. 

\noindent
\textbf{Interoperability.} Since LLM training requires a synergy of Python packages and the ecosystem is rapidly evolving, it is vital for the offloading feature to have good interoperability with other components in the same library or other libraries. SSDTrain relies on process-local alternation to PyTorch execution and can work with distributed frameworks, such as Megatron and DeepSpeed. In contrast, DeepSpeed's offloading features, e.g., ZeRO-Infinity, are available only in certain ZeRO stages. \kwc{ZeRO stage determines what is sharded. For example, stage-3 ZeRO in Fig.~\ref{fig:projected_perf_model} sharded optimizer states, gradients, and weights across the data parallel ranks.} Flexgen and LLM in a Flash have their own runtime and do not work with distributed frameworks.

\begin{table}[!t]
\centering
\begin{tabular}{ @{}l@{\hspace{\tabcolsep}}l@{\hspace{0.5\tabcolsep}}c@{\hspace{0.5\tabcolsep}}c@{\hspace{0.5\tabcolsep}}c@{\hspace{0.5\tabcolsep}}||c@{}}
\toprule
    &       & \rotatebox{68}{Flexgen} & \rotatebox{68}{LLM in a Flash} &\rotatebox{68}{ZeRO-Infinity} & \rotatebox{68}{\textbf{SSDTrain}} \\ 
\midrule  
\multicolumn{2}{@{}l}{\textbf{Training}}                           &           &  & \Checkmark & \Checkmark \\\cline{1-2}
\multirow{2}{*}{\textbf{\begin{tabular}[c]{@{}l@{}}Activation\\ offloading\end{tabular}}}        & \textbf{to main memory}        &    \Checkmark       &  \Checkmark &   Checkpoints only & \Checkmark \\\cline{2-2}
                                   & \textbf{to SSD}          &  \Checkmark         &  &   & \Checkmark \\\cline{1-2}
\multicolumn{2}{@{}l}{\textbf{Direct GPU--SSD data path}}                &           &  &  & \Checkmark\\\cline{1-2}
\multicolumn{2}{@{}l}{\textbf{Async data transfer}}                &           &  &  & \Checkmark\\\cline{1-2}
\multicolumn{2}{@{}l}{\textbf{Interoperability}} & & & & \Checkmark\\
\bottomrule
\end{tabular} 
\caption{Comparing SSDTrain with other LLM systems providing \kwc{activation} offloading features~\cite{shengFlexGenHighThroughputGenerative2023,alizadehLLMFlashEfficient2024,rajbhandariZeROinfinityBreakingGPU2021}. Without backward propagation, inference systems may discard most intermediate tensors once a layer is done. We generalize ``\textbf{Activation}'' to refer to the key-value (KV) cache in inference systems because it is reused across steps. \label{tab:salesman}}
\end{table}

\section{Design and Implementation}

\subsection{Overview of the SSDTrain System}\label{sec:ssdtrain_overview}
\begin{figure}[!t]
\centering
\includegraphics[width=\linewidth]{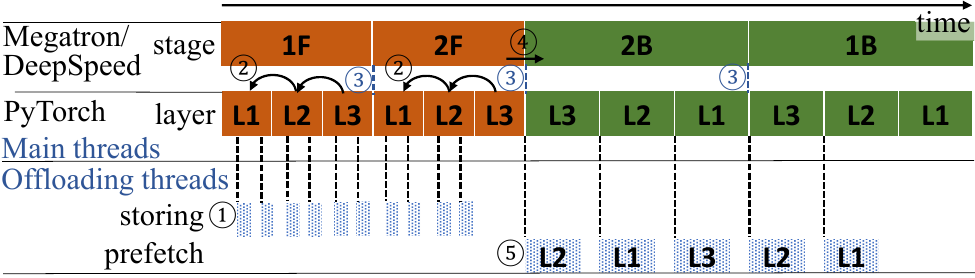}
\caption{\label{fig:pipe}SSDTrain timeline of a step of a two-micro-batch three-layer (L) model. PyTorch hooks are used to trigger tensor cache bookkeeping, tensor offloading (\textcircled{1}), and tensor loading (\textcircled{5}). In the forward (F) propagation, SSDTrain records the order of scopes (\textcircled{2}) and switches between micro-batches at the end of the stages (\textcircled{3}). SSDTrain starts loading when it is switched to the backward (B) propagation (\textcircled{4}). }
\end{figure}

SSDTrain implements a {\it tensor cache} to \kwc{manage} the offloading and reloading of tensors, facilitating the release of memory and the prefetch of tensors back to memory before they are needed for backward propagation. 
Figure~\ref{fig:pipe} demonstrates how SSDTrain works using PyTorch as an example.
SSDTrain launches its threads~(separate from PyTorch's execution threads) to store tensors~(\textcircled{1}) and to load them back~(\textcircled{5}). In forward propagation~(F), offloading of an activation starts once the operator producing it finishes~(\textcircled{1}). When activations are reused in backward propagation~(B), prefetching~(\textcircled{5}) occurs in the reverse order of layers as recorded during forward propagation~(\textcircled{2}). \kwc{If the last layer begins backward propagation immediately after its forward propagation}~(L3 \kwc{in micro-batch 2} in the example) , \kwc{SSDTrain} keeps the layer's activations \kwc{in GPU memory} instead of offloading them~(\textcircled{4}). SSDTrain keeps individual records for each micro-batch. Upon micro-batch changes~(\textcircled{2}), SSDTrain switches its record to the one corresponding to the new micro-batch.

\begin{figure}[!t]
\centering
\includegraphics[width=0.85\linewidth]{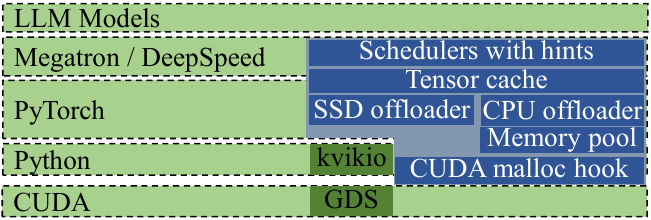}
\caption{\label{fig:software_arch} SSDTrain software architecture. The components of SSDTrain are shown as blue blocks with white text. The CUDA malloc hook is a C++ library, while others are Python code.}
\end{figure}

Figure~\ref{fig:software_arch} shows the SSDTrain software components. 
The tensor cache manages the activations and performs tensor offloading and loading. To achieve this, PyTorch hooks are used to alter PyTorch execution. Section~\ref{sec:tensor_cache} details the design and implementation of the tensor cache. SSDTrain has the SSD offloader that targets NVMe SSDs within the same node and the CPU offloader that targets host memory. Each offloader encapsulates the logic to transfer CUDA tensors to and from an offloading target. The SSD offloader leverages the GDS python binding, kvikio~\cite{nvidiaRapidsaiKvikioKvikIO2022}. Using the \texttt{LD\_PRELOAD} library interposition mechanism, CUDA malloc hook is a shared library that alters CUDA memory allocation and free API calls so that the memory is properly registered and deregistered for best GDS performance. This allows us to keep the PyTorch CUDA cached memory allocator for easy comparison with the baseline, without replicating its implementation in a PyTorch pluggable memory allocator or modifying the PyTorch runtime C++ code. The CPU offloader is for future work on clusters with massive remote SSD storage. It is backed by an allocator with pre-allocated host-pinned memory. The pool size is determined by profiling the first training step. New API calls are added to Megatron's and DeepSpeed's schedulers so that the tensor cache could get hints about stage changes and micro-batch changes, e.g., \textcircled{3} and \textcircled{4} in Figure~\ref{fig:pipe}. The following paragraph details hinted DeepSpeed's scheduler as an example.

To use SSDTrain, moderate code additions are needed in the existing script: \texttt{configure\_tensor\_cache()} in Algorithm~\ref{algo:tensor_cache_configure} shows the logic to configure tensor cache before training. The logic registers the PyTorch hooks, bookkeeps the parameters to not offload them when they are registered onto the computational graph, and monkey-patches~\cite{wikipediaMonkeyPatch2024} the schedulers. 
With the dynamicity of PyTorch, monkey-patch overrides a defined function by assigning the custom implementation to the defined \kwc{function} in a package. \texttt{deepspeed\_exec\_schedule()} shows the hints added to DeepSpeed's pipeline scheduler. Before and after the execution of each command, APIs are called to notify the tensor cache about the upcoming stage~(line 13) and the completion of an action~(line 15). Accordingly, the tensor cache can prefetch data or wait for I/O to complete. Megatron's scheduler is patched similarly.
 
SSDTrain extends naturally to distributed settings such as use with ZeRO, because frameworks like DeepSpeed and Megatron divide the workload into processes built on top of PyTorch's built-in tensor functionality.
By working below PyTorch and keeping each process' activities local, SSDTrain applies directly to distributed launches.

\subsection{Hook-Based Implementation of Tensor Cache}
\label{sec:tensor_cache}

To benefit from tensor offloading, the GPU memory that the offloaded tensors own must be released when the tensors are not in use. However, by default, PyTorch stores a reference to all the activations on the computational graph, disallowing the GPU memory to be reclaimed. The tensor cache alters the PyTorch execution so that the identifiers, not the references, of the activations are registered on the computational graph; upon PyTorch's reusing the activation tensor, the tensor cache uses the identifier from the computational graph as the key to return the requested tensor. In forward propagation, when the tensor finishes offloading, the tensor cache no longer holds a reference to it, allowing its memory to be reclaimed by Python garbage collection once the Python control flow gets out of the function scope where the tensor object is used. In the backward propagation, the tensor cache holds a reference to the tensor by loading it from the SSD before its use; when all the module scopes the tensor is referred to have been finished, the reference is no longer held, allowing its memory to be reclaimed.

In short, the tensor cache is the in-memory structure that manages the references to all activations and keeps track of activations' states, including whether they are being offloaded, the path in the file system, etc.

As Algorithm~\ref{algo:tensor_cache_hooks} shows, the tensor cache relies on the three PyTorch hook pairs to alter its execution behavior.

The forward hook pair works in the forward propagation: The start of a module triggers the forward pre-hook, and the finish of a module triggers the forward hook. 
The tensor cache maintains the current scope stack using the forward hook pair: Upon entrance to a module, the module is pushed to the stack;
when the module exits, it is popped out.

The backward hook pair is similar. 
When entering a module, the tensor cache prefetches activations in upcoming modules. Section~\ref{sec:prefetch} details prefetching.
When exiting a module, the tensor cache removes it from the scope lists of all activations. Activations no longer in use are removed, whose memory will be released by garbage collection.

\begin{algorithm}[!t]
{
\DontPrintSemicolon
    \KwIn{The tensor cache \texttt{tcache} and the LLM model \texttt{model}.}
    \SetKwFunction{FConf}{\texttt{configure\_tensor\_cache}}
    \SetKwProg{Fn}{Function}{:}{}
    \Fn{\FConf{\texttt{tcache}, \texttt{model}}}{
        \texttt{tcache.register\_hooks()}
        
        \For{\texttt{param in model.parameters()}}{
            \texttt{tcache.register\_parameters(param)}
        }

        Monkey-patch DeepSpeed's and Megatron's schedulers. 
    }

    \SetKwFor{For}{{\color{blue}for}}{{\color{blue}:}}{}
    \SetKwFunction{FSched}{\texttt{deepspeed\_exec\_schedule}}
    \Fn{\FSched{\texttt{self}, \texttt{schedule}}}{
        \For{{\color{blue}\texttt{step\_cmds in schedule}}}{
            \For{{\color{blue}\texttt{idx\_cmd, cmd in enumerate(step\_cmds)}}}{                
                \texttt{tcache.set\_stage(cmd)}

                \texttt{nxcmd = get\_next(idx\_cmd, step\_cmds)}
                
                \texttt{tcache.set\_next\_stage(nxcmd)}

                \If{\texttt{cmd} is communication and \texttt{nxcmd} is backward pass}{
                    \texttt{tcache.prefetch\_last\_module()} 
                }

                {\color{blue}\texttt{self.execute(cmd)}}

                \lIf{\texttt{cmd} is a backward pass}{
                \texttt{tcache.wait\_IO()}
                }
            }
        }
    }
}
\caption{Logic to configure tensor cache before training and DeepSpeed scheduler logic with tensor cache hints. The original code before adding hints is blue. As shown, changes to adopt tensor cache are moderate.}
\label{algo:tensor_cache_configure}
\end{algorithm}

When a tensor is to be registered onto the computational graph, the pack hook is called to produce a value to be registered instead.
When the tensor is reused, the unpack hooks are called to take in the object on the computational graph and return the original tensor.
Figure~\ref{fig:hooks} illustrates the tensor cache's activity when triggering the pack or unpack hook.
When the multiply operator $\texttt{x}\cdot\texttt{w}$ finishes (\textcircled{1}), the pack hook is called~(\textcircled{2}) on the input \texttt{x} and parameters \texttt{w}.
Tensor cache has a record of parameters and accordingly returns \texttt{w} to let it be registered on the graph as is.
The tensor will also be returned as is if the tensor is on CPU or it is too small~(line 12 in Algorithm~\ref{algo:tensor_cache_hooks}).
As line 16 in Algorithm~\ref{algo:tensor_cache_hooks} shows, the tensor cache does not offload tensors but only keeps a record when the module is to be kept in the memory or in backward propagation.
The first condition holds true when the adaptive offloading algorithm determines to keep the last few modules in GPU memory~(Section~\ref{sec:adaptive}).
The second condition is true when an activation-checkpointing-enabled function does recomputation in the backward propagation to reproduce the activations.
For tensor \texttt{x} in Figure~\ref{fig:hooks}, the tensor cache stores it to the SSDs~(\textcircled{3}) and returns a tensor identifier.
When the unpack hook is triggered~(\textcircled{B}), in the backward propagation~(\textcircled{A}), the tensor cache either waits until the prefetch finishes(\textcircled{C}), and eventually returns the tensor. 

\subsection{Deduplicating Tensors and Excluding Parameters}

Tensor cache has a \texttt{get\_id()} \kwc{method} to assign a unique identifier to each tensor.
The shortcoming of PyTorch native \texttt{id()} is that its returned value is related to the GPU memory address. As SSDTrain offloads activations, the latter will be cleared by garbage collection once the control flow goes out of its use scope.
The GPU memory address may be reused, causing identifier collision. 
To solve this, \texttt{get\_id()} combines the timestamp when it first processes the tensor with the tensor shape as the unique identifier. When \texttt{get\_id()} processes a tensor \texttt{t} for the first time, \texttt{get\_id()} adds the current timestamp as an additional attribute to the tensor's underlying storage \texttt{t.untyped\_storage()} instead of \texttt{t}.
This is because sometimes PyTorch creates new \texttt{torch.Tensor} objects representing the identical tensor.
All future \texttt{get\_id()} calls get the attribute value.
This deduplicating scheme helps prevent redundant I/Os.

PyTorch registers all needed tensors in backward propagation into the computational graph, including activations and parameters. As SSDTrain focuses on offloading activations, the tensor cache excludes the model parameters. To achieve this, before training, the tensor cache records the identifiers of all model parameters~(line 4 in Algorithm~\ref{algo:tensor_cache_configure}). As linear layers store the transpose of the parameter tensors for backward propagation, the unique identifiers of the transpose are recorded. One benefit of our \texttt{get\_id()} scheme is that the identifier for the transpose of the same parameter tensor remains consistent across steps. This is because the transpose uses the original tensor's underlying storage, to which we already assigned a timestamp before training.

\subsection{Offloading and Forwarding Tensors}
\label{sec:prefetch}

The tensor cache has two thread pools---one for storing tensors and the other for loading tensors. The jobs submitted to each thread pool are executed in first-in-first-out~(FIFO) order.

To hide the I/O latency, the tensor cache starts prefetching each activation before the corresponding module's backward propagation. 
The activations in the last module are kept in GPU memory, so they need not be prefetched.
This simple scheme suffices because, in PyTorch, the CPU submits GPU kernel launches and memory operations ahead of GPU execution. Prefetching schemes are equivalent as long as there are always I/O tasks in the GPU job queue to keep PCIe busy.

\begin{figure}[!t]
\centering
\includegraphics[width=0.91\linewidth]{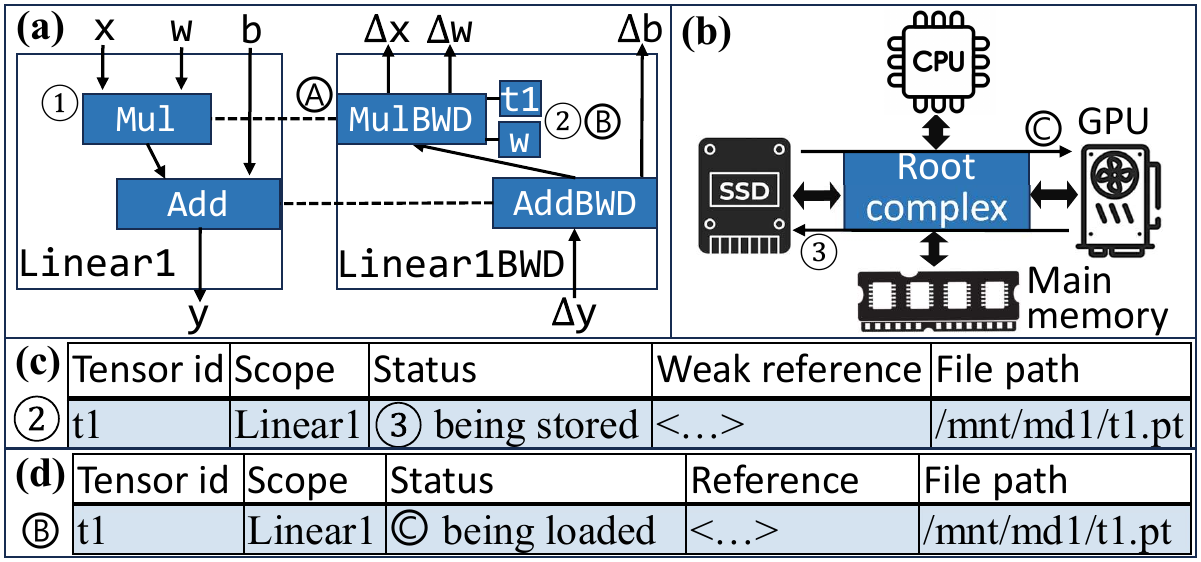}
\caption{\label{fig:hooks} Tensor cache registers pack--unpack hook pair to offload tensors and reload tensors. \textbf{(a)} shows the PyTorch computational graph. \textbf{(b)} shows the hardware data path. \textbf{(c)} and \textbf{(d)} show the tensor cache state when the pack or unpack hook is triggered. During an operator~(\textcircled{1}), PyTorch calls the pack hook with tensors to be saved for backward propagation and registers the return values on the computational graph~(\textcircled{2}). Tensor cache tracks the tensors, offloads them~(\textcircled{3}), and returns identifiers for the tensors. In an operator~(\textcircled{A}) in the backward propagation, PyTorch calls the unpack hook with the identifiers to get tensors~(\textcircled{B}). The tensor cache blocks until the requested tensors are loaded in GPU memory~(\textcircled{C}). }
\end{figure}

Upon loading a tensor, if it is still being stored, the tensor cache will return its original in-memory reference to skip loading from SSD. We call this data forwarding. For example, in Figure~\ref{fig:hooks}, when the PyTorch engine retrieves tensor \texttt{x} from the \texttt{MulBWD} node, if it is still being stored to the SSDs, it is in memory. Instead of loading the tensor, the tensor cache returns its in-memory reference by converting the weak reference to a reference and storing the obtained reference in the tensor cache for the future if it is used in other scopes.

\begin{algorithm}[!t]
{
\DontPrintSemicolon
    \KwIn{The tensor cache \texttt{tcache}, current scope \texttt{module}, tensor to pack \texttt{tensor}, and/or object to unpack \texttt{obj}.}
    \SetKwProg{Fn}{Function}{:}{}
    \SetKwFunction{FFPH}{\texttt{forward\_pre\_hook}}
    \Fn{\FFPH{\texttt{module}}}{ 
    Add \texttt{module} to \texttt{tcache}'s current scope stack.

    }
    \SetKwFunction{FFH}{\texttt{forward\_hook}}
    \Fn{\FFH{\texttt{module}}}{ 
    Pop \texttt{tcache}'s innermost scope from the current scope stack.

    }
    \SetKwProg{Fn}{Function}{:}{}
    \SetKwFunction{FBPH}{\texttt{full\_backward\_pre\_hook}\textsuperscript{*}}
    \Fn{\FBPH{\texttt{module}}}{ 
    Prefetch the tensors in the next module.
    
    }
    \SetKwFunction{FBH}{\texttt{full\_backward\_hook}\textsuperscript{*}}
    \Fn{\FBH{\texttt{module}}}{ 
    \For{each tensor \texttt{t} in \texttt{module} tracked by \texttt{tcache}}{
            Remove \texttt{module} from \texttt{t}'s record.

            Release and stop tracking \texttt{t} if no scope is using \texttt{t}.
        }
    }
    \SetKwFunction{FPCKH}{\texttt{pack\_hook}}
    \Fn{\FPCKH{\texttt{tensor}}}{
     \lIf{\texttt{tcache.is\_parameter(tensor) or tensor.is\_cpu or math.prod(tensor.size())<2**20}}{
     \Return{\texttt{tensor}}
     }

    \texttt{tid = get\_id(tensor)}

    \texttt{tcache.add\_to\_current\_scope(tid)}

    \If{\texttt{tcache.is\_current\_scope\_kept\_in\_memory() or tcache.is\_current\_in\_backward()}}{
     \texttt{tcache.keep\_in\_gpu\_memory(tid,tensor)}
     }\lElse{
     \texttt{tcache.offload(tid,tensor)}
     }

    \Return{\texttt{tid}}
    }

    \SetKwFunction{FUPCKH}{\texttt{unpack\_hook}}
    \Fn{\FUPCKH{\texttt{obj}}}{
        \lIf{\texttt{isinstance(obj,torch.Tensor)}}{
            \Return{\texttt{obj}}
        }
        \lIf{\texttt{not tcache.is\_loaded(obj)}}{
        \texttt{tcache.load\_or\_wait\_load(obj)}
        }
        \Return{\texttt{tcache.get\_loaded\_tensor(obj)}}
    }
}
\caption{Tensor cache registers PyTorch hooks to trigger actions during training.}
\label{algo:tensor_cache_hooks}
\begin{flushleft} \footnotesize * PyTorch added \texttt{full\_} prefix to the backward hook pair APIs to distinguish the current reworked design from the superseded one.
\end{flushleft}
\end{algorithm}

\begin{figure}[!t]
\centering
\includegraphics[width=0.9\linewidth]{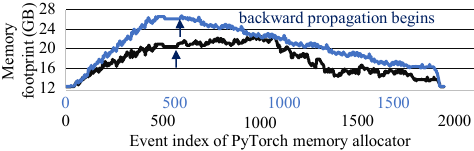}
\caption{\label{fig:snapshot} Memory footprint of one A100 in a BERT training step with offloading~(black) and without~(blue) on Table~\ref{tab:configurations}'s system. Run with offloading incurs more allocator events because of memory release and allocation caused by tensor offloading and reloading. SSDTrain reduces memory footprint at the beginning of backward propagation by 45\% and end-to-end peak memory footprint by 25\%.}
\end{figure}

\begin{figure}[!t]
\centering
\includegraphics[width=0.95\linewidth]{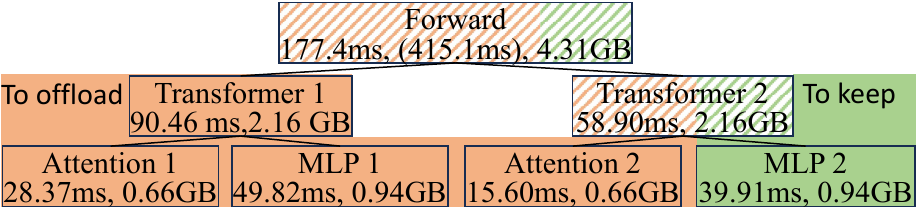}
\caption{\label{fig:adaptive} The adaptive offloading algorithm uses profiling to decide modules in which the activations are to be kept in GPU memory. The model is represented as a tree where each scope is a node. On each node, the forward computation time and data transfer size are recorded during profiling. The I/O time in the forward propagation is also recorded and shown in parenthesis in the root node.}
\end{figure}

\begin{figure}[!t]
\centering
\includegraphics[width=0.95\linewidth]{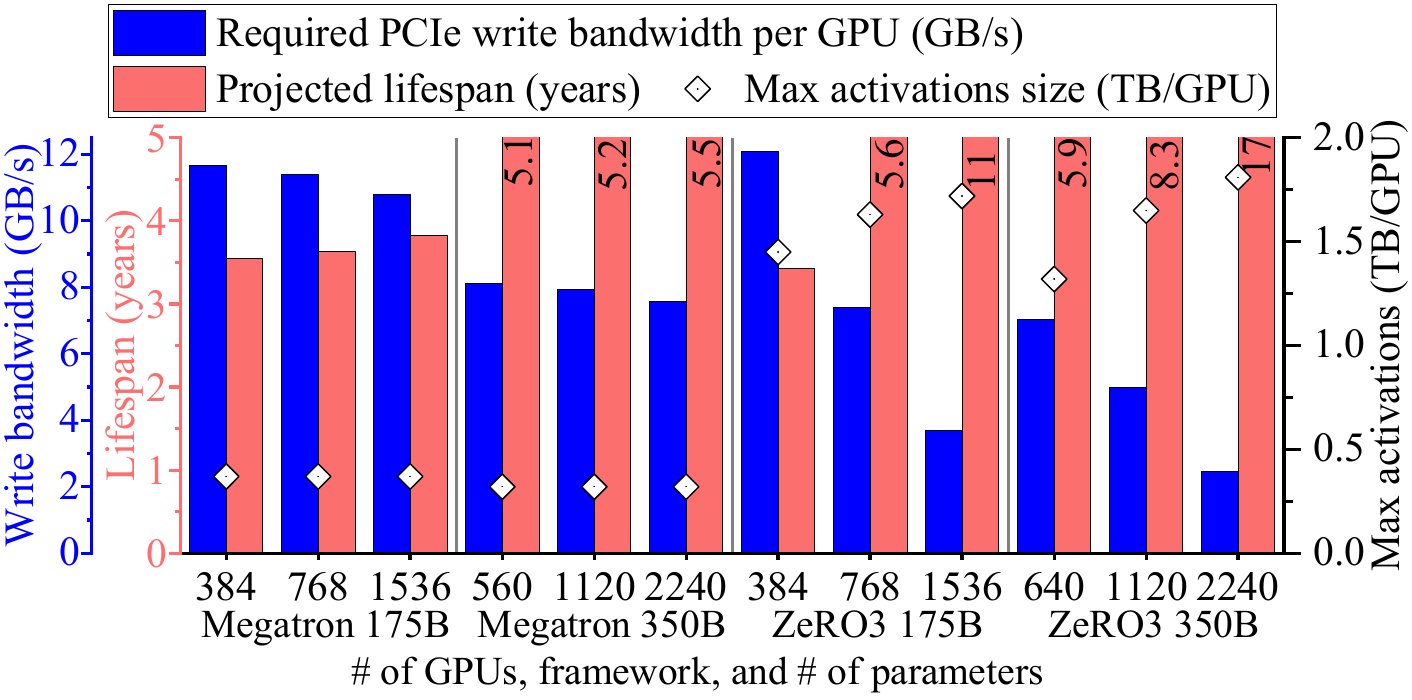}
\caption{\label{fig:projected_perf_model} Estimate of SSD lifespan~(left pink vertical axis), PCIe write bandwidth~(left blue vertical axis) and maximal activations size per GPU~(right vertical axis). Lifespans longer than 5 years are shown on top of the pink bars. The horizontal axis shows the number of GPUs, the framework, and the model size~\cite{shoeybiMegatronLMTrainingMultiBillion2020a}. ZeRO3 stands for DeepSpeed with stage-3 ZeRO, i.e., all optimizer states, gradients, and parameters are sharded \kwc{across data parallel ranks}.}
\end{figure}

\subsection{Adaptive Offloading}
\label{sec:adaptive}

One insight we got during SSDTrain is that the activation offloading should target minimizing the peak memory usage so that the same system could accommodate a configuration with larger activations without triggering out-of-memory~(OOM) errors. Offloading tensors after the peak is not helpful. In Figure~\ref{fig:snapshot}, the \kwc{blue} curve is the memory footprint without offloading; it illustrates that GPU memory usage peaks at the beginning of the backward propagation. The \kwc{black} curve shows the memory footprint with offloading, where the peak is delayed by the in-progress offloading jobs and new intermediate tensors created in backward propagation. Excessive tensor offloading may keep the tensor reference even after its last use in backward propagation, delaying the reclamation of its memory. To reduce unnecessary offloading after the peak, we devised adaptive offloading with two features.

First, when a thread is assigned a storing job, the thread will check if the tensor was forwarded. If so, the job will be canceled. Second, as illustrated in Figure~\ref{fig:adaptive}, we devise an algorithm to choose a module from which the offloading is paused. We profile a step to collect: (1) the data transfer size and computation time of each MLP block and attention block, and (2) the forward propagation's computation time, data transfer time, and total data transfer amount. Suppose module $m$ is the last module to offload in a step. The required data transfer bandwidth is to finish offloading for all the modules before $m$ and both offloading and reloading for module $m$ by the time the backward propagation of module $m$ begins. With the estimate that the backward propagation time is twice the forward propagation time, the required data transfer bandwidth can be calculated by the collected numbers. It should be no larger than the write bandwidth in the measured forward propagation.

\subsection{SSD Write Amount, Bandwidth, and Lifespan}
\label{sec:projected_life}
To confirm whether our design is viable in large-scale training systems, particularly regarding SSD endurance and required bandwidth, we conduct performance modeling to obtain the forward propagation time per training step and the size of activations produced in the process. 

We extend the performance model package \texttt{llm-analysis}~\cite{liLLMAnalysisLatencyMemory2023}.
To estimate the forward propagation time, \texttt{llm-analysis} models each transformer layer as a simple pipeline, 
$t= \max\left(\sum_{l}\max\left(t_{l,compute}, t_{l,memory}\right), t_{ZeRO,communicate}\right)$,
where $l$ denotes any layers inside a transformer layer. When ZeRO is enabled, the ZeRO communication time is assumed to be perfectly pipelined with the non-ZeRO computation and memory operations at the level of the transformer layer. 

 We model the required PCIe write bandwidth per GPU as the total amount of activations divided by half the training time. As Section~\ref{sec:adaptive} explains, some activations may be written at the early stages of the backward propagation to reduce the needed PCIe bandwidth. We also assume that the training step time $t_{step}$ is three times the forward propagation time. The lifespan is then projected as $t_{life} = S_{endurance}\cdot t_{step}/S_{activations}$ where $S_{endurance}$ is the lifetime writes allowed by the SSD endurance rating, and $S_{activations}$ is the size of activations per training step. We validated the $S_{activations}$ formula with profiled activations size in experiments in Section~\ref{sec:evaluation}. We assume four Solidigm D7-P5\kwc{620} 12.8TB~(Table~\ref{tab:ssd}) for each GPU and assume the WAF is 2.5 in JESD rating and 1 in our scenario.

With these, we obtain Figure~\ref{fig:projected_perf_model}. We use the system configurations and measured floating point throughput from Megatron-LM~\cite{shoeybiMegatronLMTrainingMultiBillion2020a}. The GPUs are A100 PCIe. Among all cases, the projected lifespan is over three years, and the PCIe write bandwidth per GPU is no greater than 12.1 GB/s.
Moreover, when the system size scales up, the required PCIe write bandwidth reduces, and the projected lifespan increases. This occurs because larger systems imply increased communication overhead and reduced computation efficiency, thus slowing down training iterations on each GPU. \kwc{Similar effects are observed when the model size scales up because larger model size leads to longer compute latency with increased 
data reuse and therefore less bandwidth requirement. Section~\ref{sec:discussion} discusses the effect of scaling up in detail. }

We also estimate the maximal size of activations each GPU produces in one step: We compute the maximal micro-batch size by assuming only two layers in a row are in GPU memory at the same time while all other activations are offloaded. Then, the activation maximal micro-batches produce in a step are the largest activations offloading could open up, which are shown as diamond marks in Figure~\ref{fig:projected_perf_model}. The maximal activations size per GPU ranges from 0.4 TB to 1.8 TB, while the micro-batch size ranges from eight to 32. Activations so large can no longer be held by the main memory~(Figure~\ref{fig:current_sys}), and therefore, SSD is the only choice as an offloading target.

To further increase SSD endurance, the data retention period can be relaxed: NAND flash gets \kwc{86}$\times$ P/E cycles when the data retention period is relaxed from three years to \kwc{one} day~\cite{caiFlashCorrectandrefreshRetentionaware2012,yucaiErrorPatternsMLC2012,liuOptimizingNANDFlashBased2012,kimBehemothFlashcentricTraining2021}. This technique was not leveraged in the reasoning of this subsection, but we discuss its impact on cost in Section~\ref{sec:discussion}.

\section{Evaluation and Discussion}
\label{sec:evaluation}
We evaluate SSDTrain and answer the following questions: 
\begin{enumerate}[Q1.]
\item How well does SSDTrain hide the I/O latency?
\item How much does SSDTrain reduce peak memory usage?
\item How \kwc{does SSDTrain effects translate into advantages as a design choice?}
\end{enumerate}

Section~\ref{sec:exp_baseline} answers Q1 and Q2 by comparing SSDTrain with execution without SSDTrain. \kwc{To answer Q3, w}e \kwc{examine the design space in Section~\ref{sec:exp_dse} and} discuss \kwc{various implications in multiple aspects} in Section~\ref{sec:discussion}.

\subsection{Experimental Setup}
\label{sec:eval_method}
We use a machine with two A100 PCIe GPUs and seven Intel P5800X SSDs, as Table~\ref{tab:configurations} specifies. The SSDs are organized into two RAID0 arrays: one with three SSDs and the other with four SSDs. Each array is the dedicated offloading target of one of the A100 GPUs. We measured the memory usage of the A100 with four SSDs during the evaluation. For consistent performance, the GPUs are locked at base frequency. The latest Megatron-DeepSpeed~\cite{microsoftMicrosoftMegatronDeepSpeedOngoing2019} is installed, incorporating DeepSpeed techniques into Megatron and ensuring interoperability.

\begin{table}[!bhtp]
\centering
\begin{tabular}{rl}
    \toprule
    \textbf{CPU} & 2$\times$ AMD EPYC 7702 64-core\\\cline{1-1}
    \textbf{Memory} & DDR4-3200 1 TB\\\cline{1-1}
    \textbf{GPU} & 2$\times$ Nvidia A100 40 GB PCIe with NVLink \\\cline{1-1}
    \textbf{SSD} & 7$\times$ Intel Optane P5800X 1.6 TB. Two RAID0 arrays. \\\cline{1-1}    \textbf{Software} &     \begin{tabular}[c]{@{}l@{}}Ubuntu 20.04.6~(kernel 5.15.0-113), CUDA 12.2~(driver\\535.183.01), PyTorch 2.2.2, DeepSpeed 0.14.2, Megatron-\\ DeepSpeed~\cite{microsoftMicrosoftMegatronDeepSpeedOngoing2019}~(latest), kvikio 24.08\end{tabular} \\
\bottomrule
\end{tabular}
\caption{Evaluation system configuration.}\label{tab:configurations}
\end{table}

We measure the system pretraining performance on three models: BERT~\cite{devlinBERTPretrainingDeep2019} as an encoder-only model, GPT~\cite{radfordLanguageModelsAre2019} as a decoder-only model, and T5~\cite{raffelExploringLimitsTransfer2023} as an encoder-decoder model. 
We use the OSCAR corpus~\cite{ortiz-suarez-etal-2020-monolingual,OrtizSuarezSagotRomary2019} as the dataset.

\kwc{Before we further explain the model setup, we clarify the batch taxonomy. Like other deep learning models, LLM model training typically uses mini-batches, smaller subsets of the training data. Before Section~\ref{sec:evaluation}, we use the terms ``mini-batch'' and ``batch'' interchangeably.

However, the introduction of data parallelism complicates this terminology: Now, the samples processed in each training step are partitioned into several groups, and each group is assigned to a data-parallel rank. To avoid confusion, in cases where data parallelism is enabled, we refer to all the samples in each training step as a \textit{global batch}, and we refer to the samples assigned to one data-parallel rank a \textit{mini-batch}. Such cases where data parallelism is enabled are only in Section~\ref{sec:discussion}.

Micro-batch is at the lowest level of the batch taxonomy. When gradient accumulation is enabled, a global batch or a mini-batch is further divided into smaller groups for concurrent processing. Similarly, when pipeline parallelism is enabled, such a phenomenon may occur. Each group of samples is called a \textit{micro-batch}. In particular, micro-batch refers to the samples processed in one operator kernel launch.}

We use the two A100 GPUs for tensor parallelism.  The number of micro-batches per step is fixed at one because without pipeline parallelism, in each training iteration, Megatron-DeepSpeed will not start a new micro-batch before both forward propagation and backward propagation of the previous micro-batch are done. A micro-batch number larger than one only brings in gradient accumulation and does not affect the activation offloading pattern. \kwc{In other words, unless stated otherwise, the micro-batch size is equivalent to global batch size throughout Section~\ref{sec:evaluation}.}
\kwc{Throughout Section~\ref{sec:evaluation}, no ZeRO technique is used. Besides, the optimizer states, i.e., what stage-1 ZeRO shards only, may be shared across other dimensions than across the data-parallel ranks: In Megatron or Megatron-DeepSpeed, this is enabled by the \texttt{--use-distributed-optimizer} argument, which we also do not enable in experiments across Section~\ref{sec:evaluation}.}
In our experiments, the hidden dimension is from 8,192 to 16,384, and we use typical hyper-parameters~\cite{devlinBERTPretrainingDeep2019,touvronLlamaOpenFoundation2023,raffelExploringLimitsTransfer2023} for hidden dimensions within this range. The attention head dimension is 128. The text sequence length is 1,024. For T5, the number of decoders is half the number of layers, rounded down. FlashAttention-2~\cite{daoFlashAttention2FasterAttention2023} is used with or without SSDTrain for optimized attention computation.

As each A100 has only 40 GB of device memory, to explore the design space closer to that in real-world training systems with A100 80 GB and later GPUs~\cite{shoeybiMegatronLMTrainingMultiBillion2020a,liuWinnerTakeAllColumnRow2023}, we make several mitigations. First, we use FP16 instead of mixed precision, eliminating the FP32 weight copy. Second, we use SGD instead of Adam as the optimizer to reduce the memory use by optimizer states. The two measures only affect accumulation operations and \kwc{weight} updates, thus imposing a constant bias in the training step time and memory usage in execution with or without SSDTrain.

\subsection{Performance and Peak Memory Usage}
\label{sec:exp_baseline}

\begin{figure}[!t]
\centering
\subcaptionbox{}
[\linewidth]{\includegraphics[scale=1.4]{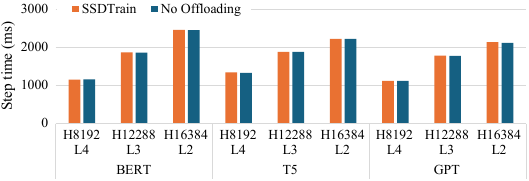}}
\subcaptionbox{}
[\linewidth]{\includegraphics[scale=1.4]{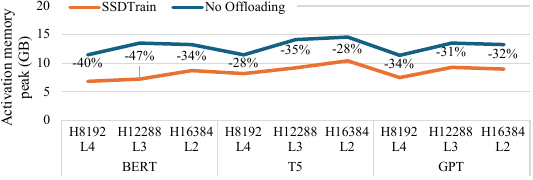}}
\caption{\label{fig:eval_perf} Comparing the step time and activations memory usage of SSDTrain with execution without tensor offloading. We test several model configurations with different hidden dimensions~(H) and number of layers~(L). Global batch size is 16.}
\end{figure}

To understand SSDTrain's impact on execution time and peak memory usage, we measure the step time of BERT, T5, and GPT and the memory peak during forward and backward propagation. The collected metrics of the system with SSDTrain and without are compared in Figure~\ref{fig:eval_perf}. For each model, we collected three scenarios with different (hidden dimension, number of layers): (8192, 4), (12288, 3) and (16384, 2). As shown, SSDTrain has almost no performance overhead in all cases. Although SSDTrain and its optimizations introduce additional CPU-executed logic, the performance comparison indicates that this logic is not on the critical path. Instead, GPU computation defines the critical path, and the CPU's role lies primarily in launching new GPU jobs before current GPU operations are complete.  Thus, the CPU is underutilized, and SSDTrain's extra work does not lead to delays in new tasks reaching the GPUs. 
Regarding the activations' memory use, SSDTrain effectively reduces the peak by 28\%--40\% in these cases.

\kwc{Notice that throughout Section~\ref{sec:evaluation}, neither ZeRO nor the Megatron's optimizer state sharding, i.e., the feature enabled by the \texttt{--use-distributed-optimizer} argument, are enabled. Both stage-1 ZeRO and Megatron's optimizer state sharding affect only the weight update stage and have no effect on SSDTrain activation offloading and reloading. As a feature for data parallelism, ZeRO may be enabled when data parallelism is enabled. Data parallelism is typically introduced when the number of GPUs exceeds 100~\cite{shoeybiMegatronLMTrainingMultiBillion2020a,nitinScalingLargeLanguage2023}.
As to be further explained in the discussion on \textbf{Impact of Upscaling} in Section~\ref{sec:discussion}, data parallelism with or without ZeRO will not negatively affect SSDTrain performance.
}

\subsection{Comparing the Activations Placement Strategies via Recompute-Offload-Keep~(ROK) Curve}
\label{sec:exp_dse}

\begin{figure}[!htbp]
\centering
\subcaptionbox{H12288 L3}
[\linewidth]{\includegraphics[scale=0.7]{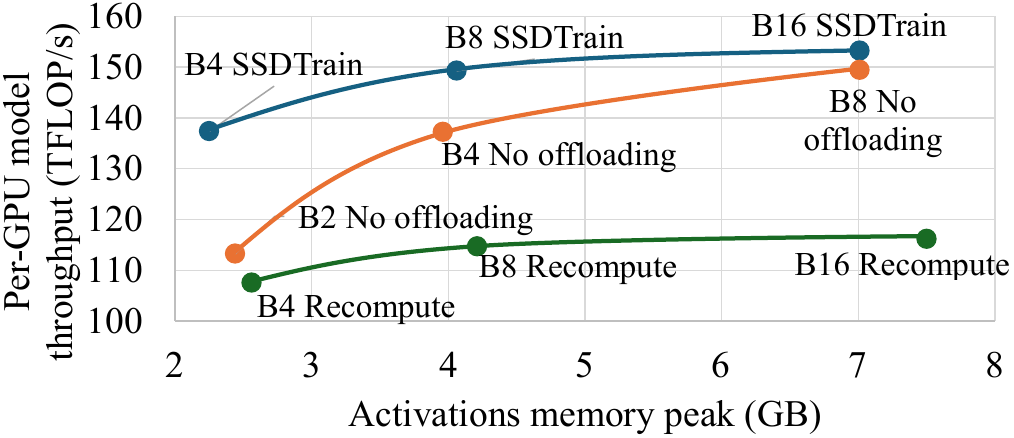}}
\subcaptionbox{H14336 L3}
[\linewidth]{\includegraphics[scale=0.7]{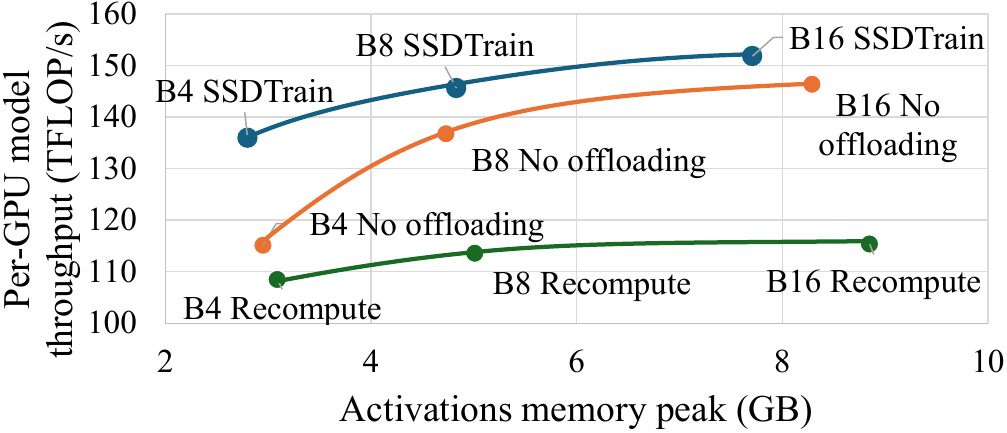}}
\caption{\label{fig:eval_dse}Recompute-offload-keep~(ROK) curve of BERT with 3 layers~(L) and hidden dimension~(H) as \textbf{(a)} 12,288 or \textbf{(b)} 14,436. Designs with a combination of \kwc{global} batch sizes~(B) and choices to offload activations, keep activations, or recompute activations are shown. \kwa{(Figure updated.)}}
\end{figure}

SSDTrain opens up offloading activations to SSDs as an option besides keeping activations in the GPU memory and activations checkpointing. We compare the three strategies here by plotting the runs on the recompute-offload-keep~(ROK) curve.  
Figure~\ref{fig:eval_dse} shows the ROK curve for training two 3-layer BERT models, one with a hidden dimension of 12,288 and the other with a hidden dimension of 14,336. In a ROK curve, each training run is represented by a point. The x-axis is the activations memory peak, and the y-axis is the model throughput. Model throughput~\cite{shoeybiMegatronLMTrainingMultiBillion2020a} refers to the number of algorithmic computations \kwc{done in unit time} regardless of software and hardware implementation, e.g., whether the activations are recomputed.
In these two cases, SSDTrain reduces the GPU activations memory peak, allowing a larger micro-batch size to attain higher throughput. Given the same micro-batch size, SSDTrain offloading attains the throughput the same as the throughput when the activations are kept in memory. Meanwhile, SSDTrain gets a lower activations memory peak than the recomputation. Compared with keeping the activations in memory, SSDTrain can double the micro-batch size with the same activations memory budget. Alternatively, people could leverage SSDTrain to run a bigger model or use fewer GPUs.

Other than the three strategies, before FlashAttention~\cite{daoFlashAttentionFastMemoryEfficient2022}, Megatron~\cite{korthikantiReducingActivationRecomputation2022} proposed selective recomputation:
noting that in the transformer layer, the operations performed by the core attention module~(the whole gray box in Figure~\ref{fig:transformer}) require less computation but create a large intermediate tensor when compared with the MLP block, the work recomputed only the core attention module.
As we adopt FlashAttention, the core attention module is done in one kernel, eliminating these intermediate tensors.  The effect of selective recomputation with FlashAttention has a negligible impact on the performance and the peak memory usage for activations.

\subsection{Discussion}
\label{sec:discussion}
\subsubsection{Examining the Modeling}
To understand the accuracy of the performance model in Section~\ref{sec:projected_life}, we compare the offloaded amount by SSDTrain with the model estimate. As shown in Table~\ref{tab:eval_activations}, the figures are close. We also compute the required PCIe write bandwidth using half of the measured training time. The PCIe write bandwidth is reduced as the hidden dimension gets larger. Typically, a model with more than 60B parameters has a hidden dimension of no less than 8K~\cite{touvronLlamaOpenFoundation2023,jordanhoffmannTrainingComputeOptimalLarge2022}. The PCIe write bandwidth of the BERT models aligns with the estimate in Section~\ref{sec:projected_life}.

\begin{table}[!htbp]
\centering
\begin{tabular}{llll}
\toprule
                              & H8192 L4  & H12288 L3 & H16384 L2 \\\midrule
\textbf{Offloaded amount}     & 10.37 GB  & 12.85 GB  & 10.75 GB  \\
\textbf{Model estimate}       & 11.13 GB  & 12.60 GB   & 11.50 GB   \\
\textbf{PCIe write bandwidth} & 18.0 GB/s & 13.8 GB/s & 8.76 GB/s \\ \bottomrule
\end{tabular}
\caption{The per-GPU offloaded tensor amount and model estimate when running BERT with different hidden dimensions~(H) and number of layers~(L). The global batch size is 16. We also compute the per-GPU PCIe write bandwidth required to fully offload the tensors. \label{tab:eval_activations}}
\end{table}

\subsubsection{Impact of Upscaling} 
When LLM systems scale up, the computation efficiency decreases due to more cross-node communication.
Section~\ref{sec:llm_scaling} demonstrates that the whole-system activations size $S_{activations}$ grows slower than the whole-system GPU throughput $C$, i.e., $S_{activations} \propto C^{\frac{5}{6}}$. Therefore, the bandwidth required to fully overlap the computation with the SSD accesses is reduced.
In short, LLM scaling is essentially a weak scaling scenario, and SSD I/O latency is easier to hide when scaled up. 

\kwc{As shown in Table~\ref{tab:eval_activations}, the required SSD throughput per GPU to fully offload tensors is negatively correlated with the hidden dimension of the LLM model, a factor of the model scale. Since most computation is GEMM, theoretically, the required SSD throughput per GPU is approximately inversely proportional to the hidden dimension of the LLM model, assuming the GPU model and computational efficiency are the same. The evaluation shows that the SSDTrain offloading performs well with two GPUs and a hidden dimension of 8K. Given that all data transfers SSDTrain offloading incurs are within the node, this configuration pressured the system more than some larger configurations, e.g., four GPUs per node and hidden dimension as 16K.

In Table~\ref{tab:upscaling_write_bandwidth}, we further project the impact of upscaling on the write bandwidth per GPU using \texttt{llm-analysis}. We follow typical parallelism configurations~\cite{shoeybiMegatronLMTrainingMultiBillion2020a,nitinScalingLargeLanguage2023} when the number of GPUs is less than 100: Initially, all GPUs are dedicated to tensor parallelism, and as the number of GPUs increases, we gradually increase the pipeline parallelism factor. In all projected cases, the write bandwidth per GPU is smaller than the corresponding original two-GPU case shown in Table~\ref{tab:eval_activations}. Notice that Table~\ref{tab:upscaling_write_bandwidth} does not study the effect of data parallelism, which is typically adopted when the number of GPUs exceeds 100. Vanilla data parallelism only affects the weight update stage and does not affect the write bandwidth because SSD offloading and reloading only happen during forward and backward propagation. A configuration with ZeRO-enabled data parallelism has no greater required write bandwidth than the corresponding configuration without data parallelism because the introduced communication operations may delay forward propagation and backward propagation.

}

\begin{table}[!htbp]
\begin{subtable}[]{\textwidth}\centering
\begin{tabular}{lrrrrr}\toprule
\textbf{Number of layers}                          & 4    & 4    & 8    & 16   & 32   \\
\textbf{Tensor   parallelism factor}               & 4    & 8    & 8    & 8    & 8    \\
\textbf{Pipeline   parallelism factor}             & 1    & 1    & 2    & 4    & 8    \\\midrule
\textbf{Write bandwidth per GPU (GB/s)}  & 17.3 & 16.0 & 16.5 & 16.8 & 17.0 \\
\bottomrule
\end{tabular}
\caption{Hidden dimension as 8,192. In all cases, the required write bandwidth per GPU is smaller than the original case's 18.0 GB/s.}
\end{subtable}\vspace{0.3cm}

\begin{subtable}[]{\textwidth}\centering
\begin{tabular}{lrrrrr}\toprule
\textbf{Number of layers}                          & 3    & 3    & 6    & 12   & 24   \\
\textbf{Tensor   parallelism factor}               & 4    & 8    & 8    & 8    & 8    \\
\textbf{Pipeline   parallelism factor}             & 1    & 1    & 2    & 4    & 8    \\\midrule
\textbf{Write bandwidth per GPU (GB/s)}  & 13.4 & 12.7 & 13.1 & 13.3 & 13.4 \\
\bottomrule
\end{tabular}
\caption{Hidden dimension as 12,288. In all cases, the required write bandwidth per GPU is smaller than the original case's 13.8 GB/s.}
\end{subtable}\vspace{0.3cm}

\begin{subtable}[]{\textwidth}\centering
\begin{tabular}{lrrrrr}\toprule
\textbf{Number of layers}                          & 2    & 2    & 4    & 8   & 16   \\
\textbf{Tensor   parallelism factor}               & 4    & 8    & 8    & 8    & 8    \\
\textbf{Pipeline   parallelism factor}             & 1    & 1    & 2    & 4    & 8    \\\midrule
\textbf{Write bandwidth per GPU (GB/s)}  & 8.55 & 8.17 & 8.47 & 8.63 & 8.69 \\
\bottomrule
\end{tabular}
\caption{Hidden dimension as 16,384. In all cases, the required write bandwidth per GPU is smaller than the original case's 8.76 GB/s.}
\end{subtable}

\caption{Projecting the SSD write bandwidth required by each A100 when scaling up the cases shown in Table~\ref{tab:eval_activations}. As the number of GPUs increases, we first increase the tensor parallelism factor with the number of layers unchanged. Then, we increase the pipeline parallelism and increase the number of layers proportionally. \kwa{(New table.)} \label{tab:upscaling_write_bandwidth}}
\end{table}

\kwc{\subsubsection{Performance Implications of Larger Micro-Batch} To further understand how larger micro-batch size improves the performance, we compare the no-offloading cases in Figure~\ref{fig:eval_dse}(a) to the same configurations with global batch size as one and break down the throughput improvement in Table~\ref{tab:break_down_batch_effect}. The improvement comes from higher kernel throughput and time-saving by weight update, where weight update saving is consistently the primary source. Such a benefit is very relevant to large-scale LLM training systems. The micro-batch size is usually set as one or two in Paxml~\cite{googlePaxmlAkaPax2022} and BLOOM~\cite{workshopBLOOM176BParameterOpenAccess2023} pretraining. For these two models, the micro-batch size is set small in exchange for smaller bubbles introduced by the pipeline parallelism. The bubble time percentage is inversely proportional to the number of micro-batches. For example, in the BLOOM training system, the tensor parallelism factor is four, and the pipeline parallelism factor is 12. In each training step, each data-parallel rank is assigned a mini-batch with 32 samples. When the micro-batch size is no less than four, the ideal pipeline bubble time percentage is no less than 11.5\%. However, the weight update and gradient accumulation cost is inversely proportional to the micro-batch size. When the micro-batch size is one or two, the cost is enormous. SSDTrain allows larger micro-batch sizes given the same activation memory budget, thus beneficial to these pipeline-parallelism-enabled training systems. 

}

\begin{table}[]\centering
\begin{tabular}{lllll}
\toprule
\textbf{Global batch size}                   & 2      & 4      & 8      & 16     \\\midrule
\textbf{Throughput improvement}       & 27.6\% & 52.2\% & 66.1\% & 71.8\% \\
\textbf{By higher compute efficiency} & 10.5\% & 21.8\% & 27.7\% & 29.4\% \\
\textbf{By weight update saving}            & 17.1\% & 30.4\% & 38.4\% & 42.4\% \\\bottomrule
\end{tabular}
\caption{Breakdown of model throughput improvements for a three-layer BERT model with a hidden dimension of 12,288, compared to a baseline with global batch size as one. \kwa{(New table.)}}
\label{tab:break_down_batch_effect}
\end{table}

\kwc{\subsubsection{Weight Offloading}

This work focuses on offloading activations. When the size of weights gets larger, it becomes more desirable to offload weights. SSDTrain can be configured to offload weights as well. As Section~\ref{sec:tensor_cache} explained, the tensor cache keeps a record of all the weights and ignores them when the pack hook is triggered. The tensor cache may be modified to offload weights in a profitable situation. For each operator, e.g., a matrix multiply, the amount of computation, weight size, and input size can be determined from the model specification without execution. SSDTrain can decide whether to offload one or both according to the GPU FP16 throughput and SSD write bandwidth. A reasonable starting strategy is to offload as much as possible while staying within SSD write bandwidth. 

Furthermore, SSDTrain could be extended to generate an optimized plan for all operators in the model before the execution by framing the decision-making process into an optimization problem and solving it. Offloading weights works when the pipeline parallelism factor is small. When the pipeline parallelism factor is large, careful planning is needed to determine what to offload because some weights are immediately reused by the later micro-batches.

Notice that offloading weights to the main memory, together with weight update to the CPU, has been explored in prior work~\cite{rajbhandariZeROinfinityBreakingGPU2021,kamahoriFiddlerCPUGPUOrchestration2024}. In the future, it may be explored to use SSDTrain together with any of the prior work to offload activations to the device memory and offload weights to the main memory. We leave the discussion to the elaboration on \textbf{Swapping and offloading} in Section~\ref{sec:ssdtrain_related}.

}

\subsubsection{Cost Analysis} 
\begin{figure}[!t]
\centering
\includegraphics[width=0.95\linewidth]{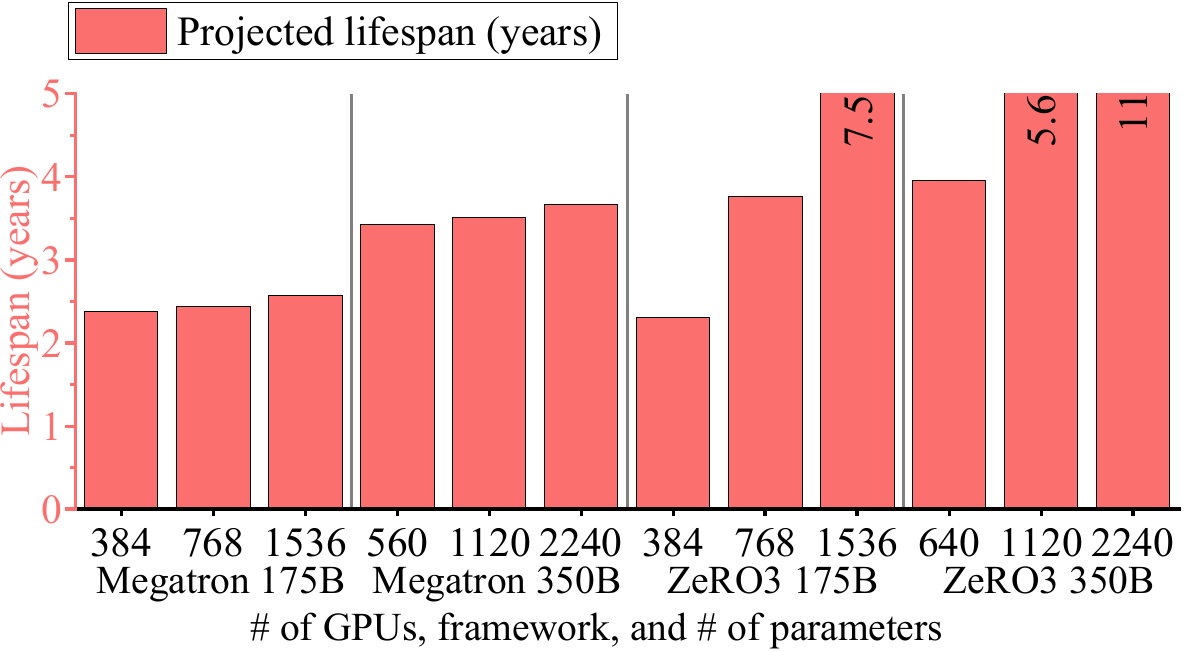}
\caption{\label{fig:projected_endurance_980pro_1TB} \kwc{Estimate of SSD lifespan in scenarios of Figure~\ref{fig:projected_perf_model} with the assumption that each A100 is paired with four Samsung 980 PRO 1TB.}}
\end{figure}

We study the SSD cost associated with adopting SSDTrain offloading in LLM systems.
To obtain the endurance in Figure~\ref{fig:projected_perf_model}, each A100 priced at US\$10K~\cite{dihuniNVIDIAA1009002100100000002021} is paired with a total of US\$6.4K worth of SSDs.
In the evaluation, we allocate seven Intel Optane P5800X for the two A100s. Although P5800X is more expensive than the models in Table~\ref{tab:ssd}, the price per PBW is comparable at US\$10.27~\cite{neweggIntelOptaneDC2021}.
We can further reduce the cost to a few percentage points by relaxing the data retention period: \kwc{For example, for all cases shown in Figure~\ref{fig:projected_perf_model}, Figure~\ref{fig:projected_endurance_980pro_1TB} shows that using four Samsung 980 PRO 1 TB for each A100 provides more than two years of SSD lifespan. The corresponding SSD cost is US\$360 per A100}~\cite{samsungSamsungVNANDSSD2021,bestbuySamsung980PRO2022}.
To have more durable storage for other data, the system may restrict the activation offloading to dedicated SSDs or utilize hardware equipped with Zoned Namespaces~(ZNS) standard~\cite{stavrinosDonBeBlockhead2021,hanZNSAdvancedZoned2021} to confine the wear within designated zones of physical blocks on the same SSD.

\kwc{Another significant cost factor is electricity. Each SSD costs around 20 watts, whereas a single GPU can easily draw several hundred watts. Taking other factors, e.g., commissioning, cooling, etc., into consideration~\cite{barrosoDatacenterComputerDesigning2019,redoakconsultingTotalCostOwnership2024}, the total cost of ownership~(TCO) of each GPU is an order of magnitude higher than its corresponding SSDs~\cite{dylanpatelNvidiaBlackwellPerf2024,sniaSNIAEnterpriseTCO2020}.}

\kwc{\subsubsection{Future Viability}
Will NVMe SSDs continue to be a good offloading target in the future? As shown in Figure~\ref{fig:gpu_trend_various}, historically, the PCIe bandwidth per lane has grown faster than the minimum requirement to keep up with the FP16 throughput, i.e., $\frac{5}{6}$ of the growth rate of FP16 throughput. The PCIe bandwidth has continued to grow rapidly, with new standards frequently released that double the bandwidth per lane of the previous version. 
Whenever a new PCIe standard is adopted, SSD vendors promptly release SSD products that provide the increased bandwidth aligned with the new PCIe standard. 
Therefore, the analysis we have done and all the conclusions we have drawn on SSDs in Chapter~\ref{ch:ssdtrain} will still be valid in the future.
}

\kwc{\subsubsection{Bringing It Altogether: System Design Decisions}

First, \textbf{Cost Analysis} shows the cost of SSDs in a system with SSDTrain enabled is an order of magnitude lower than that of GPUs. Therefore, adopting SSDs to enable SSDTrain, whether as an upgrade to existing on-premises clusters or in new on-premises machines, is profitable. The power supply should not be a problem: Typically, clusters have sufficient power redundancy to support upgrades that add new SSDs~\cite{barrosoDatacenterComputerDesigning2019}. However, adopting SSDs in cloud instances may not be profitable if high-throughput SSDs are too costly~\cite{googleGoogleCloudHyperdisk}.

Second, some clusters may have a lower \kwc{SSD-to-GPU} ratio.  
Two measures can be taken for such clusters. First, a portion of the processes on the node can use the CPU offloader (Figure~\ref{fig:software_arch}) to offload tensors to the CPUs. Second, the adaptive offloading mechanism (Figure~\ref{fig:adaptive}) measures the I/O bandwidth and determines the amount of tensors to be offloaded so as not to delay the training process. 

Let us conduct a case study on DGX H100 systems~\cite{nvidiaIntroductionNVIDIADGX2023}. Each DGX H100 node is equipped with dual-socket CPUs and eight GPUs. Within each DGX H100 node, in addition to 10 local NVMe SSDs, a significant number of PCIe lanes are allocated to storage network adapters that enable high-performance access to NVMe over Fabrics (NVMe-oF)~\cite{nvmexpressinc.NVMExpressMoves2016}. Since GDS supports both local NVMe SSDs and remote NVMe-oF SSDs, SSDTrain is compatible with both types of storage in the DGX H100 system. GDS still provides acceleration~\cite{handwikiSoftwareLustreFile2021,raviGPUDirectHDF52020} when the remote SSDs are purposed as a distributed file system, e.g., Lustre, and optimized file format is used, e.g., HDF5. Users can choose to offload activations to local SSDs when their bandwidth and capacity are sufficient. If additional bandwidth or capacity are needed, users may utilize remote SSDs when available and/or choose the host memory as an additional target, as discussed above.

}

\section{Related Work}\label{sec:ssdtrain_related}
\noindent
\textbf{Swapping and offloading.} Many LLM systems with offloading abilities are inference-only~\cite{kwonEfficientMemoryManagement2023,shengFlexGenHighThroughputGenerative2023,alizadehLLMFlashEfficient2024}. In inference, weights and KV-cache never change and are reused across iterations; researchers leverage this to enhance locality and memory efficiency. However, in LLM training, the weights are updated in each iteration, and all tensors change across the iterations. Some work avails offloading features~\cite{rajbhandariZeROinfinityBreakingGPU2021} for training but is mostly designed to accommodate larger models in a smaller system at the cost of performance. They lack the asynchronous data transfer ability to maintain performance.

Another direction is to offload data and the associated computation to the CPU~\cite{renZeROOffloadDemocratizingBillionScale2021,kamahoriFiddlerCPUGPUOrchestration2024,songPowerInferFastLarge2023}. The offloaded computation is relatively light, and the offloaded data include gradients, sparse elements in the weights, etc. 
Recognizing this direction, SSDTrain is made orthogonal because we offload the activations to SSDs via GDS to minimize the interference with the CPU. Activations are for gradient computation, which is compute-intensive and best done solely on GPUs. 

Before the massive adoption of LLMs, there is work on offloading data for deep learning~\cite{pengCapuchinTensorbasedGPU2020,wangSuperNeuronsDynamicGPU2018,baeFlashNeuronSSDEnabledLargeBatch2021,rhuVDNNVirtualizedDeep2016,huangSwapAdvisorPushingDeep2020}. Most of them offload data to main memory while some~\cite{baeFlashNeuronSSDEnabledLargeBatch2021} enable the GPU--SSD data path. LLM training is unique because massive parallelism and its implications on the memory use of optimizer states, gradients, and weights are fundamental to the design space. SSDTrain naturally supports multiple GPUs. Besides, we demonstrated its viability on clusters and introduced the ROK curve to help with the design choice. On the other hand, LLMs have such a high demand for computing power that it stimulates rapid development in specialized hardware, e.g., transformer engine~\cite{nvidiaNVIDIAH100Tensor2023}, and distributed frameworks. This is why we ensure good interoperability. In contrast, most earlier work in this direction is bound to a specific PyTorch version or a custom runtime with support to select layers.

\noindent
\textbf{Quantization and sparsity.} Some work on offloading uses quantization and/or sparsity to reduce the I/O size~\cite{shengFlexGenHighThroughputGenerative2023,alizadehLLMFlashEfficient2024,baeFlashNeuronSSDEnabledLargeBatch2021}.  To reduce computation, algorithms have been proposed to quantize parameters and introduce sparsity into the model~\cite{zaheerBigBirdTransformers2020,liuDejaVuContextual2023,kimSqueezeLLMDenseandSparseQuantization2024,dettmersSpQRSparseQuantizedRepresentation2023,frantarSparseGPTMassiveLanguage2023}. Mixture-of-Experts~(MoE)~\cite{shazeerOutrageouslyLargeNeural2017a} is in this direction as it sparsifies the token-to-neuron connection in the MLP to the token-to-expert connection. Some algorithms introduce structured sparsity, e.g., N:M~\cite{zhouLearningFinegrainedStructured2021} sparsity and 2:4~\cite{poolChannelPermutationsSparsity2021} sparsity. On the other hand, there are frameworks and specialized kernels to accelerate models with quantization and/or sparsity~\cite{galeMegaBlocksEfficientSparse2023,zhengSparTADeepLearningModel2022,galeSparseGPUKernels2020,shigangliEfficientQuantizedSparse2024}. Some kernels leverage specialized hardware, e.g., Ampere tensor core~\cite{chenDynamicFineGrainedStructured2023,mishraAcceleratingSparseDeep2021}.
These techniques are orthogonal to SSDTrain and can be used to alternate the model and accelerate the computation while using SSDTrain. Notably, given the hardware, the reuse factor to fully overlap the computation with PCIe transfer will change according to the new numerical format or sparsity access pattern. We believe that SSDTrain's adaptive offloading algorithm helps optimize the offload amounts in these cases.

\noindent
\textbf{Optimized kernels.} Previous work develops optimized kernels to accelerate LLM~\cite{daoFlashAttentionFastMemoryEfficient2022,daoFlashAttention2FasterAttention2023,nvidiaNVIDIATensorRTLLM2023}. Some kernels utilize special hardware~\cite{nvidiaNVIDIATransformerEngine2023}. SSDTrain's interoperability ensures it can be used easily with these and upcoming techniques.  

\section{Conclusion}

The growth rate of the GPU memory capacity has not been able to keep up with that of the size of LLMs,
hindering the model training process. In particular, activations---the intermediate tensors produced during forward propagation and reused in backward propagation---dominate the GPU memory use. To address this challenge, we propose SSDTrain to efficiently offload activations to high-capacity NVMe SSDs. This approach reduces GPU memory usage without impacting performance by adaptively overlapping data transfers with computation. SSDTrain is compatible with popular deep learning frameworks such as PyTorch, Megatron, and DeepSpeed and employs techniques such as tensor deduplication, forwarding, and adaptive offloading to further enhance efficiency. We extensively tested popular LLMs such as GPT, BERT, and T5. The results demonstrate that SSDTrain effectively reduces 47\% of the activation peak memory usage. At the same time, SSDTrain perfectly overlaps the I/O with the computation and incurs negligible performance overhead. We introduce the ROK curve to compare the SSDTrain offloading with two other tensor placement strategies, keeping activations in GPU memory and layerwise full recomputation.  SSDTrain achieves better memory savings than layerwise full recomputation while retaining the performance of keeping the activations in memory. \kwc{We further analyze how SSDTrain increases training throughput by increasing micro-batch size and reducing pipeline bubbles.}

\chapter{Discussion and Future Work}\label{ch:future_work}

\kwc{Before concluding this dissertation in Chapter~\ref{ch:conclusion}, this chapter provides a final discussion on the contributions made in this work and \kwc{introduces} potential future directions. Section~\ref{sec:discuss_pytorch_integration} examines the advantages and limitations of various approaches to integrate techniques into the PyTorch stack. Section~\ref{sec:future_training} explains how future work can further our investigation into data-efficient deep learning training. Lastly, Section~\ref{sec:future_tabular} \kwc{elaborates on} how the optimizations proposed in this dissertation can be applied to other data-intensive workloads, particularly tabular data analysis.}

\section{\kwc{Discussion on Integrating Techniques into the PyTorch Stack}\kwa{(New Section)}}
\label{sec:discuss_pytorch_integration}

In this dissertation, we propose and implement three projects: Hector, PyTorch-Direct, and SSDTrain. All are incorporated into the PyTorch stack in different ways. 
PyTorch-Direct wraps the zero-copy-enabled dispatch ruleset into a full-fledged unified tensor type and incorporates that into the PyTorch C++ runtime, which requires recompiling the PyTorch \kwc{source code}~(Section~\ref{sec.PyTorch_direct.Overview}).
Hector generates the kernels, compiles them as a PyTorch extension library, and loads them before training. The code generator and auxiliary logic, e.g., graph loading, are in Python~(Section~\ref{sec:hector_overview}).
SSDTrain has all logic in Python, except for an interposed library to register memory in GDS during device memory allocation and deregister the memory during deallocation~(Section~\ref{sec:ssdtrain_overview}). The software components of the three works are shown in Table~\ref{tab:components_integration}. 

Similarly to Hector, most of the literature incorporating changes into the PyTorch \kwc{runtime} \kwc{creates} Python extension \kwc{libraries} to achieve this, e.g., DeepSpeed~\cite{rasleyDeepSpeedSystemOptimizations2020}, Megatron~\cite{shoeybiMegatronLMTrainingMultiBillion2020a}, Graphiler~\cite{xieGraphilerCompilerGraph}. Similarly to PyTorch-Direct, some projects make changes to the PyTorch \kwc{source code} and recompile it to incorporate extensive modifications to the PyTorch runtime. For example, FlashNeuron~\cite{baeFlashNeuronSSDEnabledLargeBatch2021} introduces the tensor offloading mechanism into PyTorch. PopTorch~\cite{anthonybarbierAddNewKeys2022} incorporates support for GraphCore's accelerator, which requires adding a new dispatch key.

Unlike Python extension \kwc{libraries} and interposed libraries, modifying and recompiling the PyTorch \kwc{source code} usually requires consistent efforts to keep up with the latest PyTorch changes in the long run, especially when the changes are maintained in an out-of-tree repository. Merging modifications to the official PyTorch repository will alleviate such consistent efforts, if possible. Therefore, for research projects, modifying the PyTorch \kwc{source code} is advisable only when the other two methods are insufficient in adding the required functionality, e.g., adding new dispatch keys. 
In light of this, our SSDTrain project is carefully developed without modifying the PyTorch \kwc{source code}, unlike other projects such as FlashNeuron, as discussed in Section~\ref{sec:ssdtrain_related}.
As for the PyTorch-Direct project, changes in the PyTorch \kwc{source code} are required to incorporate the GPU-centric paradigm in exchange for keeping PyTorch's original programming interface. We have worked with the DGL team to integrate the particular optimized transfer scheme into the DGL repository~\cite{minjiewangReleaseV080Dmlc2022,seungwonminDocAddOfficial2021,seungwonminFeatureAddMultiGPU2021,seungwonminFeaturePerformanceGPUIntroducingUnifiedTensor2021,xinyaoDglDGLGraphpin_memory_2022} so that the optimized scheme can be activated through explicit new APIs without the need to recompile modified PyTorch \kwc{source code}.

\begin{table}[]
\centering
\begin{tabular}{lcccc} 
\toprule
& \begin{tabular}[c]{@{}c@{}}Modified and \\ recompiled\\  PyTorch \\ \kwc{source code}\end{tabular} & \begin{tabular}[c]{@{}c@{}}Python\\ extension\\\kwc{library}\end{tabular} & \begin{tabular}[c]{@{}c@{}}Interposed\\ library\end{tabular} & \begin{tabular}[c]{@{}c@{}}Python\\ code\end{tabular} \\ \midrule
\textbf{Hector}         &   &\Checkmark &   & \Checkmark  \\
\textbf{PyTorch-Direct} & \Checkmark &  &   &  \\
\textbf{SSDTrain}       &   & & \Checkmark  & \Checkmark  \\
DeepSpeed~\cite{rasleyDeepSpeedSystemOptimizations2020}        &   &\Checkmark &   & \Checkmark  \\
FlashNeuron~\cite{baeFlashNeuronSSDEnabledLargeBatch2021} & \Checkmark &  &   &  \\
Graphiler~\cite{xieGraphilerCompilerGraph}       &   &\Checkmark &   & \Checkmark  \\
Megatron~\cite{shoeybiMegatronLMTrainingMultiBillion2020a}       &   &\Checkmark &   & \Checkmark  \\
PopTorch~\cite{anthonybarbierAddNewKeys2022} & \Checkmark &  &   &  \\
\bottomrule
\end{tabular}
\caption{Comparing software components of PyTorch-Direct, Hector, SSDTrain, and existing work.\label{tab:components_integration}}
\end{table}

\section{Further Exploration in Deep Learning Training}
\label{sec:future_training}
\kwc{Cost modeling and inter-operator scheduling are two key areas to deepen our exploration in deep learning training.} Cost modeling helps choose the optimized design in the efficient frontier of the design space complicated by data efficiency.
Inter-operator scheduling helps hide the latency of memory accesses and data transfers with other operators.

\subsection{Cost Models}
\noindent
\textbf{For Hector, devise algorithms to select layouts, optimizations, and schedules according to model, input graph, and GPU architecture.} One of the most important compiler research problems is the algorithm that makes choices among candidates in the design space. It remains an open \kwc{problem} how the data-dependent sparse operations and layout choices fit in the cost model and layout choices. \kwc{Pertinently, in various applications in high-performance computing (HPC) with multiple layout and kernel choices, researchers have developed heuristics to make the optimized choice}~\cite{zhaoBridgingGapDeep2018,almasriParallelizingMaximalClique2023,kawtikwarHyLACHybridLinear2024}. Besides, the specific microarchitecture of each GPU model also makes a difference due to the architecture-specific features available, e.g., asynchronous loading to shared memory since Ampere~\cite{nvidiaControllingDataMovement2020}, and different microarchitecture characteristics in each model. Therefore, it is meaningful to investigate their impact and incorporate them into decision-making.

\noindent
\textbf{For SSDTrain, devise algorithms to pick the optimized design choice in the combined design space of both LLM parallelism strategies and tensor placement strategies.} SSDTrain demonstrates that offloading opens up design choices on the efficient frontier, given a parallelism strategy. With the memory savings from SSDTrain offloading, we may choose a new LLM parallelism strategy with higher throughput at the cost of more per-GPU memory use. \kwc{For example, as mentioned, the larger amount of activations SSDTrain allows to accommodate can be allocated to enlarge the number of micro-batches and/or to enlarge the micro-batch size.
On the other hand, pipeline parallelism brings about bubbles of idleness of the device, which could be mitigated by a larger number of micro-batches~\cite{shoeybiMegatronLMTrainingMultiBillion2020a}.
Both the throughput boost by increased micro-batch size and that by increased number of micro-batches saturate at a point, leaving the optimized strategy to allocate activations memory given parallelism configurations an intriguing question to explore.}\kwa{(Sentences moved from Section~\ref{sec:discussion}.)} \kwc{A broader and more general challenge} is how to systematically explore the combined design space of both LLM parallelism and tensor placement strategies and find the optimized design choice. In addition to throughput, TCO is an essential target. For example, it is valuable to understand the minimal SSD requirements for a particular scenario and the upgrade cost from an existing cluster configuration.

\subsection{Inter-Operator Scheduling}
\subsubsection{Leveraging CUDA Graph}
In its latest systems software stack, Nvidia provides CUDA Graph as a performant task graph runtime. CUDA Graph reduces the launch overhead of kernels and schedules and executes tasks in the graph while their dependencies are preserved. We use CUDA Graph for low-overhead inter-operator scheduling. 

\noindent
\textbf{Hide memory latency of sparse operations by enhancing intra-SM parallelism via CUDA Graph.} We have observed that both GNNs and LLMs involve a mixture of sparse operations and dense operations: for GNNs, we have broken down the models to GEMM kernels and traversal kernels; for LLMs, the layers are typically dense if neither specific design, e.g., mixture-of-experts~\cite{zhouMixtureofExpertsExpertChoice2022}, is performed nor pruning is done, but the output of the activation layers is typically sparse by its nature.

The mixture of dense and sparse operations allows us to hide the memory latency of sparse operations by running dense and sparse operations in parallel. In particular, we will break down sparse and dense operations into smaller kernels and schedule them so that both dense and sparse kernels are run on the same SM simultaneously. For example, GEMM and SpMM can be broken down by partitioning the input matrices into blocks and performing matrix multiplication among block pairs before reduction. To reduce launch overhead, we use the CUDA graph to manage task dependencies and execute the series of kernels.

\noindent
\textbf{For GNNs, optimize data movement in mini-batch training.} Graphs not fitting into GPU memory must stay in host memory or even storage during RGNN execution. In each step, the subgraphs are sampled and transferred to the GPU. With knowledge of graph semantics, data layout, and operator-specific schedules, Hector can help improve the scheduling of sampling and data transfer and generate CUDA kernels that gather data from the host memory on the fly~\cite{minLargeGraphConvolutional2021}.

\subsubsection{Leveraging Warp Specialization}
During backward propagation, the system needs to compute the gradient of both the weights and the input for each layer. This doubles the cost compared to forward propagation. On the other hand, the computation of the two gradients uses identical tensors, creating an opportunity yet to be leveraged to reuse data across the calculation of the two gradients.

\section{\kwc{Applying Techniques to Tabular Data Analysis}\kwa{(New Section)}}
\label{sec:future_tabular}
Data tables have been widely adopted in data analytics and machine learning pipelines. Data analytics aims to gain insights from massive data, where data tables are a core data structure. In SQL database systems, tables are essential elements to organize raw data and outputs of each query; in many data processing libraries and languages, such as pandas and R, data tables are the fundamental class as well. In machine learning pipelines, data tables hold the data, at least during preprocessing, before input to the machine learning model. The preprocessing stages involve ETL (extract, transform, and load) and feature engineering. \kwc{Preprocessing may} reoccur in data streaming scenarios or when iteratively refining the algorithm. Data processing takes a substantial amount of time: 80\%—90\% of the work time of a data scientist is dedicated to processing data~\cite{kaggle2017KaggleMachine, crowdflower2017DataScientist2017}. 

Thanks to the high bandwidth of device memory and a massive number of processing units, GPUs could greatly help analytical workloads that typically involve many simple homogeneous operations. Aligned with this direction, many GPU-optimized databases have been established recently, involving Brytlyt, Kinetica, OmniSci (formerly MapD), SQream, etc.~\cite{DoingRealityCheck2018} 
Nvidia released the RAPIDS Python suite to allow developers to run end-to-end data analytics and data science pipelines on the GPU~\cite{nvidiaRAPIDSGPUAcceleratedData2018}. Central to it is the cudf package, which is the CUDA equivalent of the data table Python package pandas. In cudf, in-memory data tables are in columnar format. Other packages in RAPIDS, e.g., BlazingSQL, cuGraph, etc., enable SQL queries, graph analytics, etc., by using cudf to store data in data tables. 

Similarly to deep learning training, tabular data analysis is data-intensive~\cite{caoGPUDatabaseSystems2023}. Data table operations typically have small arithmetical intensity, e.g., comparing the values of two columns and light arithmetic computation of a few cells for each row. Besides, real-world tabular data analysis usually involves massive data, rendering the limited GPU HBM memory capacity a problem~\cite{xiangyaoyuGPUDatabasesNew2024}.

The techniques proposed in this dissertation can also be applied to tabular data analysis. \kwc{As an} example, \kwc{the following} explains how code generation with flexible data access schemes proposed by the Hector project could help tabular data analysis with indexes.
Index is an essential optimization in tabular data analysis~\cite{optimizdbaDatabaseOptimizationTechniques2018,oracleMySQLMySQL842024}: Index stores the presorted results of a column or multiple columns and the mapping from the result to the row index in the original table. Many data table operations can be accelerated using the index to save computation.
Nevertheless, GPU-accelerated tabular data analysis software has limited support on indices~\cite{IndexTable2023,sqreamSQreamsUniqueArchitecture2024}.
By introducing a Hector-like code generator with optional indirect addressing by index, the software gets 1)~optimized kernel development cost \kwc{without the need to maintain kernel variants of the same operator} and 2)~free of intermediate data tables when doing indirect addressing.
Such optimization is aligned with kernel fusion for tabular data analysis~\cite{caoGPUDatabaseSystems2023,palkarEvaluatingEndendOptimization2018}, and the optimizations to avoid materialization of intermediate data tables~\cite{nakandalaTensorCompilerUnified2020}. However, none of the\kwc{se} existing projects support the index.

\chapter{Conclusion}\label{ch:conclusion}

Due to both demand from the workload and hardware advancements, it becomes increasingly critical to ensure data efficiency in deep learning training. 
Data inefficiency in deep learning training arises from the data-intensive nature of workloads and the oversimplification inherent in the PyTorch computing stack.
To effectively mitigate data inefficiency for deep learning training, this dissertation analyzes data inefficiency in representative deep training tasks, specifically in GNNs and LLMs. It then proposes novel runtime and code generation techniques to mitigate these challenges and implements these optimizations seamlessly within the PyTorch stack while maintaining strong programmability and interoperability.

First, this dissertation \kwc{devises} the Hector IR and code generator. By introducing domain-specific high-level abstraction and code generation, Hector systematically addresses significant performance challenges due to the inherent memory intensiveness, the gap between the programming interface and kernel APIs, and the high kernel optimization cost due to the kernel coupling with layout and heterogeneity.

Then, this dissertation \kwc{designs and implements} PyTorch-Direct to incorporate the GPU-centric PCIe data transfer paradigm in PyTorch for GNN training. PyTorch-Direct significantly reduces CPU utilization, resulting in higher end-to-end training performance.

Last, LLM training systems are increasingly constrained by GPU memory, with activations being one of the primary culprits. This dissertation \kwc{creates} the SSDTrain activations offloading framework with a direct GPU–SSD data path and good interoperability.

This dissertation proves that code generation and runtime techniques \kwc{can} effectively mitigate data inefficiency in deep learning training.

\appendix

\backmatter

\bibliographystyle{IEEE_ECE}
\bibliography{ref/Hector/final_kuns_manually_revised_based_on_zotero_generated,ref/Hector/ref, ref/Hector/thesis, ref/Proposal/AICE2023,ref/Proposal/IIDAI2023,ref/Proposal/NextSteps,ref/Proposal/CUDASchedulingPolicy,ref/SSDTrain/manual,ref/SSDTrain/ManuallyValidatedFlashtrainfromZotero,ref/PyTorchDirect/ref,ref/Background,ref/PyDArXiv/example_paper,ref/FutureWork}

\end{document}